\documentclass[fleqn,usenatbib,useAMS]{Papers}
\usepackage{newtxtext,newtxmath}
\usepackage[T1]{fontenc}
\usepackage{subfloat}
\usepackage{makecell}
\usepackage[authoryear]{natbib}
\usepackage{float}
\usepackage{graphicx}	
\usepackage{amsmath}	

\title[Effect of mass cascade on halo deformation, energy, size, and density]{Inverse mass cascade in dark matter flow and effects on halo deformation, energy, size, and density profiles}

\author[Z. Xu]{
Zhijie (Jay) Xu,$^{1}$\thanks{E-mail: zhijie.xu@pnnl.gov; zhijiexu@hotmail.com}
\\
$^{1}$Physical and Computational Sciences Directorate, Pacific Northwest National Laboratory; Richland, WA 99352, USA
}
\date{Accepted XXX. Received YYY; in original form ZZZ}

\pubyear{2022}

\begin{document}
\label{firstpage}
\pagerange{\pageref{firstpage}--\pageref{lastpage}}
\maketitle

\begin{abstract}
Inverse mass cascade is a key feature of the intermediate statistically steady state for self-gravitating collisionless dark matter flow (SG-CFD). This paper focus on effects of mass cascade on halo energy, momentum, dispersion, size, and density. Halo with fast mass accretion has an expanding core. Mass cascade forms a new layer of mass that deforms the original halo and induces nonzero radial flow (outwards in core and inwards in outer regions). The inward/outward flow leads to an extra length scale (scale radius) that is not present in isothermal profile. Halo concentration c=3.5 can be derived for fast growing halos. For cusp-core controversy, a double-power-law density is proposed as a result of nonzero radial flow. The inner/outer density are controlled by halo deformation rate and halo growth, respectively. The slower deformation at center, the steeper density. For fast growing halos, radial flow at center is simply Hubble flow that leads to the existence of central core. Mass cascade leads to nonzero halo surface energy/tension and radial flow that enhances dispersion in outer region. An effective exponent of gravity $n_e$=-1.3 (not -1) is obtained due to halo surface energy. Evolution of halo size follows a geometric Brownian motion and lognormal distribution. The Brownian motion of particles in evolving halos leads to Fokker-Planck equations for particle distribution that is dependent on the radial and osmotic flow. Complete solutions of particle distribution are presented based on a simple model of osmotic flow. The proposed model agrees with simulation for various halo group sizes. With reference pressure/density defined at center, equation of state can be established for relative pressure/density. Pressure, density, and dispersion at halo center are presented. The core size $x_c$ is obtained where Hubble flow is dominant. Simple closures are proposed for self-consistent halo density.
\end{abstract}

\begin{keywords}
\vspace*{-15pt}
Dark matter; N-body simulations; Theoretical models
\end{keywords}

\begingroup
\let\clearpage\relax
\tableofcontents
\endgroup

\section{Introduction}
\label{sec:1}
The self-gravitating collisionless fluid dynamics (SG-CFD) is the study of motion of collisionless matter under the influence of its own gravity. A typical example is the large-scale gravitational collapse of collisionless system \citep{Lukic:2007-The-halo-mass-function--High-r}. The self-organization of self-gravitating collisionless matter leads to the formation and evolution of large-scale structures due to the gravitational instability. Highly localized and virialized halos are major manifestation of nonlinear gravitational collapse \citep{Neyman:1952-A-Theory-of-the-Spatial-Distri,Cooray:2002-Halo-models-of-large-scale-str} and the building blocks of large-scale structures. 

By contrast, incompressible hydrodynamics also develops instability if Reynolds number is sufficiently high, where turbulence starts to initiate and develop. The "eddies", building blocks of turbulence, are formed at different length scales and interacting with each other, as described by a famous poem :"Big whirls have little whirls, That feed on their velocity; And little whirls have lesser whirls, And so on to viscosity" \citep{Richardson:1922-Weather-Prediction-by-Numerica}. Large eddies feed smaller eddies, which feed even smaller eddies, and then lead to viscous dissipation at the smallest scale, i.e. the concept of a direct energy cascade. While direct energy cascade is the key feature of three-dimensional turbulence, two-dimensional turbulence possesses a range of scales over which kinetic energy is transferred from small to large scales , i.e. an inverse energy cascade \citep{Kraichnan:1967-Inertial-Ranges-in-2-Dimension}. 

The similarity between "eddies" in turbulence and "halos" in dark matter flow (SG-CFD) allows a new poem by simply replacing "whirls" with "halos". "Little halos have big halos, That feed on their mass; And big halos have greater halos, And so on to growth". This picture describes the inverse mass cascade in dark matter flow \citep{Xu:2021-Inverse-mass-cascade-mass-function}. There exists a broad spectrum of halo size. Small halos are created, interacting, and merging with other halos. Halos pass their mass onto larger and larger halos, until halo mass growth becomes dominant over mass propagation. 

While "eddy" is not a well-defined object in turbulence literature, "halos" are well-defined dynamical objects, whose abundance and internal structure have been extensively studied over several decades. The abundance of halos is described by a halo mass function, a fundamental quantity to model structure formation and evolution. The seminal Press-Schechter (PS) model \citep{Press:1974-Formation-of-Galaxies-and-Clus,Bond:1991-Excursion-Set-Mass-Functions-f} allows one to predict the shape and evolution of mass function. This model relies on a threshold value of density contrast that can be analytically derived from the nonlinear collapse of a spherical top hat over-density \citep{Tomita:1969-Formation-of-Gravitationally-B,Gunn:1972-Infall-of-Matter-into-Clusters}. Further improvement was achieved by extending the PS formalism to elliptical collapse \citep{Sheth:2001-Ellipsoidal-collapse-and-an-im,Sheth:1999-Large-scale-bias-and-the-peak-}. In addition, halo mass function can be interpreted as an intrinsic distribution to maximize system entropy during statistically steady state of dark matter flow \citep{Xu:2021-The-maximum-entropy-distributi,Xu:2021-Mass-functions-of-dark-matter-}.

The internal structure of halos is primarily described by the halo density profile, another important quantity for structure formation and evolution \citep{Del_Popolo:2009-Density-profiles-of-dark-matte}. Structure of halos can be studied both analytically and numerically with \textit{N}-body simulations \citep{Moore:1998-Resolving-the-structure-of-col,Klypin:2001-Resolving-the-structure-of-col}. The spherical collapse model relates assumed power-law density with the initial density fluctuations, which can be dependent on the effective index of the power spectrum from linear theory. This simple similarity model leads to an isothermal density profile for virialized halos. However, high-resolution \textit{N}-body simulations of structure formation have shown that the simulated halos have a density shallower than the isothermal profile at smaller radius and steeper at larger radius \citep{Navarro:1997-A-universal-density-profile-fr,Navarro:2004-The-inner-structure-of-ACDM-ha}. Many effects might contribute to this deviation. The effect of halo mass cascade (accretion) is one of the most critical effect that is absent, which renders this simple similarity model invalid. We will discuss the effect of mass cascade on halo density profile in detail (see Section \ref{sec:3.3}). 

By revisiting fundamental ideas of turbulence, the inverse mass/energy cascade can be mathematically formulated and briefly reviewed here \citep{Xu:2021-Inverse-mass-cascade-mass-function,Xu:2021-Inverse-and-direct-cascade-of-}. Mass cascade is local, two-way, and asymmetric in mass space. Halos inherit/pass their mass mostly from/to halos of similar size. The net mass transfer proceeds in a "bottom-up" fashion. Two distinct ranges can be identified, i.e. a propagation range with a scale-independent rate of mass transfer $\varepsilon _{m} $ and a deposition range with cascaded mass consumed to form and grow halos. A fundamental merging frequency $f_{0} \sim m_{p}^{\left(\lambda -1\right)} a^{-\tau _{0} } $ between two single mergers of elementary mass $m_{p}$ can be identified, where \textit{a} is the scale factor, $m_{p} $ is the particle mass, $\lambda $ and $\tau _{0} $ are two key mass cascade parameters that may be dependent on the exact cosmology model. The waiting time $\tau _{g} $ (halo lifespan) for halos to pass their mass to larger halos scales as $\tau _{g} \sim m_{h}^{-\lambda } a^{\tau _{0} } $. Consequently, the everlasting inverse mass cascade with a scale-independent mass transfer rate $\varepsilon _{m} \sim a^{-\tau _{0} } $ in the propagation range is a distinct feature of the intermediate statistically steady state of dark matter flow. Entire mass cascade was also formulated as random-walk of halos in mass space \citep{Xu:2021-Inverse-mass-cascade-mass-function}. This results in a heterogeneous diffusion model with position-dependent diffusivity, where mass function can be analytically derived without relying on any specific collapse model. 

In addition, the elementary step of mass cascade, i.e a two-body collapse \citep{Xu:2021-A-non-radial-two-body-collapse}, the evolution of halo mean flow, velocity dispersion \citep{Xu:2022-The-mean-flow--velocity-disper}, and halo momentum and energy \citep{Xu:2022-The-evolution-of-energy--momen} were studied in separate papers, along with the correlation-based statistical theory for correlation and structure functions in dark matter flow \citep{Xu:2022-The-statistical-theory-of-2nd,Xu:2022-The-statistical-theory-of-3rd,Xu:2022-Two-thirds-law-for-pairwise-ve}. This is an important topic with potential relevance to dark matter particle mass and properties \citep{Xu:2022-Postulating-dark-matter-partic}, MOND (modified Newtonian dynamics) theory \citep{Xu:2022-The-origin-of-MOND-acceleratio}, and baryonic-to-halo mass relation \citep{Xu:2022-The-baryonic-to-halo-mass-rela}.

This paper focus on the effects of inverse mass cascade on halo energy, momentum, size, and internal structure. Especially, it is still not clear why halos that form in SG-CFD have nearly universal profiles. We will demonstrate that the radial flow lead to an extra length scale (scale radius) for density profile where the radial flow is at its maximum. A double-power-law density is a natural result with inner density dominated by halo deformation rate and outer density controlled by halo growth. There exists a limiting halo concentration for large halos as a result of vanishing linear moment. The effects of mass cascade on velocity dispersion and surface energy are explicitly discussed and presented. Stochastic models for halo size and particle motion in halos are also discussed along with the equation of state for halos. The rest of this paper is organized as follows: Section \ref{sec:2} introduces the simulation and numerical data, followed by the effects of mass cascade on halo properties in Section \ref{sec:3}. Stochastic models for halo size and random-walk of collisionless particles in halos are presented in Section \ref{sec:4} with complete solutions provided. 

\section{N-body simulations and numerical data}
\label{sec:2}
The numerical data for this work is publicly available and generated from \textit{N}-body simulations carried out by the Virgo consortium. A comprehensive description of simulation data can be found in \citep{Frenk:2000-Public-Release-of-N-body-simul,Jenkins:1998-Evolution-of-structure-in-cold}. The same set of simulation data has been widely used in a number of different studies from clustering statistics \citep{Jenkins:1998-Evolution-of-structure-in-cold} to the formation of halo clusters in large scale environments \citep{Colberg:1999-Linking-cluster-formation-to-l}, and testing models for halo abundance and mass functions \citep{Sheth:2001-Ellipsoidal-collapse-and-an-im}. More details on simulation parameters are provided in Table \ref{tab:1}.

Two relevant datasets from this N-boby simulation, i.e. halo-based and correlation-based statistics of dark matter flow, can be found at Zenodo.org  \citep{Xu:2022-Dark_matter-flow-dataset-part1, Xu:2022-Dark_matter-flow-dataset-part2}, along with the accompanying presentation slides, "A comparative study of dark matter flow \& hydrodynamic turbulence and its applications" \citep{Xu:2022-Dark_matter-flow-and-hydrodynamic-turbulence-presentation}. All data files are also available on GitHub \citep{Xu:Dark_matter_flow_dataset_2022_all_files}.

\begin{table}
\caption{Numerical parameters of N-body simulation}
\begin{tabular}{p{0.25in}p{0.05in}p{0.05in}p{0.05in}p{0.05in}p{0.05in}p{0.4in}p{0.1in}p{0.4in}p{0.4in}} 
\hline 
Run & $\Omega_{0}$ & $\Lambda$ & $h$ & $\Gamma$ & $\sigma _{8}$ & \makecell{L\\(Mpc/h)} & $N$ & \makecell{$m_{p}$\\$M_{\odot}/h$} & \makecell{$l_{soft}$\\(Kpc/h)} \\ 
\hline 
SCDM1 & 1.0 & 0.0 & 0.5 & 0.5 & 0.51 & \centering 239.5 & $256^{3}$ & 2.27$\times 10^{11}$ & \makecell{\centering 36} \\ \hline 
\end{tabular}
\label{tab:1}
\end{table}

\section{Effects of mass cascade on halo properties}
\label{sec:3}
\subsection{Halo density profiles}
\label{sec:3.1}
\begin{figure}
\includegraphics*[width=\columnwidth]{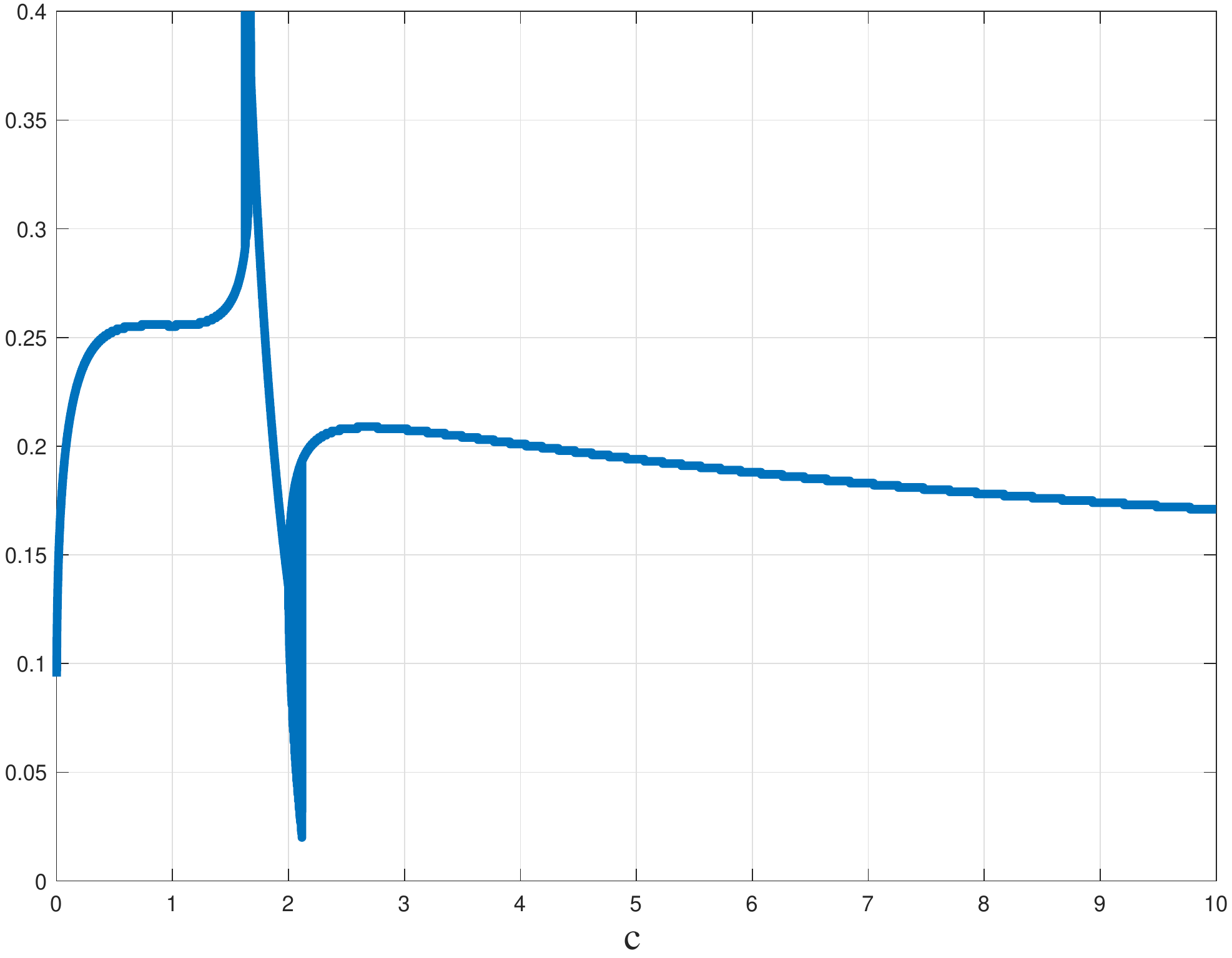}
\caption{The variation of shape parameter $\alpha$ of an Einasto profile with the halo concentration parameter c. Both NFW and Einasto profiles are assumed to have the same density at halo surface. This $c$-$\alpha$ relation is obtained by numerically solving Eq. \eqref{ZEqnNum720481}. Note that there is a discontinuity at c=2. Concentration parameter $\alpha$ is on the order of 0.2 and slowly decreases with increasing c for small halos.} 
\label{fig:S1}
\end{figure}

The formation of halos is a complex, hierarchical, and nonlinear process. However, the radial density profile $\rho _{h} \left(r\right)$ of halos can be robustly fitted by relatively simple functions from cosmological \textit{N}-body simulations. This section briefly reviews the NFW profile \citep{Navarro:1997-A-universal-density-profile-fr}, Einasto profile \citep{Einasto:1984-Structure-of-Superclusters-and} and power-law density profile (see Appendix \ref{appendix:a}). Especially, the isothermal profile is a direct result of infinitesimal lifetime or extremely fast mass accretion with vanishing radial flow (see Fig. \ref{fig:2} and Eq. \eqref{ZEqnNum852508}). 
Both NFW and Einasto profiles involve a halo concentration parameter $c={r_{h}/r_{s} }$, where $r_{h} $ and $r_{s}$ are the halo size and scale radius. Simulations have shown that the concentration $c={r_{h}/r_{s} } $ can be dependent on both halo mass and redshift. The evolution of \textit{c} depends very much on the mass accretion rate and the faster the halo grows, the slower \textit{c} increases. A constant value of \textit{c} is expected for large halos with extremely fast mass accretion and short lifespan, where $c\approx 4$ was estimated as a limiting value for large halos from \textit{N}-body simulations \citep{Zhao:2009-Accurate-Universal-Models-for-,Correa:2015-The-accretion-history-of-dark-}. The inner structures of these halos are still being dynamically adjusted due to fast mass accretion. 
Figure \ref{fig:S1} plots the variation of shape parameter $\alpha $ of an Einasto profile with the concentration \textit{c} by numerically solving Eq. \eqref{ZEqnNum720481}. The shape parameter $\alpha $ decreases from 0.2 to 0.155 for concentration \textit{c} varying from 4 to 10, i.e. $\alpha $ increases with increasing halo mass that is consistent with simulations \citep{Gao:2008-The-redshift-dependence-of-the}.

\subsection{Effects of mass cascade on halo deformation and radial flow}
\label{sec:3.2}
\begin{figure}
\includegraphics*[width=\columnwidth]{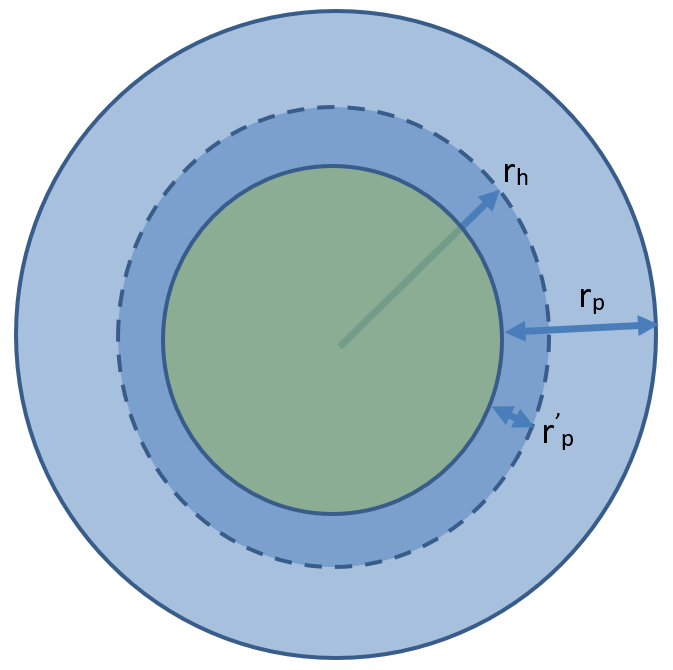}
\caption{Schematic plot of halo mass accretion and size change during an infinitesimal time interval $dt$. Original halo has a size $r_{h} $ at time \textit{t} (the dash line in the plot). By the time $t+dt$, the halo size will increase from $r_{h} $ to $r_{h} -r_{p}^{'} +r_{p} $ due to the mass accretion. The original halo at time \textit{t} deforms to the new size of  $r_{h} -r_{p}^{'} $ (green). Halo mass accretion/cascade creates a new layer of mass around the original halo with a thickness of $r_{p} $. This layer of mass potentially deforms the original halo to a new size (green) due to the gravitational interaction, which creates a non-zero radial flow of mass. Special case $r_{p}^{'} =0$ (no radial flow) leads to an isothermal density profile.} 
\label{fig:1}
\end{figure}

Large halos with extremely short lifespan should have fast mass accretion rate. At the same redshift \textit{z}, these halos are dynamical objects with a constant mean density regardless of their masses. The fast halo mass accretion with short lifespan during mass cascade affects the halo density profiles by creating a non-zero radial flow. To quantitatively formulate this idea, let's first consider the time variation of mass of these halos,  
\begin{equation} 
\label{ZEqnNum872917} 
m_{h}^{} \left(a\right)=\frac{4}{3} \pi r_{h}^{3} \Delta _{c} \bar{\rho }_{0} a^{-3} ,         
\end{equation} 
where $\Delta _{c} =18\pi ^{2} $ is a critical density ratio that can be obtained from spherical collapse model or a two-body collapse model \citep[see][Eq. (89)]{Xu:2021-A-non-radial-two-body-collapse}. Here $\bar{\rho }_{0} $ is the background density at the current epoch of $a=1$. The halo size $r_{h} \left(a\right)$ is defined as halo virial radius. Equation \eqref{ZEqnNum872917} implies that circular velocity at surface of halos satisfies (with $H^{2} a^{3} ={8\pi G\bar{\rho }_{0}/3} =H_{0}^{2} $ for matter dominant model), 
\begin{equation} 
\label{ZEqnNum837340} 
v_{cir}^{2} =\frac{Gm_{h} }{r_{h} } =4\pi ^{2} \frac{r_{h}^{2} }{t^{2} } =\left(3\pi Hr_{h} \right)^{2},
\end{equation} 
where $H\left(a\right)$ and $H_{0} $ are the Hubble parameter at scale factor \textit{a} and Hubble constant at the current epoch. The following relation for variation of halo size $r_{h} $ with \textit{a} can be obtained from Eq. \eqref{ZEqnNum872917},   
\begin{equation} 
\label{ZEqnNum728607} 
\frac{\partial \ln r_{h} }{\partial \ln a} =\frac{1}{3} \frac{\partial \ln m_{h} }{\partial \ln a} +1=\frac{a}{3m_{h} } \frac{\partial m_{h} }{\partial a} +1.        
\end{equation} 
The time variation of typical halos of mass $m_{h} $ can be expressed in terms of time scale $\tau _{g}$ \citep[see][Eq. (6)]{Xu:2021-Inverse-mass-cascade-mass-function}, 
\begin{equation}
\label{ZEqnNum808457} 
\frac{\partial m_{h} }{\partial a} =\frac{1}{Ha} \frac{\partial m_{h} }{\partial t} =\frac{m_{p} }{\tau _{g} Ha} ,  \end{equation} 
where $\tau _{g} \left(m_{h} ,a\right)$ is the mean waiting time (lifespan) of a given halo for merging with a single merger of mass $m_{p} $ and passing its mass to large scale. After inserting Eq. \eqref{ZEqnNum808457} into Eq. \eqref{ZEqnNum728607}, 

\begin{equation} 
\label{ZEqnNum587991} 
\frac{\partial \ln r_{h} }{\partial \ln a} =\frac{m_{p} }{3\tau _{g} Hm_{h} } +1=\frac{1}{3\tau _{g} Hn_{p} } +1,       
\end{equation} 

\noindent where $n_{p} ={m_{h}/m_{p} } $ is the number of particles in that halo. Next consider the halo density at the surface of halos (as shown in Fig. \ref{fig:1}),
\begin{equation} 
\label{ZEqnNum227253} 
\rho _{h} \left(r=r_{h} \right)=\frac{N_{s} m_{p} }{4\pi r_{h}^{2} r_{p} } ,          
\end{equation} 
where $N_{s} $ is the number of elementary mass $m_{p} $ in spherical shell of thickness $r_{p} $.  

Figure \ref{fig:1} illustrates how mass cascade changes the original halo size during an infinitesimal time interval $dt$. The original halo has a size $r_{h} $ at time \textit{t} (the dashed line). By the time $t+dt$, the halo size will increase from $r_{h} $ to $r_{h} -r_{p}^{'} +r_{p} $ due to mass cascade. The original halo surface (dashed line) shrinks to a smaller size of  $r_{h} -r_{p}^{'} $ (solid line around green circle). First, halo mass cascade (accretion) creates a new layer of mass around the original halo with a thickness of $r_{p} -r_{p}^{'} $. Second, this layer of mass deforms the original halo (dark blue) to a new size (green) due to gravitational interaction. This deformation creates a non-zero inward radial flow of mass. For isothermal profile with vanishing radial flow, $r_{p}^{'} =0$ such that mass accretion does not deform the original halo. This is only possible for extremely fast mass accretion such that deformation is relatively much slower. 

The time variation of halo radius $r_{h} $ due to mass accretion can be expressed as (with Eq. \eqref{ZEqnNum227253}),
\begin{equation} 
\label{ZEqnNum258955} 
\frac{\partial r_{h} }{\partial a} =\frac{1}{Ha} \frac{\partial r_{h} }{\partial t} =\frac{r_{p} -r_{p}^{'} }{N_{s} \tau _{g} Ha} =\frac{m_{p} }{\tau _{g} Ha} \frac{1}{4\pi r_{h}^{2} \rho _{h} \left(r_{h} \right)} \left(1-\frac{r_{p}^{'} }{r_{p} } \right),      
\end{equation} 
where $N_{s} \tau _{g} $ is the total time it takes to form the new layer of mass. Equivalently, we have 
\begin{equation}
\label{ZEqnNum410355} 
\frac{\partial \ln r_{h} }{\partial \ln a} =\frac{m_{p} }{4\pi r_{h}^{3} } \frac{\alpha _{h} }{\tau _{g} H\rho _{h} \left(r_{h} \right)} ,         
\end{equation} 
where the incremental change in halo size is $dr_{h} =r_{p} -r_{p}^{'} =\alpha _{h} r_{p} $. The halo deformation parameter $\alpha _{h} =1-{r_{p}^{'}/r_{p} } $ is introduced to reflect the effect of mass cascade on halo deformation. Mass density at halo surface can be obtained by comparing Eq. \eqref{ZEqnNum410355} with Eq. \eqref{ZEqnNum587991},
\begin{equation}
\label{ZEqnNum913739} 
\rho _{h} \left(r_{h} \right)=\frac{m_{h} }{4\pi r_{h}^{3} } \frac{3\alpha _{h} }{1+3\tau _{g} Hn_{p} }. \end{equation} 

From inverse mass cascade \citep[see][Eqs. (8) and (51)]{Xu:2021-Inverse-mass-cascade-mass-function}, we can estimate that on average, 
\begin{equation} 
\label{ZEqnNum464837} 
\tau _{f} \left(m_{h}^{L} ,a\right)=\frac{1-\lambda }{1-{2\tau _{0}/3} } t\sim t,         
\end{equation} 
where the time scale $\tau _{f} =\tau _{g} n_{p} $ is the time it takes to form the entire halo. Therefore,  
\begin{equation} 
\label{ZEqnNum821124} 
\tau _{g} Hn_{p} =\tau _{f} H=\frac{2\left(1-\lambda \right)}{3-2\tau _{0} } .         
\end{equation} 
With the help of Eq. \eqref{ZEqnNum821124}, Eqs. \eqref{ZEqnNum808457} and \eqref{ZEqnNum587991} give the time variation of halo mass and halo size,
\begin{equation}
\frac{\partial \ln m_{h} }{\partial \ln a} =\frac{3-2\tau _{0} }{2\left(1-\lambda \right)} \quad \textrm{and} \quad \frac{\partial \ln r_{h} }{\partial \ln a} =\frac{9-2\tau _{0} -6\lambda }{6\left(1-\lambda \right)},     
\label{ZEqnNum952904}
\end{equation}

\noindent both of which are dependent on two mass cascade parameters $\lambda $ and $\tau _{0} $. Now, the time variation of power-law density profile can be derived with Eqs. \eqref{ZEqnNum859547} and \eqref{ZEqnNum952904}, 
\begin{equation} 
\label{ZEqnNum808315} 
\frac{\partial \ln \rho _{h} \left(r\right)}{\partial \ln a} =-3+\frac{m\left(9-2\tau _{0} -6\lambda \right)}{6\left(1-\lambda \right)} .        
\end{equation} 

By comparing Eq. \eqref{ZEqnNum913739} with the power-law density profile in Eq. \eqref{ZEqnNum789386}, it is found that large halos should have a density profile of $m=3-\alpha _{h} $ if parameters $\lambda ={2/3} $ and $\tau _{0} =1$.  For an isothermal profile with $m=2$, it is necessary that $\alpha _{h} =1$ or $r_{p}^{'} =0$ such that mass accretion will not affect the halo internal structure. This is the limiting situation where halo mass accretion is extremely fast such that halos have no time to relax through radial deformation. The other limit is that $\alpha _{h} =0$ or $r_{p}^{'} =r_{p} $ (halo size gained from mass accretion exactly cancels the decrease in halo size due to the deformation) such that $m=3$ which is the maximum exponent for a power-law halo density profile. 

We have scaling laws of $m_{h} \sim a^{{3/2} } \sim t$, $r_{h} \sim a^{{3/2} } \sim t$, $\rho _{h} \left(r=r_{h} \right)\sim r_{h}^{-2} \sim a^{-3} $, and $\rho _{h} \left(r\right)\sim a^{0} $ from Eqs. \eqref{ZEqnNum952904} and \eqref{ZEqnNum808315} for an isothermal profile. The halo density $\rho _{h} \left(r\right)$ is time-invariant as a result of $\alpha _{h} =1$ such that the halo density at any radius \textit{r} is fully determined at the moment that shell of halo is formed and will not change thereafter (Eq. \eqref{ZEqnNum808315}). This is the key feature of a power-law density profile that is different from NFW and Einasto profiles.

By comparing the density at surface of halo (Eq. \eqref{ZEqnNum913739}) with the NFW density in Eq. \eqref{ZEqnNum859547}, the halo concentration parameter $c$ can be related to deformation parameter $\alpha _{h}$, 
\begin{equation} 
\label{ZEqnNum412141} 
\frac{\bar{\rho }_{h} \left(a\right)}{\rho _{h} \left(r_{h} ,a\right)} =3\left[1n\left(1+c\right)-\frac{c}{1+c} \right]\left(1+\frac{1}{c} \right)^{2} =\frac{1}{\alpha _{h} } \left(\frac{9-2\tau _{0} -6\lambda }{3-2\tau _{0} } \right).  
\end{equation} 
The concentration parameter \textit{c} is closely dependent on the halo deformation via $\alpha _{h} $ and on the mass cascade via parameters $\tau _{0} $ and $\lambda $. For large halos with $c=4$, and parameters $\lambda ={2/3} $ and $\tau _{0} =1$, the deformation parameter $\alpha _{h} \approx 0.79$. For small halos with $c=10$ and mass cascade parameter $\lambda ={2/3} $ and $\tau _{0} =1$, $\alpha _{h} \approx 0.56$. It is expected that the deformation parameter $\alpha _{h} $ increases with the halo mass (smaller halos have relatively greater deformation and smaller $\alpha _{h} $). The concentration-mass relation (the mass dependence of \textit{c}) might be related to the mass dependence of both $\alpha _{h} $ and geometry parameter $\lambda $ that should be further explored.

\subsection{Effects of radial flow on halo density distribution}
\label{sec:3.3}
The inverse mass cascade creates a new layer of mass that deforms the original halo to a new size (green in Fig. \ref{fig:1}). This creates a non-zero radial flow that can be analyzed using the continuity equation. Let's start from a general expression of mass $m_{r} \left(r,a\right)$ ,

\begin{equation}
m_{r} \left(r,a\right)=m_{h} \left(a\right)\frac{F\left(x\right)}{F\left(c\right)} \quad \textrm{and} \quad x\left(r,a\right)=\frac{r}{r_{s} \left(a\right)} =\frac{cr}{r_{h} \left(a\right)},   
\label{ZEqnNum632204}
\end{equation}

\noindent where $m_{r} \left(r,a\right)$ is the mass in a sphere of radius \textit{r}, $x\left(r,a\right)$ is a reduced spatial-temporal variable that lumps the position \textit{r} and scale factor \textit{a} into a single variable. This general expression can represent both NFW (Eq. \eqref{eq:A4}) and Einasto (Eq. \eqref{ZEqnNum807556}), or any other density profiles via different functions $F\left(x\right)$. In principle, function $F\left(x\right)$ can be an arbitrary unknown function that satisfies $F\left(0\right)=0$. For example, $F\left(x\right)=x$ for an isothermal profile. Equations \eqref{ZEqnNum588557} and \eqref{ZEqnNum807556} give expressions of $F\left(x\right)$ for NFW and Einasto profiles. The halo density, potential, and velocity dispersion can all be determined in terms of unknown function $F\left(x\right)$. The halo density profile reads  
\begin{equation} 
\label{ZEqnNum482501} 
\rho _{h} \left(r,a\right)=\frac{1}{4\pi r^{2} } \frac{\partial m_{r} \left(r,a\right)}{\partial r} =\frac{m_{h} \left(a\right)}{4\pi r_{h}^{3} } \frac{c^{3} F^{'} \left(x\right)}{x^{2} F\left(c\right)} ,      
\end{equation} 
and the logarithmic slope of the halo density reads
\begin{equation} 
\label{ZEqnNum922312} 
\frac{\partial \ln \rho _{h} }{\partial \ln x} =\frac{\partial \ln F^{'} \left(x\right)}{\partial \ln x} -2=x\frac{F^{''} \left(x\right)}{F^{'} \left(x\right)} -2. 
\end{equation} 
Evidently, the halo deformation parameter satisfies \\
$\alpha _{h} ={cF^{'} \left(c\right)/F\left(c\right)}$\\ 
by comparing the density at halo surface to Eq. \eqref{ZEqnNum913739} with $\lambda ={2/3} $ and $\tau _{0} =1$.  Time variation of $\rho _{h} \left(r,a\right)$ can be obtained from Eq. \eqref{ZEqnNum482501},
\begin{equation} 
\label{ZEqnNum505562} 
\frac{\partial \rho _{h} \left(r,a\right)}{\partial t} =\frac{1}{4\pi r^{2} } \frac{\partial ^{2} m_{r} \left(r,a\right)}{\partial r\partial t} .        
\end{equation} 
The mass continuity equation for a spherical halo in spherical coordinate simply reads,
\begin{equation} 
\label{ZEqnNum114919} 
\frac{\partial \rho _{h} \left(r,a\right)}{\partial t} +\frac{1}{r^{2} } \frac{\partial \left[r^{2} \rho _{h} \left(r,a\right)u_{r} \left(r,a\right)\right]}{\partial r} =0,       
\end{equation} 
where $u_{r} \left(r,a\right)$ is the mean radial flow velocity. From Eqs. \eqref{ZEqnNum505562} and \eqref{ZEqnNum114919}, the mass $m_{r} \left(r,a\right)$ is related to the radial flow velocity as,
\begin{equation} 
\label{ZEqnNum861530} 
\frac{\partial m_{r} \left(r,a\right)}{\partial t} =-4\pi r^{2} u_{r} \left(r,a\right)\rho _{h} \left(r,a\right).   \end{equation} 
With $m_{r} \left(r,a\right)$ from Eq. \eqref{ZEqnNum632204} and $\rho _{h} \left(r,a\right)$ from Eq. \eqref{ZEqnNum482501}, the radial flow velocity reads
\begin{equation} 
\label{ZEqnNum255276} 
u_{r} =-\frac{1}{4\pi r^{2} } \frac{\partial \ln m_{r} }{\partial \ln t} \frac{m_{r} \left(r,a\right)}{\rho _{h} \left(r,a\right)t} =-\frac{r_{s} \left(t\right)}{t} \frac{\partial \ln m_{r} }{\partial \ln t} \frac{F\left(x\right)}{F^{'} \left(x\right)} .     
\end{equation} 
While from Eq. \eqref{ZEqnNum632204} for $m_{r} \left(r,a\right)$, we have
\begin{equation} 
\label{ZEqnNum295393} 
\frac{\partial \ln m_{r} }{\partial \ln t} =\frac{\partial \ln m_{h} }{\partial \ln t} -\frac{xF^{'} \left(x\right)}{F\left(x\right)} \frac{\partial \ln r_{s} }{\partial \ln t} -\frac{cF^{'} \left(c\right)}{F\left(c\right)} \frac{\partial \ln c}{\partial \ln t} .       
\end{equation} 
Substituting Eqs. \eqref{ZEqnNum295393} into \eqref{ZEqnNum255276}, the radial flow has a very simple expression,
\begin{equation} 
\label{ZEqnNum853896} 
u_{r} \left(r,a\right)=\left[x\frac{\partial \ln r_{s} }{\partial \ln t} +\left(\frac{\partial \ln F\left(c\right)}{\partial \ln t} -\frac{\partial \ln m_{h} }{\partial \ln t} \right)\frac{F\left(x\right)}{F^{'} \left(x\right)} \right]\frac{r_{s} }{t} ,     
\end{equation} 
which is a general equation for mean radial flow with a time-varying concentration \textit{c}. For small halos with a stable core and extremely slow mass accretion (${\partial r_{s}/\partial t} \approx 0$ and ${\partial m_{h}/\partial t} \approx 0$) and constant $m_{r} \left(r_{s} ,a\right)$, we shall expect that $F\left(c\right)\propto m_{h} $ is almost a constant (from Eq. \eqref{ZEqnNum632204}) and $u_{r} \left(r,a\right)=0$ (Eq. \eqref{ZEqnNum853896}) that is consistent with the stable clustering hypothesis (small halos are virialized and well bound structures). 

For the other limiting situation, i.e. large halos with extremely fast mass accretion and an expanding core, the concentration \textit{c} is relatively a constant. Equation \eqref{ZEqnNum853896} reduces to
\begin{equation} 
\label{ZEqnNum849591} 
u_{r} \left(r,a\right)=\frac{1}{c} \left[x-\frac{\partial \ln m_{h} }{\partial \ln r_{h} } \frac{F\left(x\right)}{F^{'} \left(x\right)} \right]\frac{\partial r_{h} }{\partial t} .        
\end{equation} 
In principle, the non-zero halo growth rate ${\partial r_{h}/\partial t} $ should lead to a non-zero mean radial flow. A special case is the isothermal profile with $F\left(x\right)=x$ and $m_{h} \propto r_{h} $, where $u_{r} \left(r,a\right)=0$, i.e. a vanishing radial flow for isothermal profile even if ${\partial r_{h}/\partial t} \ne 0$. The radial flow $u_{r} \left(r,a\right)$ is a function of reduced position \textit{x} only and scaled by the rate of halo growth ${\partial r_{h}/ \partial t} $. We introduce a dimensionless radial flow velocity $u_{h} \left(x\right)$ (normalized by the core expanding speed ${r_{s}/t} $) as
\begin{equation} 
\label{ZEqnNum278808} 
\begin{split}
u_{h} \left(x\right)=\frac{cu_{r} \left(r,a\right)t}{r_{h} } &=\frac{u_{r} \left(r,a\right)}{{r_{s}/t} }\\ &=\Big[\underbrace{x}_{1}-\underbrace{\frac{\partial \ln m_{h} }{\partial \ln r_{h} } \frac{F\left(x\right)}{F^{'} \left(x\right)} }_{2}\Big]\frac{\partial \ln r_{h} }{\partial \ln t}.
\end{split}
\end{equation} 
Clearly, the halo growth rate ${\partial r_{h}/\partial t} $ affects the mean radial flow in Eq. \eqref{ZEqnNum278808}. For large halos with a constant value of $c$, ${\partial r_{h}/\partial t} ={r_{h}/ t={v_{cir}/\left(2\pi \right)} } $ (Eq. \eqref{ZEqnNum837340}) does not varying with time and $u_{r} \left(r,a\right)$ is self-similar and only dependent on the reduced variable $x$. The total radial flow can be decomposed into two contributions: 1) the outward flow due to the halo growth where $u_{r} \left(r,a\right)={r/t} $ (term 1), and 2) the inward flow due to the halo deformation (term 2). 

\begin{figure}
\includegraphics*[width=\columnwidth]{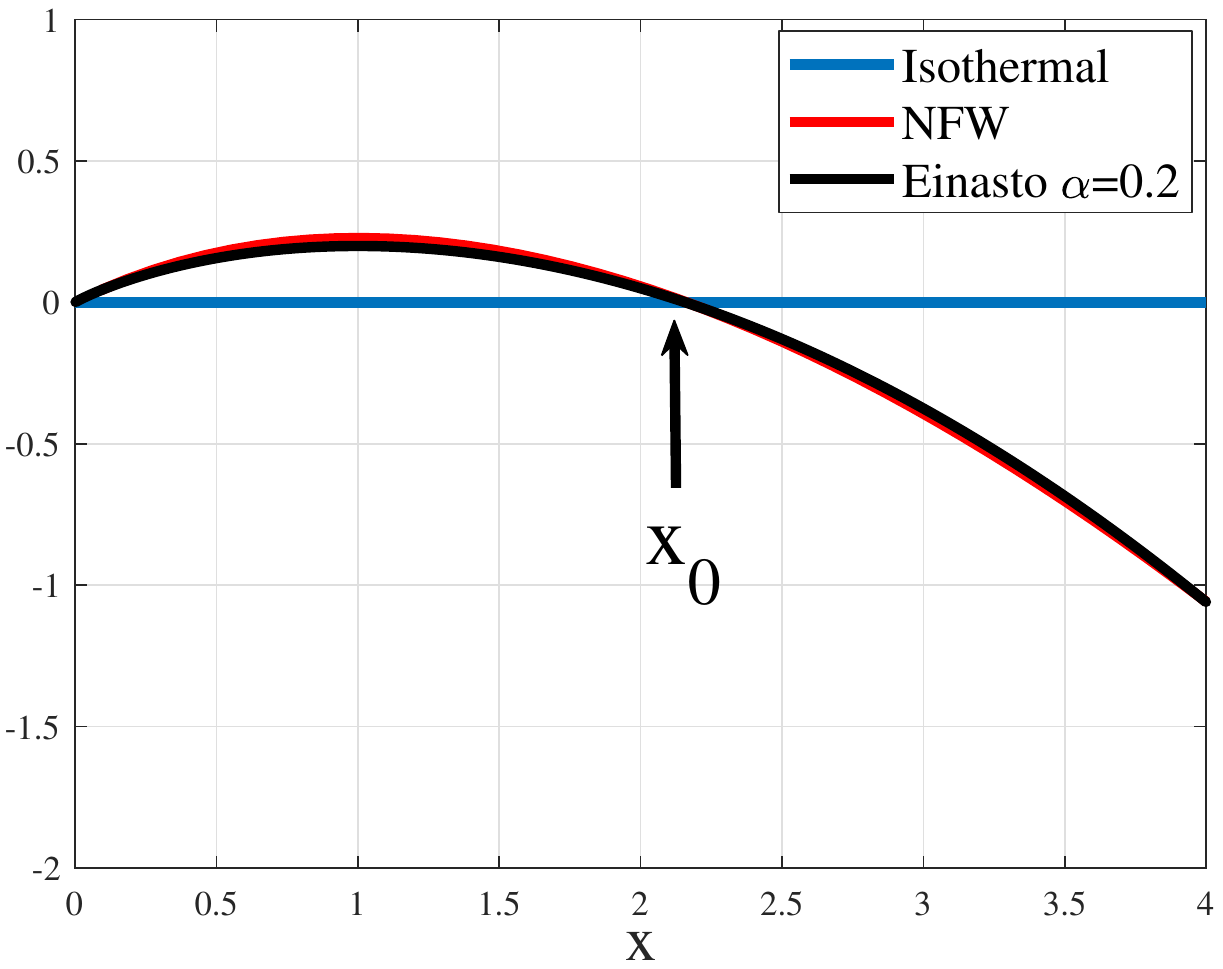}
\caption{The normalized mean radial flow $u_{h} \left(x\right)$ (Eq. \eqref{ZEqnNum278808}) for three different density profiles. The isothermal halo density corresponds to a vanishing radial flow, i.e.halos with extremely fast mass accretion and no internal deformation. The NFW and Einasto profiles ($c=4$ and $\alpha =0.2$) lead to very similar radial flow with out-flow ($u_{h} \left(x\right)>0$) for the core region and in-flow ($u_{h} \left(x\right)<0$) for the outer region. The maximum radial flow is at $x=1$ or $r=r_{s} $. The mass $m_{r} \left(r,a\right)$ inside the radius \textit{r} decreases with time for $x<x_{0} $ and increases with time for $x>x_{0} $ (Eq. \eqref{ZEqnNum861530}). The circular velocity is at its maximum at $x=x_{0} $. The difference between two profiles is that $u_{h} \left(x\to 0\right)={x/2} $ for NFW and $u_{h} \left(x\to 0\right)={2x/3} $ for Einasto profile.}
\label{fig:2}
\end{figure}

By comparing the density at surface of halos (Eqs. \eqref{ZEqnNum482501} and \eqref{ZEqnNum913739}) and using the help of Eqs. \eqref{ZEqnNum821124} and \eqref{ZEqnNum952904}, the constraints and boundary conditions for radial flow velocity are,
\begin{equation}
\begin{split}
&u_{h} \left(x=0\right)=0, \quad u_{h} \left(x=c\right)=c\left(1-\frac{1}{\alpha _{h} } \right)\frac{\partial \ln r_{h} }{\partial \ln t},\\
&\textrm{and}\\
&x_{0} =\frac{\partial \ln m_{h} }{\partial \ln r_{h} } \left. \frac{F\left(x\right)}{F^{'} \left(x\right)} \right|_{x=x_{0} },  
\label{ZEqnNum871868}
\end{split}
\end{equation}

\noindent where $u_{h} \left(x_{0} \right)=0$. Figure \ref{fig:2} plots the normalized radial velocity $u_{h} \left(x\right)$ for three different density profiles. The NFW and Einasto profiles (for $c=4$ and $\alpha =0.2$) lead to a very similar radial flow velocity with out-flow ($u_{h} \left(x\right)>0$) for core region ($x<x_{0} $) and in-flow ($u_{h} \left(x\right)<0$) for outer region ($x>x_{0} $) of halos, where $u_{h} \left(x_{0} \right)=0$. The maximum radial flow is at $x=1$ or $r=r_{s} $. The total mass $m_{r} \left(x,a\right)$ decreases with time for $x<x_{0} $ and increases with time for $x>x_{0} $ (from Eq. \eqref{ZEqnNum861530}). The difference between two density profiles is that $u_{h} \left(x\to 0\right)={x/2} $ for NFW profile and $u_{h} \left(x\to 0\right)={2x/3} $ for Einasto profile.

The dimensionless peculiar radial flow can be obtained by subtracting the Hubble flow,
\begin{equation} 
\label{eq:27} 
u_{p} \left(x\right)=u_{h} \left(x\right)-\frac{2}{3} x=\frac{1}{3} x-\frac{F\left(x\right)}{F^{'} \left(x\right)} .        
\end{equation} 
Especially for an isothermal profile with $F\left(x\right)=x$, $u_{p} \left(x\right)=-{2x/3} $, which can be a good approximation of peculiar radial flow. This is consistent with the stable clustering hypothesis, i.e. the peculiar radial flow 
\begin{equation}
\label{eq:28} 
u_{p} \left(x\right)\frac{r_{s} }{t} =-\frac{2}{3} x\frac{r_{h} }{ct} =-\frac{2}{3} \frac{cr}{r_{h} } \cdot \frac{r_{h} }{ct} =-Hr.       
\end{equation} 

\subsection{The angle of incidence for mass cascade}
\label{sec:3.4}
An interesting quantity is the angle $\theta _{vr} $ (the angle of incidence) between particle peculiar velocity and its position vector from halo center of mass, 
\begin{equation} 
\label{ZEqnNum340811} 
\begin{split}
\cot \left(\theta _{vr} \right)&=\frac{u_{p} \left(x\right)}{v_{c} \left(r,a\right)} \frac{r_{h} }{ct}\\ &=\frac{x}{2\pi c} \left(\frac{1}{3} -\frac{1}{\alpha _{h} } \cdot \frac{F^{'} \left(c\right)}{F^{'} \left(x\right)} \cdot \frac{cF\left(x\right)}{xF\left(c\right)} \right)\sqrt{\frac{xF\left(c\right)}{cF\left(x\right)} } ,    
\end{split}
\end{equation} 
where circular velocity (normalized) at any radius \textit{r} of the halo is 
\begin{equation}
\label{ZEqnNum537303} 
v_{nc}^{2} \left(r,a\right)=\frac{v_{c}^{2} \left(r,a\right)}{v_{cir}^{2} } =\frac{Gm_{r} \left(r,a\right)}{rv_{cir}^{2} } =\frac{cF\left(x\right)}{F\left(c\right)x} .       
\end{equation} 
Interestingly, $v_{nc}^{2} \left(x,a\right)$ is at its maximum when $x=x_{0} $ where $u_{h} \left(x_{0} \right)=0$ (using Eq. \eqref{ZEqnNum871868}). Quantity $\cot \left(\theta _{vr} \right)$ quantifies the ratio of radial motion (radial momentum) to the circular motion (angular momentum). With $\alpha _{h} ={cF^{'} \left(c\right)/F\left(c\right)} $, $\cot \left(\theta _{vr} \right)$ at halo surface ($x=c$) and halo center ($x=0$) are obtained from Eq. \eqref{ZEqnNum340811} as  
\begin{equation} 
\label{eq:31} 
\cot \left(\theta _{vr} \left(x=c\right)\right)=\frac{1}{2\pi } \left(\frac{1}{3} -\frac{1}{\alpha _{h} } \right) 
\end{equation} 
and 
\begin{equation}
\label{ZEqnNum184216} 
\cot \left(\theta _{vr} \left(x=0\right)\right)={\mathop{\lim }\limits_{x\to 0}} \frac{1}{2\pi c} \left(\frac{1}{3} -\frac{F\left(x\right)}{xF^{'} \left(x\right)} \right)\sqrt{\frac{x^{3} F\left(c\right)}{cF\left(x\right)} } .      
\end{equation} 
When $x=c$ at halo surface, $\cot \left(\theta _{vr} \right)=-{1/\left(3\pi \right)} $ and $\theta _{vr} \approx 96.06^{o} $ for $\alpha _{h} =1$ (isothermal profile) and $\theta _{vr} \approx 98.44^{o} $ for $\alpha _{h} =0.79$ (NFW profile with c=4 from Eq. \eqref{ZEqnNum412141}). From Eq. \eqref{ZEqnNum184216}, angle $\theta _{vr} $ should gradually decrease to $\theta _{vr} =90^{o}$ ($\cot \left(\theta _{vr} \right)=0$ with $u_{p} \left(x\right)\to 0$) at the core region for any $F\left(x\right)\propto x^{m} $ ($m\le 3$) with $x\to 0$. This can be also demonstrated by a two-body gravitational collapse (TBCM) model \citep[see][Eq. (105)]{Xu:2021-A-non-radial-two-body-collapse}. 

By taking derivative of $u_{h} \left(x\right)$ (Eq. \eqref{ZEqnNum278808}) with respect to \textit{x}, 
\begin{equation} 
\label{ZEqnNum273026} 
\frac{\partial u_{h} \left(x\right)}{\partial x} =\left[1-\frac{\partial \ln m_{h} }{\partial \ln r_{h} } +\frac{\partial \ln m_{h} }{\partial \ln r_{h} } \frac{F\left(x\right)F^{''} \left(x\right)}{F^{'2} \left(x\right)} \right]\frac{\partial \ln r_{h} }{\partial \ln t} .      
\end{equation} 
Particularly for a NFW profile,
\begin{equation} 
\label{eq:34} 
\frac{\partial u_{h} \left(x\right)}{\partial x} =\left[\ln \left(1+x\right)-\frac{x}{1+x} \right]\frac{1-x^{2} }{x^{2} } .        
\end{equation} 
It can be verified that for both NFW and Einasto profiles, the conditions of maximum flow ($\left. {\partial u_{h} \left(x\right)/\partial x} \right|_{x=1} =0$) and logarithmic slope of -2 ($\left. {\partial \ln \rho _{h}/\partial \ln x} \right|_{x=1} =-2$) at scale radius ($x=1$ or $r=r_{s} $) requires $\left. F^{''} \left(x\right)\right|_{x=1} =0$ (Eq. \eqref{ZEqnNum922312}). Hence, ${\partial \ln m_{h} /\partial \ln r_{h} } =1$ from Eq. \eqref{ZEqnNum273026}, which confirms $\tau _{0} ={3\lambda/2} $ for large halos (from Eq. \eqref{ZEqnNum952904}). 

Existence of an extra length scale $r_{s}$ in density profile origins from mass cascade induced radial flow. The in-flow in halo outer region and the out-flow in the inner region creates a maximum mass flow rate at scale radius $r_{s} $ (or $x=1$) and introduces an extra length scale $r_{s}$ for halo density profile, which does not exist for a scale-free isothermal density profile (Fig. \ref{fig:2}). 

We are especially interested in the logarithmic slope of the unknown function $F^{'} \left(x\right)$ that directly impacts the halo density profile (Eq. \eqref{ZEqnNum482501}). It can be obtained from the mean radial flow $u_{h} \left(x\right)$ using Eqs. \eqref{ZEqnNum278808} and \eqref{ZEqnNum273026},
\begin{equation}
\label{ZEqnNum696290} 
\frac{\partial \ln F^{'} \left(x\right)}{\partial \ln x} =\frac{\frac{\partial u_{h} \left(x\right)}{\partial x} -\frac{\partial \ln r_{h} }{\partial \ln t} +\frac{\partial \ln m_{h} }{\partial \ln t} }{\frac{\partial \ln r_{h} }{\partial \ln t} -\frac{u_{h} \left(x\right)}{x} } .       
\end{equation} 
Specially, for matter dominant system, we expect,
\begin{equation}
\frac{\partial \ln r_{h} }{\partial \ln t} =\frac{\partial \ln m_{h} }{\partial \ln t} =1, \quad\frac{\partial \ln F^{'} \left(x\right)}{\partial \ln x} =\frac{xF^{''} \left(x\right)}{F^{'} \left(x\right)} =\frac{{\partial u_{h}/\partial x} }{1-{u_{h}/x} }.    
\label{ZEqnNum445988}
\end{equation}

\noindent The logarithmic slope of halo density profile (from Eq. \eqref{ZEqnNum922312}) reads
\begin{equation} 
\label{ZEqnNum852508} 
\frac{\partial \ln \rho _{h} }{\partial \ln x} =\frac{\partial \ln F^{'} \left(x\right)}{\partial \ln x} -2=\frac{{\partial u_{h}/\partial x} }{1-{u_{h}/x} } -2. 
\end{equation} 
We have simple expressions for NFW and Einasto profiles, \begin{equation}
\frac{\partial \ln F^{'} \left(x\right)}{\partial \ln x} =\frac{1-x}{1+x} \quad \textrm{and} \quad \frac{\partial \ln F^{'} \left(x\right)}{\partial \ln x} =2-2x^{\alpha }.     
\label{eq:38}
\end{equation}

To provide some insights into the long-standing cusp-core controversy (core/cusp problem), a double-power-law density profile can be proposed as a natural result of Eq. \eqref{ZEqnNum852508}. The inner halo density is determined by the velocity gradient (halo deformation rate) $\gamma _{h} =\left. {\partial u_{h}/\partial x} \right|_{x=0} $ such that inner halo density follows a power-law
\begin{equation} 
\label{ZEqnNum973033} 
\rho _{h} \left(r<r_{s} \right)\propto r^{{\left(3\gamma _{h} -2\right)/\left(1-\gamma _{h} \right)} }  
\end{equation} 
that is dependent on parameter $\gamma _{h} $ only. The smaller $\gamma _{h} $ (slower deformation at the halo center) leads to a steeper density profile. The baryonic feedback processes may enhance the deformation rate $\gamma _{h}$ at halo center and lead to the formation of core structure.  

In addition, for a matter dominant universe, the radial flow should be exactly the Hubble flow if both gravitational and pressure forces are not present in halos. If the potential and pressure are symmetric functions of \textit{r} and regular at origin \textit{r}=0, the gravitational and pressure forces should vanish at origin such that $u_{r} \left(r\to 0\right)=Hr$. We expect the initial velocity of mass shells at the center of halo is simply the Hubble flow for halos with fast mass accretion, 
\begin{equation}
\label{ZEqnNum801048} 
u_{r} \left(r\to 0\right)=u_{h} \left(x\to 0\right)\frac{r_{s} }{t} =\gamma _{h} x\frac{r_{s} }{t} =\gamma _{h} \frac{r}{t} =Hr,      
\end{equation} 
such that $\gamma _{h} =Ht={2/3} $.  This means a central core with $\rho _{h} \left(r<r_{s} \right)\propto r^{0} $ from Eq. \eqref{ZEqnNum973033} does exist for large halos with fast mass accretion, in agreement with the finding that large halos can be better fitted by an Einasto profile \citep{Klypin:2016-MultiDark-simulations--the-sto}. 

For outer halo region (especially $r\gg x_{0} $ in Fig. \ref{fig:2}), we can approximate (using Eq. \eqref{ZEqnNum871868})
\begin{equation}
\frac{\partial u_{h} }{\partial x} \approx \frac{c\left(1-{1/\alpha _{h} } \right)}{c-x_{0} } \quad \textrm{and} \quad 1-\left. \frac{u_{h} }{x} \right|_{x=c} \approx \frac{1}{\alpha _{h} },      
\label{eq:41}
\end{equation}

\noindent Such that from Eq. \eqref{ZEqnNum852508}
\begin{equation}
\label{ZEqnNum367971} 
\rho _{h} \left(r>r_{s} \right)\propto r^{\frac{c\left(\alpha _{h} -1\right)}{c-x_{0} } -2} ,   
\end{equation} 
with a power-law density profile steeper than the isothermal profile of -2 for the outer halo region. Equations \eqref{ZEqnNum973033} and \eqref{ZEqnNum367971} provide a double-power-law density with inner density controlled by halo deformation rate parameter $\gamma _{h} $ and outer density controlled by the halo growth via a halo deformation parameter $\alpha _{h} $ and concentration \textit{c}.

In principle, accurate halo density profiles can be obtained only if the normalized mean flow $u_{h} \left(x\right)$ is known. Without loss of generality, the Taylor expansion of $u_{h} \left(x\right)$ around the center (up to the third order) can be given by
\begin{equation} 
\label{ZEqnNum621572} 
u_{h} \left(x\right)=a_{0} +\gamma _{h} x+a_{2} x^{2} +a_{3} x^{3} ,        
\end{equation} 
with three unknown coefficients. To satisfy the boundary conditions \eqref{ZEqnNum871868} and the constraint $u_{h} \left(0\right)=0$ and $\left. {\partial u_{h} /\partial x} \right|_{x=1} =0$, we have $a_{2} $ and $a_{3} $ expressed as
\begin{equation} 
\label{ZEqnNum909055} 
a_{0} =0, a_{2} =-\frac{\left(c^{2} -3\right)\gamma _{h} +3-{3/\alpha _{h} } }{2c^{2} -3c} , a_{3} =\frac{\left(c-2\right)\gamma _{h} +2-{2/\alpha _{h} } }{2c^{2} -3c} .   
\end{equation} 
The unknown function $F\left(x\right)$ can be analytically solved from Eq. \eqref{ZEqnNum278808} and we have the solution,
\begin{equation} 
\label{ZEqnNum676651} 
\begin{split}
&F\left(x\right)=\left(\frac{x^{3} }{x-u_{h} \left(x\right)} \right)^{\frac{1}{2\left(1-\gamma _{h} \right)} }\\
&\cdot \exp \left\{\frac{{a_{2}/\left(\gamma _{h} -1\right)} }{\sqrt{4a_{3} \left(\gamma _{h} -1\right)-a_{2}^{2} } }\textrm{atan}\left[\frac{a_{2} +2a_{3} x}{\sqrt{4a_{3} \left(\gamma _{h} -1\right)-a_{2}^{2} } } \right]\right\},
\end{split}
\end{equation} 
with which the density profile can be obtained from Eq. \eqref{ZEqnNum482501}. 

We have shown that a complete description of $u_{h} \left(x\right)$ or $F\left(x\right)$ requires at least three parameters, the deformation rate $\gamma _{h} $ at the center of halo, halo deformation parameter $\alpha _{h}$ at the surface of halo, and concentration \textit{c} for the size of halos. The location $x_{0}$ where $u_{h} \left(x_{0} \right)=0$ is estimated to be (from Eq. \eqref{ZEqnNum621572})
\begin{equation} 
\label{eq:46} 
x_{0} =\frac{-a_{2} -\sqrt{a_{2}^{2} -4a_{3} \gamma _{h} } }{2a_{3} } ,         
\end{equation} 
with limiting values
\begin{equation}
x_{0} =\frac{3}{2}\quad \textrm{for} \quad \gamma _{h} \to 0,\quad x_{0} =\frac{2c-3}{c-2}\quad \textrm{for} \quad \gamma _{h} \to \infty.    
\label{eq:47}
\end{equation}

\noindent The radial flow at $x=1$ and its derivative at $x=c$ are,
\begin{equation}
\begin{split}
&u_{h} \left(x=1\right)=\frac{\left(c-1\right)^{2} \gamma _{h} -\left(1-{1/\alpha _{h} } \right)}{c\left(2c-3\right)}\\  
&\textrm{and}\\ 
&\left.\frac{\partial u_{h} }{\partial x} \right|_{x=c} =\frac{\left(c-3\right)\gamma _{h} +6\left(1-{1/\alpha _{h} } \right)}{{\left(2c-3\right)/\left(c-1\right)} }.
\end{split}
\label{eq:48}
\end{equation}

\noindent An even simpler case is a Taylor expansion of $u_{h} \left(x\right)$ up to the second order (i.e. $a_{3} =0$) that will lead to solutions,
\begin{equation}
a_{0} =0, \quad a_{2} =-{\gamma _{h}/2}, \quad \textrm{and} \quad \gamma _{h} ={\left(1-{1/\alpha _{h} } \right)\left(1-{c/2}\right)}   \label{eq:49}
\end{equation}

\noindent from Eq. \eqref{ZEqnNum909055}. We have $x_{0} =2$ from Eq. \eqref{ZEqnNum621572} for expansion of $u_{h} \left(x\right)$ up to the second order. 

Alternatively, function $F\left(x\right)$ can be modelled directly with the following constraints:
\begin{equation}
\begin{split}
&F\left(0\right)=0, \quad F\left(x\to 0\right)=x^{1/\left(1-\gamma _{h} \right)},\\
&\frac{cF^{'} \left(c\right)}{F\left(c\right)} =\alpha _{h},\quad \textrm{and} \quad \left. \frac{\partial ^{2} F}{\partial x^{2} } \right|_{x=1} =0. 
\end{split}
\label{ZEqnNum434585}
\end{equation}

\noindent Solutions of all relevant quantities can be easily obtained with either $u_{h} \left(x\right)$ or $F\left(x\right)$ explicitly modelled. Therefore, halo density profile can be found by the correctly modeling of either dimensionless radial flow $u_{h} \left(x\right)$ or unknown function $F\left(x\right)$.

In short, the logarithmic slope of density profile is continuously dependent on the mean radial flow $u_{h} \left(x\right)$ (Eq. \eqref{ZEqnNum852508}). An accurate model of $u_{h} \left(x\right)$ due to mass cascade can improve halo density models. Since the matter density spectrum is closely related to halo density profiles and mass functions, effects of mass cascade and mass accretion on the density spectrum can be further investigated. 

\subsection{Limiting concentration c from momentum/kinetic energy}
\label{sec:3.5}
\begin{figure}
\includegraphics*[width=\columnwidth]{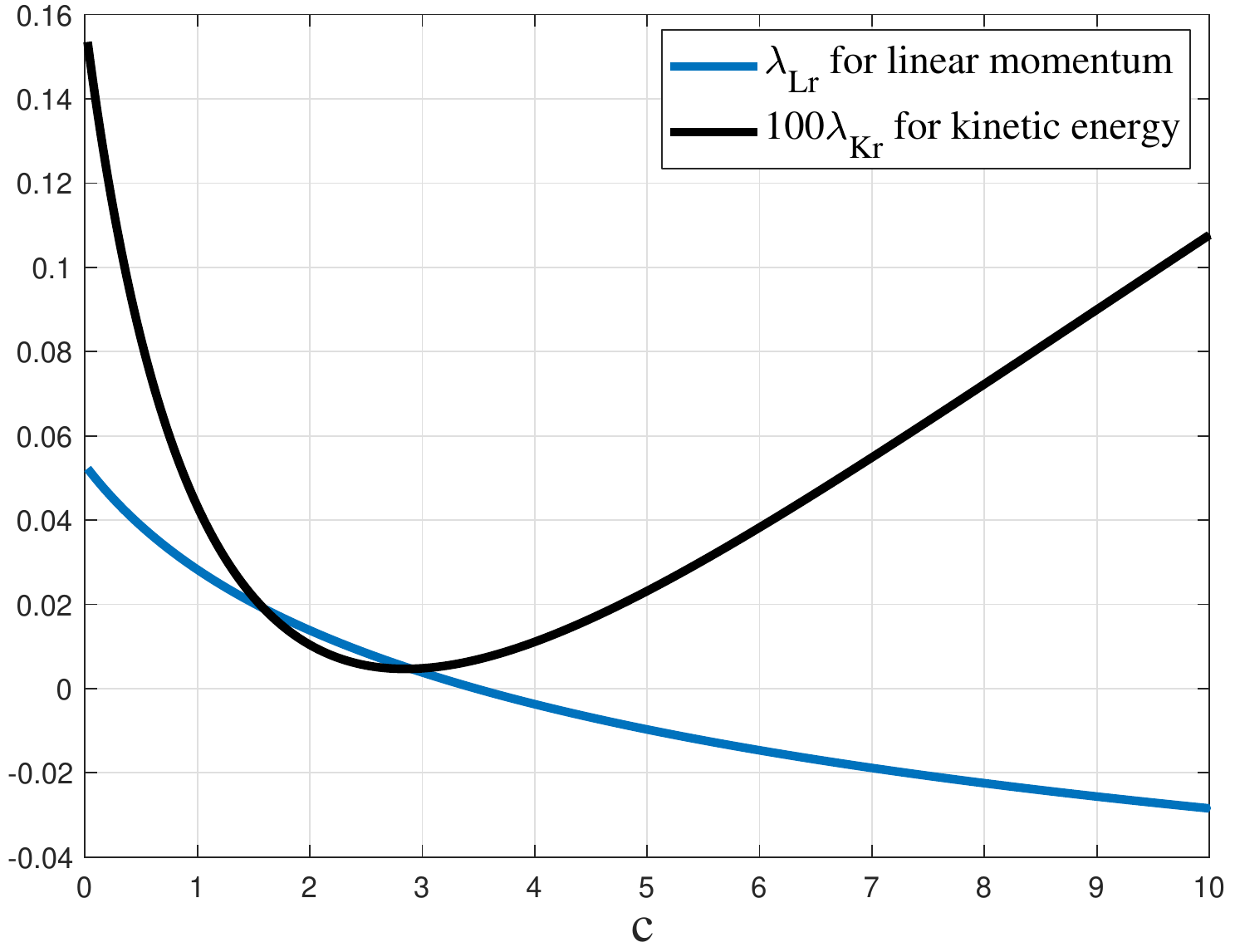}
\caption{The variation of normalized radial momentum $\lambda _{Lr} $ and kinetic energy $\lambda _{Kr} $ with the concentration parameter \textit{c} for a NFW profile. A limiting value of $c=3.48$ can be found for large halos with $\lambda _{Lr} =0$, where the linear radial momentum vanishes. The normalized radial kinetic energy is also close to its minimum at the limiting value of \textit{c}.}
\label{fig:3}
\end{figure}

The limiting value of concentration $c\approx 4$ for large halos with fast mass accretion was estimated from \textit{N}-body simulations. It is possible to analytically derive this limiting value by requiring a vanishing radial momentum for large halos with fast mass accretion. With Eq. \eqref{ZEqnNum849591} for mean flow $u_{r} \left(r,a\right)$ and Eq. \eqref{ZEqnNum482501} for density $\rho _{h} \left(r,a\right)$, the radial linear momentum is
\begin{equation}
\label{ZEqnNum808595} 
\begin{split}
L_{hr} \left(a\right)&=\int _{0}^{r_{h} }u_{r} \left(r,a\right)4\pi r^{2} \rho _{h} \left(r,a\right)dr\\
&=\frac{m_{h} v_{cir} }{2\pi cF\left(c\right)} \left(cF\left(c\right)-2\int _{0}^{c}F\left(x\right)dx \right).
\end{split}
\end{equation} 
With unknown function $F\left(x\right)=\ln \left(1+x\right)-{x/\left(1+x\right)} $ (Eq. \eqref{ZEqnNum588557}) for NFW profile, the radial linear momentum reduces to
\begin{equation}
\label{eq:52} 
L_{hr} \left(a\right)=\frac{m_{h} v_{cir} }{2\pi cF\left(c\right)} \left(\frac{c\left(4+3c\right)}{1+c} -\left(4+c\right)\ln \left(1+c\right)\right).      
\end{equation} 
The critical value of $c$ for individual halos can be identified from the condition of a vanishing linear momentum (like spherical shells at turn-around point with a zero velocity in spherical collapse model). Therefore, by requiring $L_{hr} \left(a\right)=0$, Eq. \eqref{ZEqnNum808595} becomes
\begin{equation}
\label{ZEqnNum722864} 
cF\left(c\right)=2\int _{0}^{c}F\left(x\right)dx ,      
\end{equation} 
where it was found that $c=3.48$ for a NFW profile (Fig. \ref{fig:3}). With sufficiently fast mass accretion rate, halos keep growing with a vanishing radial momentum $L_{hr} $. For halos with $c>3.48$, this self-similar solution leads to a negative radial linear momentum $L_{hr} <0$ indicating an overall in-flow of momentum. 

Next, the radial kinetic energy is given by,
\begin{equation} 
\label{eq:54}
\begin{split}
&K_{hr} \left(a\right)=\frac{1}{2} \int _{0}^{r_{h} }u_{r}^{2} \left(r,a\right)4\pi r^{2} \rho _{h} \left(r,a\right)dr\\
&=\frac{m_{h} v_{cir}^{2} }{8\pi ^{2} c^{2} F\left(c\right)} \left(c^{2} F\left(c\right)-4\int _{0}^{c}xF\left(x\right)dx +\int _{0}^{c}\frac{F^{2} \left(x\right)}{F^{'} \left(x\right)} dx \right).
\end{split}
\end{equation} 
Specifically, the radial kinetic energy for a NFW profile is given by the expression of,
\begin{equation} 
\label{eq:55} 
\begin{split}
&K_{hr} =\frac{m_{h} v_{cir}^{2} }{8\pi ^{2} c^{2} F\left(c\right)} \left[\frac{c\left(13c^{2} +7c-10\right)}{\left(1+c\right)}\right.\\
&\left.+2\ln \left(1+c\right)\left(5-10c-5c^{2}+\left(3+4c+c^{2} +2\ln \left(-c\right)\right)\ln \left(1+c\right)\right)+\right.\\
&\left.+8\ln \left(1+c\right)\textrm{Polylog}\left(2,1+c\right)-8\textrm{Polylog}\left(3,1+c\right)+8\textrm{Zeta}\left(3\right)\right],
\end{split}
\end{equation} 
which involves polylogarithm and zeta functions. In general, we can express both radial linear momentum and radial kinetic energy in terms of the circular velocity $v_{cir} $ (Eq. \eqref{ZEqnNum837340}) with two coefficients that are functions of \textit{c},  
\begin{equation}
L_{hr} \left(a\right)=\lambda _{Lr} \left(c\right)m_{h} v_{cir} \quad \textrm{and} \quad K_{hr} =\lambda _{Kr} \left(c\right)m_{h} v_{cir}^{2}.     
\label{eq:56} 
\end{equation}

\noindent With $c=3.48$ for NFW profile, we have $\lambda _{Lr} =0$ and $\lambda _{Kr} =7\times 10^{-5} $. Figure \ref{fig:3} plots the variation of two coefficients with concentration \textit{c}. Note that $\lambda _{Kr}$ is rescaled by 100 times to be plotted in the same plot as $\lambda _{Lr} $. Halos with $c>3.48$ have increasing kinetic energy with \textit{c}. Large halos with fast mass accretion have a vanishing radial linear momentum with $c=3.48$ and a (almost) minimum radial kinetic energy for all different concentration \textit{c}. Large halos with fast mass accretion tend to grow with vanishing radial momentum and minimum radial kinetic energy.

\subsection{Effects of radial flow on halo velocity dispersion}
\label{sec:3.6}
The Jeans' equation coupled with inverse mass cascade can be used to study the effect of radial flow on velocity dispersion. First, the gravitational potential in terms of unknown function $F\left(x\right)$ reads  
\begin{equation}
\label{ZEqnNum207666} 
\phi _{h} \left(r,a\right)=-G\int _{r}^{\infty }\frac{m_{r} \left(y,a\right)}{y^{2} } dy =-v_{cir}^{2} \frac{c}{F\left(c\right)} \int _{x}^{\infty }\frac{F\left(y\right)}{y^{2} } dy .     
\end{equation} 
A shifted gravitation potential can be introduced to satisfy $\phi _{h}^{*} \left(r=0,a\right)=0$,
\begin{equation} 
\label{ZEqnNum269276} 
\begin{split}
\phi _{h}^{*} \left(r,a\right)&=\phi _{h}^{} \left(r,a\right)+v_{cir}^{2} \frac{c}{F\left(c\right)} \int _{0}^{\infty }\frac{F\left(y\right)}{y^{2} } dy\\
&=v_{cir}^{2} \frac{c}{F\left(c\right)} \int _{0}^{x}\frac{F\left(y\right)}{y^{2} } dy.
\end{split}
\end{equation} 
The full dynamic Jeans' equation along radial direction is usually written as 
\begin{equation} 
\label{ZEqnNum535400} 
\begin{split}
\underbrace{\frac{\partial u_{r} }{\partial t} +u_{r} \frac{\partial u_{r} }{\partial r} }_{1}&+\frac{1}{\rho _{h} } \frac{\partial \left(\rho _{h} \sigma _{r}^{2} \right)}{\partial r} +\frac{2}{r} \beta _{h} \sigma _{r}^{2} \left(r,a\right)\\
&=-\frac{\partial \phi _{h} \left(r,a\right)}{\partial r} =-\frac{Gm_{r} \left(r,a\right)}{r^{2} } , 
\end{split}
\end{equation} 
where $\sigma _{r}^{2} \left(r,a\right)$ is the radial velocity dispersion. The anisotropy of velocity dispersion is defined through an anisotropy parameter $\beta _{h} =1-{\sigma _{t}^{2}/\sigma _{r}^{2} } $, where $\sigma _{t}^{2} \left(r,a\right)$ is the tangential velocity dispersion. For isotropic velocity dispersion, we have $\sigma _{t} =\sigma _{r} $ and $\beta _{h} =0$. The full dynamic Jeans' equation \eqref{ZEqnNum535400} relates non-zero radial flow to halo velocity dispersion $\sigma _{r}^{2}$ for a non-rotating spherical halo. The dynamics of a rotating halo with finite angular momentum is much more complicated and presented in a separate paper \citep{Xu:2022-The-mean-flow--velocity-disper}.  

Term 1 from the radial flow is often neglected for small virialized halos ($u_{r} =0$) and the radial velocity dispersion can be solved by the static Jeans equation with known $m_{r} \left(r,a\right)$ or a given density profile \citep{Binney:1982-M-L-and-Velocity-Anisotropy-fr}. However, large halos with fast mass accretion are dynamic objects, where mass cascade/accretion leads to a non-zero mean radial flow $u_{r} $ that will contribute significantly to velocity dispersion (especially to the outer region of halos). Here we attempt to solve an inverse problem, i.e. solving for the velocity dispersion $\sigma _{r}^{2} $ with a known mean radial flow $u_{r} $. After substituting the expressions for $m_{r} \left(r,a\right)$ (Eq. \eqref{ZEqnNum632204}) and $u_{r} \left(r,a\right)$ (Eq. \eqref{ZEqnNum849591}) into Jeans' equation \eqref{ZEqnNum535400} with chain rule from $x={r/r_{s} \left(t\right)} $,
\begin{equation}
\frac{\partial }{\partial t} =\frac{\partial }{\partial x} \frac{\partial x}{\partial t} =-\frac{x}{t} \frac{\partial \ln r_{s} }{\partial \ln t} \frac{\partial }{\partial x} \quad \textrm{and} \quad \frac{\partial }{\partial r} =\frac{\partial }{\partial x} \frac{\partial x}{\partial r} =\frac{1}{r_{s} } \frac{\partial }{\partial x},    
\label{ZEqnNum253473}
\end{equation}

\noindent The original Jeans' equation becomes
\begin{equation} 
\label{eq:61}
\begin{split}
&\underbrace{\sigma _{r}^{2} \frac{\partial \ln \left(\rho _{h} \sigma _{r}^{2} \right)}{\partial \ln x} }_{1}\\
&=x\underbrace{\frac{r_{s}^{2} }{t^{2} } \left[\frac{\partial u_{h} }{\partial x} \left(x\frac{\partial \ln r_{s} }{\partial \ln t} -u_{h} \right)+u_{h} \left(1-\frac{\partial \ln r_{s} }{\partial \ln t} \right)\right]}_{2}-\underbrace{v_{c}^{2} }_{3}.
\end{split}
\end{equation} 
An equivalent equation in terms of the function $F\left(x\right)$ (using Eq. \eqref{ZEqnNum482501} for density) reads 
\begin{equation}
\label{ZEqnNum325109} 
\frac{c^{2} }{x\rho _{h} v_{cir}^{2} } \frac{\partial \rho _{h} \sigma _{r}^{2} }{\partial x} =\frac{1}{4\pi ^{2} } \frac{F\left(x\right)^{2} F^{''} \left(x\right)}{xF^{'} \left(x\right)^{3} } -\frac{\rho _{h} \left(x\right)}{\bar{\rho }_{h} \left(a\right)} \frac{3F\left(x\right)}{xF^{'} \left(x\right)} ,      
\end{equation} 
where $v_{c}^{} $ is the circular velocity at radius \textit{r} (Eq. \eqref{ZEqnNum537303}) and $\bar{\rho }_{h} \left(a\right)$ is the average halo density\textit{.} Here, ${\partial \ln r_{s}/\partial \ln t} =1$ from mass cascade was used (Eq. \eqref{ZEqnNum952904}). Term 1 comes from the pressure gradient due to radial velocity dispersion, term 2 is due to the nonzero radial flow, and term 3 comes from gravity. 

For small halos with a stable core, the stable clustering hypothesis is valid and $u_{h} =0$ (term 2 vanishes), where the pressure (term 1) exactly balances the gravity (term 3) everywhere. While for the other limiting situation, i.e. large halos with fast mass accretion, the Hubble flow ($u_{h} \left(x\right)={2x/3} $) at halo center leads to a central core with a finite core density $\rho _{h} \left(0\right)\equiv \rho _{h} \left(x=0\right)$ (Eq. \eqref{ZEqnNum973033}). For core region with $u_{h} \left(x\right)={2x/3} $, Eq. \eqref{ZEqnNum325109} can be transformed to 
\begin{equation}
\label{ZEqnNum786564} 
\frac{c^{2} }{x\rho _{h} \left(x\right)v_{cir}^{2} } \frac{\partial \left(\rho _{h} \sigma _{r}^{2} \right)}{\partial x} =\frac{1}{\Delta _{c} } -\frac{\rho _{h} \left(x\right)}{\bar{\rho }_{h} \left(a\right)} =-\delta \left(x\right)\frac{\bar{\rho }\left(a\right)}{\bar{\rho }_{h} \left(a\right)} ,      
\end{equation} 
where $\Delta _{c} =18\pi ^{2} $ is the critical density ratio, $\delta \left(x\right)$ is overdensity, and $\bar{\rho }\left(a\right)$ is background density. The pressure in core region can be approximated by a parabolic function of \textit{x} (from Eq. \eqref{ZEqnNum786564}),
\begin{equation}
\label{ZEqnNum253695} 
p_{h} \left(x\right)=\rho _{h} \left(x\right)\sigma _{r}^{2} \left(x\right)=p_{h} \left(x=0\right)-\frac{1}{2} J_{c} x^{2} ,       
\end{equation} 
where constant $J_{c} $ in the unit of pressure is (from Eq. \eqref{ZEqnNum786564}),
\begin{equation}
\label{ZEqnNum594491} 
J_{c} =\left(\frac{\rho _{h} \left(0\right)}{\bar{\rho }_{h} } -\frac{1}{18\pi ^{2} } \right)\frac{\rho _{h} \left(0\right)v_{cir}^{2} }{c^{2} } \approx \frac{\rho _{h}^{2} \left(0\right)v_{cir}^{2} }{\bar{\rho }_{h} c^{2} } .      
\end{equation} 
A core size $x_{c} $ where Hubble flow is dominant can defined by setting $p_{h} \left(x_{c} \right)=0$ in Eq. \eqref{ZEqnNum253695},
\begin{equation}
\label{ZEqnNum946518} 
x_{c} =\sqrt{\frac{2\bar{\rho }_{h} \left(a\right)}{\rho _{h} \left(0\right)} } \frac{c\sigma _{r} \left(0\right)}{v_{cir} } . 
\end{equation} 

Next, let's work on the velocity dispersion profile. The general expression for the radial dispersion $\sigma _{r}^{2} $ reads (from Eq. \eqref{ZEqnNum325109})
\begin{equation} 
\label{ZEqnNum154562} 
\begin{split}
\frac{\partial }{\partial x} \left[\frac{F^{'} \left(x\right)\sigma _{r}^{2} }{x^{2} } \right]&=\frac{F^{'} \left(x\right)}{x^{2} } \frac{r_{s}^{2} }{t^{2} } \left[\frac{\partial u_{h} }{\partial x} \left(x\frac{\partial \ln r_{s} }{\partial \ln t} -u_{h} \right)\right.\\
&\left.+u_{h} \left(1-\frac{\partial \ln r_{s} }{\partial \ln t} \right)\right]-\frac{Gm_{h} }{r_{h} } \frac{cF\left(x\right)F^{'} \left(x\right)}{F\left(c\right)x^{4}}.
\end{split}
\end{equation} 
If we apply the evolution of halo size $r_{h} \sim t$ or ${\partial \ln r_{h}/\partial \ln t} =1$ from mass cascade, the integration of Eq. \eqref{ZEqnNum154562} leads to an explicit expression for radial dispersion normalized by circular velocity $\sigma _{nr}^{} ={\sigma _{r}/v_{cir} =} \sigma _{r} {t/\left(2\pi r_{h} \right)} $ (with Eqs. \eqref{ZEqnNum837340} and \eqref{ZEqnNum273026}),
\begin{equation} 
\label{ZEqnNum807329} 
\begin{split}
\sigma _{nr}^{2}&=\underbrace{\frac{x^{2} }{4\pi ^{2} c^{2} F^{'} \left(x\right)} \int _{x}^{\infty }\frac{F\left(x\right)^{2} }{x^{2} } \left(\frac{1}{F^{'} \left(x\right)} \right)^{'}  dx}_{1}\\
&\quad\quad\quad\quad\quad\quad+\underbrace{\frac{cx^{2} }{F^{'} \left(x\right)} \int _{x}^{\infty }\frac{F\left(x\right)F^{'} \left(x\right)}{F\left(c\right)x^{4} }  dx}_{2}
\end{split}
\end{equation} 
with two separate contributions from radial flow (term 1) and from gravitational potential (term 2), respectively. Term 1 is usually neglected for small virialized halos with $u_{r} =0$, but can be important for large halos with fast mass accretion. Here we require the pressure term $\left. \left(\rho _{h} \sigma _{r}^{2} \right)\right|_{x=\infty } =0$ at infinity when integrating Eq. \eqref{ZEqnNum154562}.  
For an isothermal profile with $F\left(x\right)=x$, the normalized dispersion has a constant value of $\sigma _{nr}^{2} ={1/2} $. For NFW profile with $F\left(x\right)=\ln \left(1+x\right)-{x/\left(1+x\right)} $, two contributions can be derived explicitly from Eq. \eqref{ZEqnNum807329}. Term 1 reads
\begin{equation} 
\label{ZEqnNum410484}
\begin{split}
\sigma_{nr1}^{2}=&-\frac{\left(1+x\right)^{2}}{36\pi ^{2} x^{2} c^{2}} \Big\{3x^{2} -4\pi ^{2} x^{3} +12x^{3} \textrm{Polylog} \left(2,1+x\right)\\
&-\Big[3\left(1+x\right)^{2} \left(2x-1\right)\ln\left(1+x\right)+6x-12x^{2}\\
&-12x^{3} \ln x-12i\pi x^{3}\Big]\ln\left(1+x\right)\Big\},
\end{split}
\end{equation} 
with an approximation of\\
$\sigma _{nr1}^{2} \approx \left(\frac{1}{18} -\frac{1}{3\pi^{2} } \right)\frac{x}{c^{2} } $ for $x\to 0$.\\ 
Term 2 becomes 
\begin{equation} 
\label{ZEqnNum447542}
\begin{split}
&\sigma _{nr2}^{2} =\frac{c\ln \left(1+x\right)}{2xF\left(c\right)} +\frac{c}{2F\left(c\right)} \Big\{-1-9x-7x^{2}\\
&+\left[-2-8x-4x^{2} +x^{3} \right]\ln \left(1+x\right)\\+ 
&\Big[\pi ^{2} +6poly\log \left(2,-x\right)-\ln x+3\left(\ln \left(1+x\right)\right)^{2} \Big]x\left(1+x\right)^{2} \Big\},
\end{split}
\end{equation} 
with the approximation\\
$\sigma _{nr2}^{2} \approx -\frac{c}{2F\left(c\right)} x\ln \left(x\right)$ for $x\to 0$.\\

\begin{figure}
\includegraphics*[width=\columnwidth]{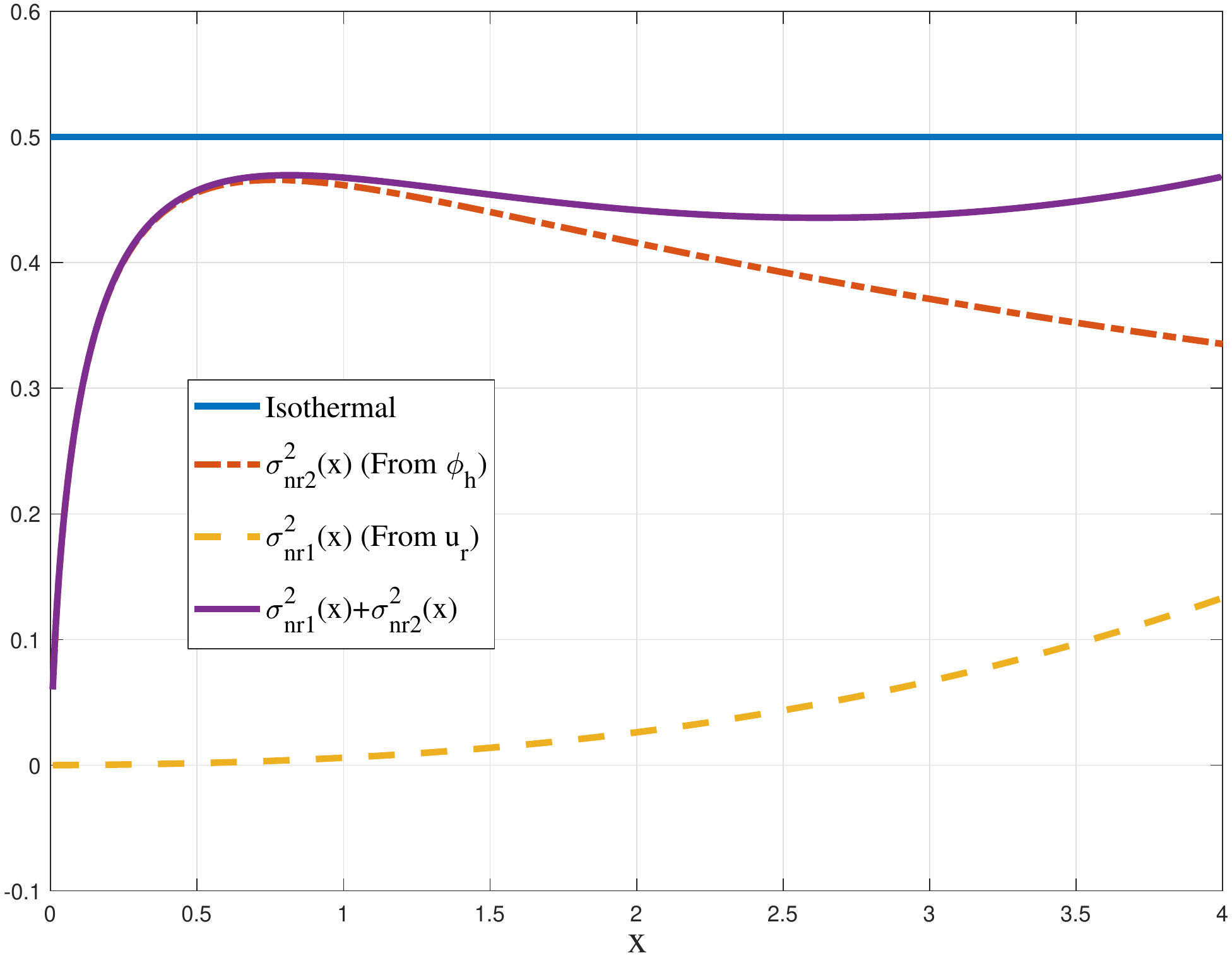}
\caption{The normalized radial velocity dispersion $\sigma _{nr}^{2} \left(x\right)$ for an isothermal profile (a constant value of 1/2) and NFW profile (varying with \textit{x} for $c=4$) with two contributions, i.e. $\sigma _{r1}^{2} \left(x\right)$ from the mean radial flow (Eq. \eqref{ZEqnNum410484}) and $\sigma _{r2}^{2} \left(x\right)$ from the gravitational potential (Eq. \eqref{ZEqnNum447542}), respectively. The radial flow tends to enhance the radial random motion and is only significant for large \textit{x} in halo outer region.}
\label{fig:4}
\end{figure}

Figure \ref{fig:4} plots the variation of total radial velocity dispersion $\sigma _{nr}^{2} \left(x\right)=\sigma _{nr1}^{2} \left(x\right)+\sigma _{nr2}^{2} \left(x\right)$ for an isothermal profile and NFW profile ($c=4$). Two separate contributions are also presented in the same plot, i.e. $\sigma _{nr1}^{2} \left(x\right)$ from the mean radial flow and $\sigma _{nr2}^{2} \left(x\right)$ from the gravitational potential, respectively. The first contribution from mean radial flow tends to enhance the radial velocity dispersion and is only significant at a large \textit{x} for the outer region of halo.

Now we have complete solutions of radial pressure and potential for halos with a NFW profile with effect of radial flow or mass cascade included. Both are normalized by circular velocity $v_{cir} $ and read (using Eq. \eqref{ZEqnNum482501} for density and Eq. \eqref{ZEqnNum207666} for potential)
\begin{equation} 
\label{ZEqnNum595258} 
p_{nr} \left(x\right)=\frac{\rho _{h} \sigma _{r}^{2} }{\bar{\rho }_{h} v_{cir}^{2} } =\frac{c^{3} F^{'} \left(x\right)}{3F\left(c\right)x^{2} } \sigma _{nr}^{2} \left(x\right),        
\end{equation} 
\begin{equation} 
\label{ZEqnNum686764} 
\phi _{nh} \left(x\right)=\frac{\phi _{h} }{v_{cir}^{2} } =-\frac{c\ln \left(1+x\right)}{xF\left(c\right)}. \end{equation} 

\begin{figure}
\includegraphics*[width=\columnwidth]{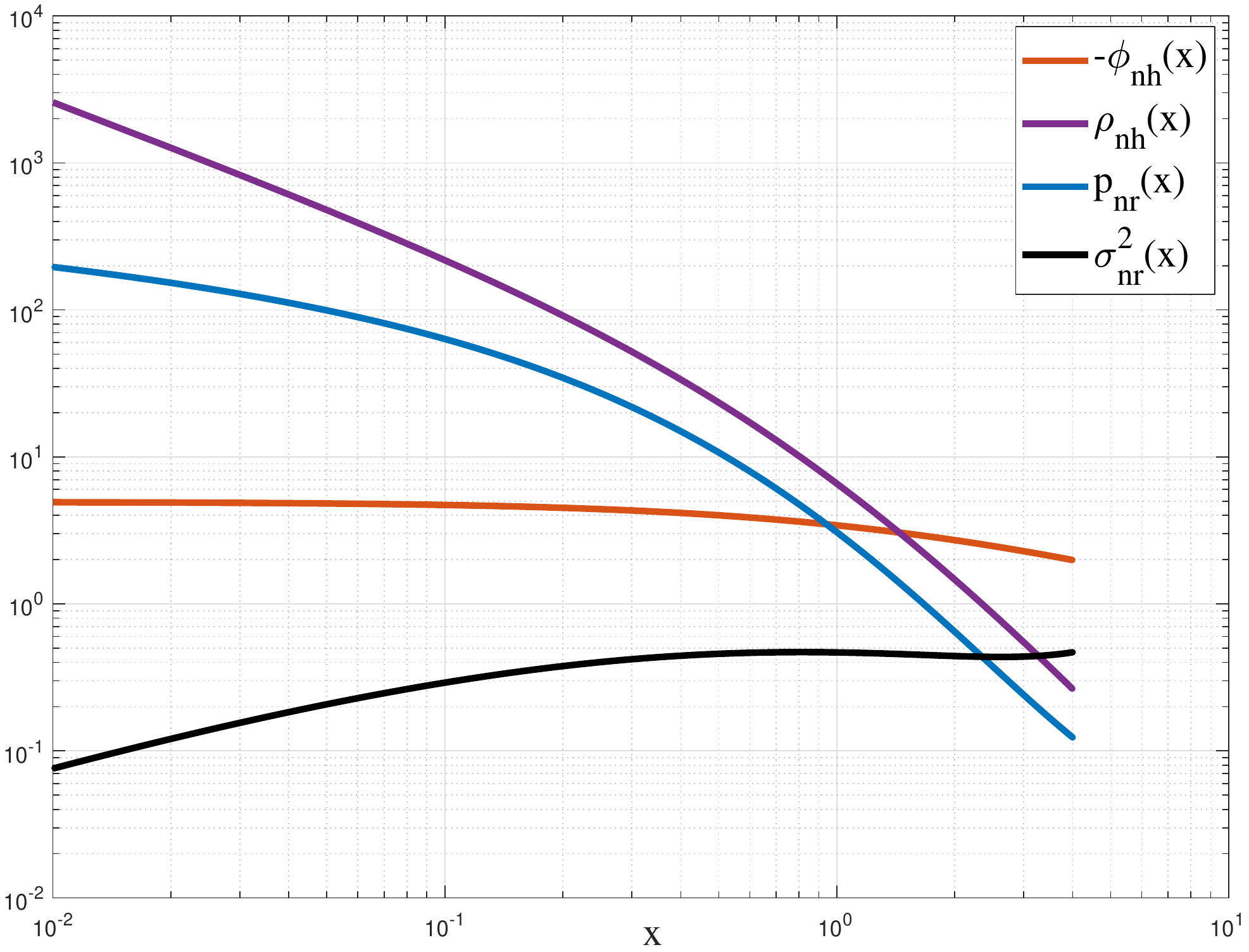}
\caption{Normalized density, pressure, gravitational potential, and radial velocity dispersion for a NFW profile with a nonzero radial flow.}
\label{fig:s2}
\end{figure}
\noindent Figure \ref{fig:s2} plots the normalized pressure $p_{nr} \left(x\right)$ (Eq. \eqref{ZEqnNum595258}), gravitational potential $\phi _{nh} \left(x\right)$ (Eq. \eqref{ZEqnNum686764} and radial velocity dispersion $\sigma _{nr}^{2} \left(x\right)$ (Eq. \eqref{ZEqnNum807329}) for a NFW profile with $c=4$. The halo density $\rho _{nh} \left(x\right)$ is normalized by the halo mean density $\bar{\rho }_{h} $. The density $\rho _{nh} \left(x\to 0\right)\sim x^{-1} $ and pressure $p_{nr} \left(x\to 0\right)\sim -\log \left(x\right)$. This leads to an Equation of State $p_{nr} \left(x\to 0\right)\sim a+b\log \left(\rho _{nh} \right)$ for NFW profile. Both pressure and density fields are divergent and irregular at the center of halo for NFW profile. 

A convenient formula for the logarithmic slope of radial pressure can be derived from the full Jeans' equation (Eq. \eqref{ZEqnNum325109}) for ${\partial \ln r_{h}/\partial \ln t} =1$, where
\begin{equation} 
\label{eq:73} 
\frac{\partial \ln p_{nr} }{\partial \ln x} =\frac{x^{2} -xu_{h} }{4\pi ^{2} c^{2} \sigma _{nr}^{2} } \frac{\partial u_{h} }{\partial x} -\frac{v_{nc}^{2} }{\sigma _{nr}^{2} } .        
\end{equation} 
Similarly, two contributions can be identified (from the mean radial flow $u_{h} $ and gravitational potential, respectively). At scale radius $r_{s} $, $\left. {\partial u_{h}/\partial x} \right|_{x=1} =0$ and the logarithmic slope is exactly ${-v_{nc}^{2}/\sigma _{nr}^{2} } $. The slope equals -2 everywhere for an isothermal profile.  

With expressions for all relevant halo quantities explicitly derived, the scaling of these quantities in core region is summarized in Table \ref{tab:2} that is fully determined by the deformation rate parameter $\gamma _{h} $. It should be noted that circular velocity and velocity dispersion follow same scaling ($v_{c}^{2} \left(r\right)\sim \phi _{h}^{*} \left(r\right)\sim \sigma _{r}^{2} \left(r\right)$) if $\gamma _{h} <{1/2} $, regardless of the value of $\gamma _{h}$. 

\begin{table}
\centering
\caption{Scaling at center of halo for different deformation rate parameter $\gamma_{h}$}
\begin{tabular}{p{0.6in}m{0.5in}m{0.45in}m{0.45in}m{0.45in}}
\hline 
&{$y$} & Isothermal  & NFW & Einasto\\ 
\hline 
$\gamma _{h} \ge 0$ &  & $\gamma _{h} =0$ & $\gamma _{h} ={1/2} $ & $\gamma _{h} ={2/3} $\\
\hline 
$F\left(x\right)\propto x^{y} $ (Eq. \eqref{ZEqnNum434585}) & {$y=\frac{1}{1-\gamma _{h} } $} & $y=1$ & $y=2$ & $y=3$ 
\\ \hline 
$u_{h} \left(x\right)\propto x^{y} $ (Eq. \eqref{ZEqnNum278808}) & {$y=1$} & $y=1$ & $y=1$ & $y=1$ \\ \hline 
$v_{c}^{2} \left(x\right)\propto x^{y} $ (Eq. \eqref{ZEqnNum537303}) & {$y=\frac{\gamma _{h} }{1-\gamma _{h} } $} & $y=0$ & $y=1$ & $y=2$ \\ \hline 
$\rho _{h} \left(x\right)\propto x^{y} $ (Eq. \eqref{ZEqnNum482501}) & {$y=\frac{3\gamma _{h} -2}{1-\gamma _{h} } $} & $y=-2$ & $y=-1$ & $y=0$ \\ \hline 
$\phi _{h}^{*} \left(x\right)\propto x^{y} $ (Eq. \eqref{ZEqnNum269276}) & $y=\frac{\gamma _{h} }{1-\gamma _{h} } $ & diverge \newline at $r=0$ & $y=1$ & $y=2$ \\ \hline 
$\sigma _{r}^{2} \left(x\right)\propto x^{y} $ (Eq. \eqref{ZEqnNum807329}) & \makecell{$\frac{1}{2} <\gamma _{h} <\frac{2}{3} $\\ $y=\frac{2-3\gamma _{h} }{1-\gamma _{h} } $} & N/A & N/A & $y=0$ \\ \hline 
$\sigma _{r}^{2} \left(x\right)\propto x^{y} $ (Eq. \eqref{ZEqnNum807329}) & {$\gamma _{h} ={1/2} $} & N/A & $-x\ln \left(x\right)$ & N/A\\ \hline 
$\sigma _{r}^{2} \left(x\right)\propto x^{y} $ (Eq. \eqref{ZEqnNum807329}) & \makecell{$\gamma _{h} <\frac{1}{2}$\\ $y=\frac{\gamma _{h} }{1-\gamma _{h}} $} & $y=0$ & N/A & N/A \\ \hline 
\end{tabular}
\label{tab:2}
\end{table}

\subsection{Effects of mass cascade on halo energies and surface tension }
\label{sec:3.7}
For complete effects of mass cascade on halo properties, total energies of entire halos are studied in this section. It was found that contributions from the radial flow to velocity dispersion could be important and should not be neglected for large halos. In contrast to small halos, large halos with fast mass accretion are dynamic objects with an expanding core and size. Multiplying the continuity Eq. \eqref{ZEqnNum114919} by the mean radial flow $u_{r} $ and the Jeans' equation \eqref{ZEqnNum535400} by density $\rho _{h} $, and adding two equations together, we have the equation
\begin{equation}
\label{ZEqnNum754617} 
\frac{\partial \left(\rho _{h} u_{r} \right)}{\partial t} +\frac{1}{r^{2} } \frac{\partial \left(\rho _{h} r^{2} u_{r}^{2} \right)}{\partial r} +\frac{\partial \left(\rho _{h} \sigma _{r}^{2} \right)}{\partial r} +\rho _{h} \frac{Gm_{r} \left(r,a\right)}{r^{2} } =0.     
\end{equation} 
Multiplying Eq. \eqref{ZEqnNum754617} by $4\pi r^{3} $ and integrating with respect to $r$ from 0 to $r_{h} $ leads to an exact  energy equation for non-rotating halos ,
\begin{equation} 
\label{ZEqnNum532404} 
I_{h} +S_{u} +S_{\sigma } -2K_{u} -6K_{\sigma } -\Phi _{h} =0,        
\end{equation} 
with all terms here normalized by $m_{h} v_{cir}^{2} $. This is a generalized version of virial theorem, as the standard virial theorem does not include the contributions from a nonzero radial flow through surface energy $S_{u} $ and kinetic energy $K_{u}$. Since $r_{h} =r_{h} \left(t\right)$, the integration of the first term in Eq. \eqref{ZEqnNum754617} can be separated into two contributions using the Leibniz's rule,
\begin{equation} 
\label{eq:76} 
\int _{0}^{r_{h} }4\pi r^{3} \frac{\partial \left(\rho _{h} u_{r} \right)}{\partial t} dr =\frac{\partial G_{h} }{\partial t} -4\pi r_{h}^{3} \rho _{h} \left(r_{h} \right)u_{r} \left(r_{h} \right)\frac{\partial r_{h} }{\partial t} ,      
\end{equation} 
where a halo virial quantity (radial momentum)
\begin{equation} 
\label{eq:77} 
G_{h} =\int _{0}^{r_{h} }4\pi r^{3} \rho _{h} u_{r} dr =\frac{m_{h} \left(t\right)v_{cir}^{2} t}{4\pi ^{2} c^{2} F\left(c\right)} \left[c^{2} F\left(c\right)-3\int _{0}^{c}xF\left(x\right)dx \right] 
\end{equation} 
is defined as the first order moment of radial flow. The virial quantity for peculiar velocity that excludes the Hubble flow reads
\begin{equation} 
\label{eq:78} 
\begin{split}
G_{hp}^{} &=\int _{0}^{r_{h} }4\pi r^{3} \rho _{h} \left(u_{r} -Hr\right)dr\\
&=\frac{m_{h} \left(t\right)v_{cir}^{2} t}{4\pi ^{2} c^{2} F\left(c\right)} \left[\frac{1}{3} c^{2} F\left(c\right)-\frac{5}{3} \int _{0}^{c}xF\left(x\right)dx \right].
\end{split}
\end{equation} 
For comparison, $L_{hr} \left(a\right)$ is the (zeroth order) linear momentum of radial flow. The (normalized) time derivative of the virial quantity ($I_{h} $) is obtained as, 
\begin{equation} 
\label{ZEqnNum811693} 
I_{h} =\frac{1}{m_{h} v_{cir}^{2} } \frac{\partial G_{h} }{\partial t} =\left[\frac{1}{2\pi ^{2} } -\frac{3}{2\pi ^{2} c^{2} F\left(c\right)} \int _{0}^{c}xF\left(x\right)dx \right].      
\end{equation} 
The surface energy terms include the contribution $S_{u} $ from the surface pressure due to radial flow at halo surface, 
\begin{equation} 
\label{eq:80}
\begin{split}
S_{u}&=\frac{4\pi r_{h}^{3} \rho _{h} \left(r_{h} \right)}{m_{h} v_{cir}^{2} } \left[u_{r}^{2} \left(r_{h} \right)-u_{r} \left(r_{h} \right)\frac{\partial r_{h} }{\partial t} \right]\\
&=\frac{1}{4\pi ^{2} } \left[\frac{F\left(c\right)}{cF^{'} \left(c\right)} -1\right]=\frac{{1/\alpha _{h} -1} }{4\pi ^{2} } 
\end{split}
\end{equation} 
and the contribution $S_{\sigma }$ from the surface pressure due to velocity dispersion at halo surface. Since the radial velocity dispersion has two contributions (Eq. \eqref{ZEqnNum807329}), we have two corresponding contributions to the pressure term $S_{\sigma }$, i.e. from the radial flow ($S_{\sigma 1}$) and from the gravitational potential ($S_{\sigma 2}$), respectively,
\begin{equation}
\label{eq:81} 
S_{\sigma } =\frac{4\pi r_{h}^{3} \rho _{h} \left(r_{h} \right)\sigma _{r}^{2} \left(r_{h} \right)}{\left(m_{h} v_{cir}^{2} \right)} =S_{\sigma 1} +S_{\sigma 2} ,        
\end{equation} 
\begin{equation} 
\label{eq:82} 
S_{\sigma 1} =\frac{c}{4\pi ^{2} F\left(c\right)} \int _{c}^{\infty }\frac{F^{2} \left(x\right)}{x^{2} } \left(\frac{1}{F^{'} \left(x\right)} \right) ^{'} dx 
\end{equation} 
or
\begin{equation} 
\label{eq:83} 
S_{\sigma 1} =\frac{c}{4\pi ^{2} F\left(c\right)} \left\{\left(\left. \frac{F^{2} \left(x\right)}{x^{2} F^{'} \left(x\right)} \right|_{c}^{\infty } \right)-\int _{c}^{\infty }\left[\frac{2F\left(x\right)}{x^{2} } -\frac{2F^{2} \left(x\right)}{F^{'} \left(x\right)x^{3} } \right] \right\}dx,     
\end{equation} 
\begin{equation} 
\label{eq:84} 
S_{\sigma 2} =\frac{c^{4} }{F^{2} \left(c\right)} \int _{c}^{\infty }\frac{F\left(x\right)F^{'} \left(x\right)}{x^{4} }  dx.
\end{equation} 
The total kinetic energy of a halo includes the contribution directly from radial flow,
\begin{equation} 
\label{eq:85} 
\begin{split}
K_{u}&=\lambda _{Kr} =\frac{\int _{0}^{r_{h} }4\pi r^{2} \rho _{h} u_{r}^{2} dr }{2m_{h} v_{cir}^{2} }\\
&=\frac{1}{8\pi ^{2} } -\frac{\int _{0}^{c}xF\left(x\right)dx }{2\pi ^{2} c^{2} F\left(c\right)} +\frac{1}{8\pi ^{2} c^{2} F\left(c\right)} \int _{0}^{c}\frac{F^{2} \left(x\right)}{F^{'} \left(x\right)} dx.
\end{split}
\end{equation} 
Similarly, the second contribution of kinetic energy is from velocity dispersion (random motion) that again includes contributions from radial flow ($K_{\sigma 1} $) and from gravitational potential ($K_{\sigma 2} $), respectively, according to Eq. \eqref{ZEqnNum807329},
\begin{equation} 
\label{eq:86} 
K_{\sigma } =\frac{1}{2m_{h} v_{cir}^{2} } \int _{0}^{r_{h} }4\pi r^{2} \rho _{h} \sigma _{r}^{2} dr= K_{\sigma 1} +K_{\sigma 2} ,       
\end{equation} 
\begin{equation} 
\label{eq:87} 
K_{\sigma 1} =\frac{1}{8\pi ^{2} c^{2} F\left(c\right)} \int _{0}^{c}x^{2} \left[\int _{x}^{\infty }\frac{F\left(y\right)^{2} }{y^{2} } \left(\frac{1}{F^{'} \left(y\right)} \right) ^{'} dy\right] dx 
\end{equation} 
or if ${\mathop{\lim }\limits_{x\to \infty }} {F\left(x\right)^{2}/x^{2} F^{'} \left(x\right)} =0$, 
\begin{equation} 
\label{eq:88} 
\begin{split}
K_{\sigma 1} =-\frac{1}{24\pi ^{2} c^{2} F\left(c\right)} &\left\{\int _{0}^{c}\left(\frac{F\left(x\right)}{F^{'} \left(x\right)} +2x\right)F\left(x\right)dx\right.\\
&\left.+c^{3} \int _{c}^{\infty }\left[\frac{2F\left(x\right)}{x^{2} } -\frac{2F^{2} \left(x\right)}{F^{'} \left(x\right)x^{3} }\right] dx\right\}.
\end{split}
\end{equation} 
The kinetic energy of velocity dispersion due to gravitational interaction is
\begin{equation} 
\label{eq:89} 
K_{\sigma 2} =\frac{c^{4} }{6F{}^{2} \left(c\right)} \int _{c}^{\infty }\frac{F\left(x\right)F^{'} \left(x\right)}{x^{4} } dx +\frac{c}{6F{}^{2} \left(c\right)} \int _{0}^{c}\frac{F\left(x\right)F^{'} \left(x\right)}{x}  dx.    
\end{equation} 
The total gravitational potential of a halo is given by, 
\begin{equation}
\label{ZEqnNum765989} 
\Phi _{h} =-\frac{1}{m_{h} v_{cir}^{2} } \int _{0}^{r_{h} }4\pi r^{2} \rho _{h} \frac{Gm_{r} }{r}  dr=-\frac{c}{F{}^{2} \left(c\right)} \int _{0}^{c}\frac{F\left(x\right)F^{'} \left(x\right)}{x}  dx.    
\end{equation} 
In principle, we can derive explicit expressions for all these terms for a halo with a known function of $F\left(x\right)$. For example, some of these terms for a NFW profile are presented here, 
\begin{equation} 
\label{eq:91} 
I_{h} =\frac{\left[5c^{2} +6c-2\left(1+c\right)\left(c+3\right)\log \left(1+c\right)\right]\left(c-3\right)}{8\pi ^{2} c^{2} \left(1+c\right)\left[\log \left(1+c\right)-{c/ \left(1+c\right)} \right]} ,      
\end{equation} 
\begin{equation} 
\label{eq:92} 
S_{u} =\frac{\left[\log \left(1+c\right)-{c/ \left(1+c\right)} \right]\left(1+c\right)^{2} -c^{2} }{4\pi ^{2} c^{2} } ,       
\end{equation} 
\begin{equation} 
\label{eq:93} 
\Phi _{h} =-\frac{c\left[c\left(2+c\right)-2\left(1+c\right)\ln \left(1+c\right)\right]}{2\left[\ln \left(1+c\right)-{c/ \left(1+c\right)} \right]^{2} \left(1+c\right)^{2} } .       
\end{equation} 

\begin{figure}
\includegraphics*[width=\columnwidth]{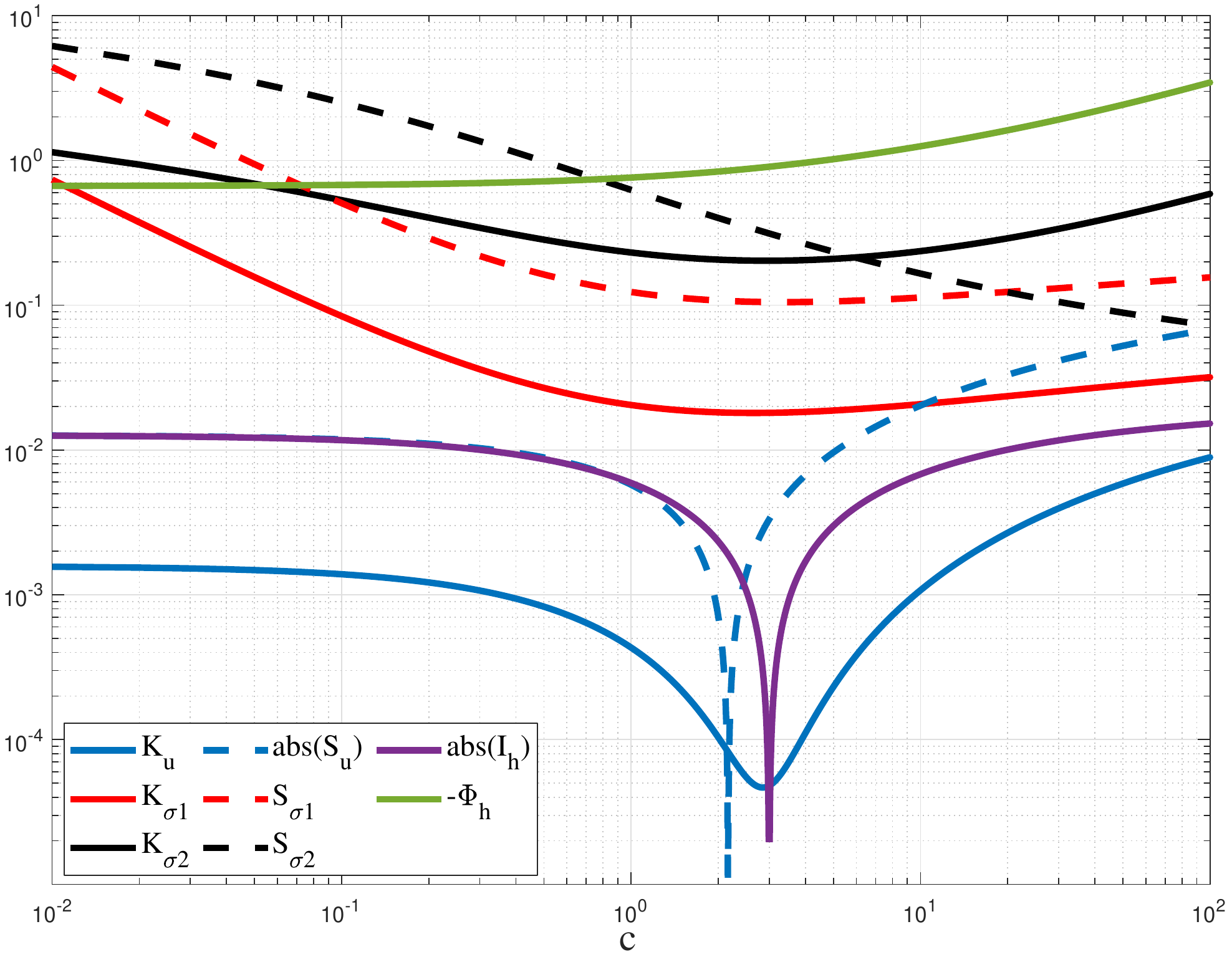}
\caption{The log-log variation of various halo energies with halo concentration parameter \textit{c} for NFW profile. The dynamic term $I_{h} \left(c\right)$ is positive for small \textit{c} and negative for large \textit{c}, while the surface energy $S_{u} \left(c\right)$ due to radial flow is negative for small \textit{c} and changing to be positive for large \textit{c}. The kinetic energy $K_{u} +K_{\sigma 1} $ due to the radial flow is small compared to the kinetic energy purely due to the velocity dispersion $K_{\sigma 2} $. However, the surface energy $S_{u} +S_{\sigma 1} $ is comparable to $S_{\sigma 2} $. This can be explained by the fact that radial flow is only significant in the outer region of halos.}
\label{fig:5}
\end{figure}

Figure \ref{fig:5} plots the variation of these energy terms with concentration \textit{c} for halos with a NFW profile. The dynamic term $I_{h} \left(c\right)$ is positive for small \textit{c} and negative for large \textit{c } with a critical concentration around $c=3$ where $I_{h} =0$. The surface energy $S_{u}$ due to radial flow is negative for small \textit{c} and changing to be positive for large \textit{c}. The total kinetic energy $K_{u} +K_{\sigma 1}$ due to radial flow is small compared to the kinetic energy due to random motion or velocity dispersion $K_{\sigma 2}$. However, the total surface energy $S_{u} +S_{\sigma 1}$ (due to radial flow) is comparable to $S_{\sigma 2}$ (due to velocity dispersion) and should not be neglected. This can be explained by the fact that the mean radial flow is only significant in the outer region of halos. 

Among these terms, $I_{h} $ due to radial flow is relatively small. The other three terms from radial flow should not be neglected and contribute to the total energy balance with $S_{\sigma 1} >K_{\sigma 1} >S_{u} $. By neglecting the dynamic term $I_{h} $ and using Eq. \eqref{ZEqnNum532404}, we write the surface energy of large halos with fast mass accretion,
\begin{equation} 
\label{ZEqnNum267268} 
\begin{split}
S_{eh}&=\left(S_{u} +S_{\sigma 1} +S_{\sigma 2} \right)m_{h} v_{cir}^{2}\\
&=\left[2\left(K_{u} +3K_{\sigma 2} +3K_{\sigma 1} \right)+\Phi _{h} \right]m_{h} v_{cir}^{2} , 
\end{split}
\end{equation} 
which is the extra energy required to create the expanding halo surface. The halo surface energy is also the difference between the total energy of halo with and without halo surface. The equivalent halo surface tension can be introduced as the surface energy per unit area, 
\begin{equation} 
\label{ZEqnNum212204} 
S_{th} =\frac{S_{eh} }{2A_{h} } =\frac{2\left(K_{u} +3K_{\sigma 2} +3K_{\sigma 1} \right)+\Phi _{h} }{8\pi r_{h}^{2} } m_{h} v_{cir}^{2} ,      
\end{equation} 
where $A_{h} =4\pi r_{h}^{2} $ is the surface area of a halo. For large halos with the limiting concentration $c=3.48$, the normalized surface tension (from Eq. \eqref{ZEqnNum212204}) is estimated to be around 
\begin{equation}
\begin{split}
&S_{nth} =\frac{2A_{h} S_{th} }{m_{h} v_{cir}^{2} } =\frac{S_{eh} }{m_{h} v_{cir}^{2} } \approx 0.3 \quad \textrm{for NFW profile}\\ 
&\textrm{and}\\
&S_{nth} =\frac{2A_{h} S_{th} }{m_{h} v_{cir}^{2} } =\frac{S_{eh} }{m_{h} v_{cir}^{2} } =0.5 \quad \textrm{for isothermal profile.}  \label{ZEqnNum336244}
\end{split}
\end{equation}
\noindent The normalized surface tension $S_{nth}$ is a constant regardless of halo mass and time, which contributes to the effective potential exponent of entire N-body system $n_e=-1.3\neq-1$ \citep[see][Fig. 1b]{Xu:2022-The-evolution-of-energy--momen} (also Eq. \eqref{ZEqnNum797157}).  

An equation analog to the Young--Laplace equation can be written to relate the pressure difference across halo surface to halo radius, or equivalently halo surface curvature,
\begin{equation} 
\label{eq:97} 
\begin{split}
\Delta P_{h}&=\frac{2S_{th} }{r_{h} } =\frac{S_{eh} }{A_{h} r_{h} }\\ &=\frac{1}{3} \left[S_{u} +S_{\sigma 1} +S_{\sigma 2} \right]\bar{\rho }_{h} v_{cir}^{2}=\frac{1}{3} S_{nth} \bar{\rho }_{h} v_{cir}^{2} .  
\end{split}
\end{equation} 
The pressure difference across surface is $\Delta P_{h} \approx 0.1\bar{\rho }_{h} v_{cir}^{2} $, which is also approximately the pressure right at halo surface (pressure is zero on the outside of halo). For a non-spherical halo, pressure may be different at different location depending on the local curvature. 

We can introduce surface density for a given halo as $\rho _{sur} ={N_{s} m_{p}/A_{h} } $, where $N_{s} $  is the number of particles on surface. Halo surface tension (an inherent property of halo surface) may be fully described by the surface density $\rho _{sur} $, gravitational constant \textit{G}, and halo size $r_{h} $,
\begin{equation} 
\label{eq:98} 
S_{th} =\alpha _{st} \left(G\right)^{\alpha _{1} } \left(\rho _{sur}^{} \right)^{\alpha _{2} } \left(r_{h} \right)^{\alpha _{3} } ,         
\end{equation} 
where $\alpha _{st} $ is a numerical constant. A simple dimensional analysis leads to expression
\begin{equation}
\label{ZEqnNum184683} 
S_{th} =\alpha _{st} G\rho _{sur}^{2} r_{h} ,          
\end{equation} 
where halo surface density reads (inserting Eq. \eqref{ZEqnNum336244} into \eqref{ZEqnNum184683})
\begin{equation} 
\label{ZEqnNum292775} 
\rho _{sur} =\sqrt{\frac{S_{nth} }{8\pi \alpha _{st} } } \frac{m_{h} }{r_{h}^{2} } .          
\end{equation} 
For halos with $\lambda ={2/3} $ and $\tau _{0} =1$, we have $m_{h} \propto r_{h} $, halo surface tension $S_{th} \propto r_{h}^{-1} $, halo surface density $\rho _{sur} \propto r_{h}^{-1}$, and thickness of halo surface layer $r_{p} ={\rho _{sur}/\rho _{h} \left(r_{h} \right)} \propto r_{h}^{} $ from Eq. \eqref{ZEqnNum227253}. A complete list of dependence of these parameters on the mass cascade parameters $\lambda $ and $\tau _{0} $ is presented in Table \ref{tab:3}.

\begin{table}
\centering
\caption{The dependence of power-law exponent \textit{m} ($\mathrm{\sim}$ $a^{m} $) on mass cascade}
\begin{tabular}{p{0.4in}p{0.4in}p{0.5in}p{0.4in}p{0.35in}p{0.35in}} \hline 
$m_{h} $ & $r_{h} $ & $\rho _{sur} $ & $S_{th} $\textbf{} & $v_{cir}^{2} $ & $r_{p} $ \\ \hline 
$\frac{3-2\tau _{0} }{2\left(1-\lambda \right)} $ & $\frac{9-2\tau _{0} -6\lambda }{6\left(1-\lambda \right)} $ & $\frac{12\lambda -9-2\tau _{0} }{6\left(1-\lambda \right)} $ & $\frac{6\lambda -3-2\tau _{0} }{2\left(1-\lambda \right)} $ & $\frac{3\lambda -2\tau _{0} }{3\left(1-\lambda \right)} $ & $\frac{9-2\tau _{0} -6\lambda }{6\left(1-\lambda \right)} $ \\ \hline 
Eq. \eqref{ZEqnNum952904} & Eq. \eqref{ZEqnNum952904} & Eq. \eqref{ZEqnNum292775} & Eq. \eqref{ZEqnNum212204} & Eq. \eqref{ZEqnNum837340} & Eq. \eqref{ZEqnNum227253} \\ \hline 
\end{tabular}
\label{tab:3}
\end{table}

From Table \ref{tab:3}, the thickness of surface layer $r_{p} $ is proportional to halo size $r_{h}$, i.e. $r_{p} \propto r_{h}$, regardless of the exact values of $\lambda$ and $\tau _{0}$ that may depend on the exact cosmology. This hints a geometric Brownian process (incremental change $r_{p}$ proportional to the current value $r_{h}$) for halo size in Section \ref{sec:4.1}.  

Finally, an effective exponent for gravitational interaction $n_{e} $ can be introduced based on the virial theorem for halos with fast mass accretion and expanding size, 
\begin{equation} 
\label{ZEqnNum797157} 
n_{e} =\frac{2\left(K_{u} +3K_{\sigma 2} +3K_{\sigma 1} \right)}{\Phi _{h} } =\frac{S_{nth} }{\Phi _{h} } -1.
\end{equation} 
With $c=3.48$ and $\Phi _{h} \approx -1$, the effective exponent $n_{e} \approx -1.3$. The deviation of $n_{e} \approx -1.3$ from -1 (the actual potential exponent is -1 for $V\left(r\right)\approx r^{-1} $) reflects the effects of surface energy/tension from inverse mass cascade. This can be directly confirmed by N-body simulations \citep[see][Fig. 1b]{Xu:2022-The-evolution-of-energy--momen}.
 
\section{Stochastic models for halo size and density profile}
\label{sec:4}
\subsection{Stochastic model for halo size evolution}
\label{sec:4.1}
The random walk of halos in mass space was applied to derive the double-$\lambda$ halo mass function \citep{Xu:2021-Inverse-mass-cascade-mass-function}. Similarly, stochastic models can be developed for halo size and particle distributions that describe the halo internal structure. The halo structure (distribution of particles) is highly dependent on the evolution of halo size, and therefore on the mass cascade. 

In mass cascade, the halo waiting time $\tau _{gr} $ is a random variable and follows an exponential distribution with mean $\tau _{g} =\left\langle \tau _{gr} \right\rangle \propto m_{h}^{-\lambda }$ \citep[see][Eq. (45)]{Xu:2021-Inverse-mass-cascade-mass-function}. The evolution of halo mass $m_{h}^{} $ can be modeled by a stochastic process with perturbations due to the randomness in halo waiting time $\tau _{gr}$,
\begin{equation} 
\label{ZEqnNum315498} 
\frac{\partial m_{h}^{} }{\partial t} =\frac{m_{p} }{\tau _{gr} } =\frac{m_{h}}{\tau _{g} n_{p}} \cdot \frac{\tau _{g}}{\tau _{gr}}.         
\end{equation} 
Equation \eqref{ZEqnNum315498} becomes deterministic equation by replacing the random waiting time $\tau _{gr}$ with the mean waiting time $\tau _{g}$ (see Eq. \eqref{ZEqnNum808457}). By introducing a random variable $\xi_{gr}$ and using Eq. \eqref{ZEqnNum821124},
\begin{equation}
\xi _{gr} \left(t\right)=\frac{\tau _{g}}{\tau _{gr}} -1\quad \textrm{and} \quad \tau _{g} n_{p} =\frac{3\left(1-\lambda \right)}{3-2\tau _{0} } t,   
\label{eq:103}
\end{equation}

\noindent we will have a stochastic differential equation for $m_{h}^{} $ 
\begin{equation}
\label{ZEqnNum513835} 
\frac{\partial \ln m_{h}^{} }{\partial \ln t} =\frac{3-2\tau _{0} }{3\left(1-\lambda \right)} \left(1+\xi _{gr} \left(t\right)\right),        
\end{equation} 
where $\xi _{gr} $ is approximately a Gaussian random variable with a zero mean. Equation \eqref{ZEqnNum513835} reduces to Eq. \eqref{ZEqnNum952904} for $\xi _{gr} \to 0$. As shown in Eqs. \eqref{ZEqnNum808457} and \eqref{ZEqnNum410355}, the original evolution of halo mass and size can be generalized to stochastic models
\begin{equation} 
\label{eq:105} 
\frac{\partial \ln m_{h}^{} }{\partial \ln t} =\frac{m_{p} t}{m_{h} \tau _{gr} }  
\end{equation} 
and 
\begin{equation} 
\label{ZEqnNum173628} 
\frac{\partial \ln r_{h} }{\partial \ln t} =\frac{m_{p} }{4\pi r_{h}^{3} } \frac{\alpha _{h} t}{\tau _{gr} \rho _{h} \left(r_{h} \right)} =\frac{\alpha _{h} \bar{\rho }_{h} }{3\rho _{h} \left(r_{h} \right)} \frac{m_{p} t}{m_{h}^{} \tau _{gr} } =\frac{\partial \ln m_{h}^{} }{\partial \ln t} \frac{\alpha _{h} \bar{\rho }_{h} }{3\rho _{h} \left(r_{h} \right)} ,    
\end{equation} 
where the average waiting time $\tau _{g} $ is simply replaced by a random waiting time $\tau _{gr} $. Finally, we have the stochastic equation for halo size $r_h$ from Eqs. \eqref{ZEqnNum513835} and \eqref{ZEqnNum173628},
\begin{equation}
\label{ZEqnNum278210} 
\frac{\partial \ln r_{h}^{} }{\partial \ln t} =\frac{\partial \ln m_{h} }{\partial \ln t} \frac{\alpha _{h} \bar{\rho }_{h} }{3\rho _{h} \left(r_{h} \right)} =\frac{3-2\tau _{0} }{3\left(1-\lambda \right)} \frac{\alpha _{h} \bar{\rho }_{h} }{3\rho _{h} \left(r_{h} \right)} \left(1+\xi _{gr} \left(t\right)\right).     
\end{equation} 
Halos evolving with a vanishing noise term in Eq. \eqref{ZEqnNum278210} always satisfy Eq. \eqref{ZEqnNum872917} with a mean halo density $\bar{\rho }_{h} =\Delta _{c} \bar{\rho }_{0} a^{-3} $. However, the existence of noise term $\xi _{gr} $ in Eq. \eqref{ZEqnNum278210} may drive halos away from Eq. \eqref{ZEqnNum872917}. At any instant, from Eqs. \eqref{ZEqnNum173628}, we should have 
\begin{equation}
\label{ZEqnNum641162} 
\frac{\partial \ln m_{h} }{\partial \ln r_{h} } =\frac{3\rho _{h} \left(r_{h} \right)}{\alpha _{h} \bar{\rho }_{h} } =\frac{3\rho _{sur} }{\alpha _{h} r_{p} \bar{\rho }_{h} } =\frac{3\left(3-2\tau _{0} \right)}{9-2\tau _{0} -6\lambda }  
\end{equation} 
to be always valid ($\tau_0$ and $\lambda$ are constants in mass cascade), where $\bar{\rho }_{h} \left(a\right)$ is the average halo density and $\rho _{sur}=r_{p}\rho _{h}(r_h)$ is the surface density. Therefore, these stochastic models (Eqs. \eqref{ZEqnNum513835} and \eqref{ZEqnNum278210}) describe randomly evolving mass and size of halos with fast mass accretion (i.e. constant concentration $c$ and/or shape parameter $\alpha$) and satisfying condition \eqref{ZEqnNum641162} at any instant \textit{t}, i.e. $m_{h} \propto r_{h} $ for $\tau _{0} =1$ and $\lambda ={2/3} $. Finally, the halos size $r_{h} \left(t\right)$ should evolve as (from Eqs. \eqref{ZEqnNum278210} and \eqref{ZEqnNum641162}),
\begin{equation}
\frac{dr_{h} }{dt} =b_{rh} \frac{r_{h} \left(t\right)}{t} \left(1+\xi _{gr} \left(t\right)\right) \quad \textrm{with} \quad b_{rh} =\frac{9-2\tau _{0} -6\lambda }{9\left(1-\lambda \right)}    
\label{ZEqnNum824858}
\end{equation}

\noindent from Eqs. \eqref{ZEqnNum278210} and \eqref{ZEqnNum641162}, which is a geometric Brownian motion with a multiplicative noise. The parameter $b_{rh} $ is from mass cascade and $b_{rh} =1$ for $\tau _{0} =1$ and $\lambda ={2/3} $. 

The evolution of halo size can be also understood as a result of fluctuating halo surface with a random velocity proportional to the velocity dispersion at the halo surface, i.e. the radial velocity dispersion $\sigma _{r} \left(r=r_{h} \right)$ discussed in Section \ref{sec:3}. The stochastic differential equation with an initial halo size of $r_{h} \left(t=t_{i} \right)=r_{h0} $ reads,  
\begin{equation} 
\label{ZEqnNum979638} 
\frac{dr_{h} \left(t\right)}{dt} =b_{rh} \frac{r_{h} \left(t\right)}{t} +\beta _{rh} \sigma _{r} \left(r_{h} \right)\bar{\xi }_{rh} \left(t\right),        
\end{equation} 
which is consistent with Eq. \eqref{ZEqnNum824858}. The covariance of the noise term satisfies $\left\langle \overline{\xi }_{rh} \left(t\right)\overline{\xi }_{rh} \left(t^{'} \right)\right\rangle ={\delta \left(t-t^{'} \right)/H} $. 

It is shown that the velocity dispersion at halo surface $\sigma _{r} \left(r_{h} \right)=\alpha _{rh} \left(c\right)v_{cir} =3\pi \alpha _{rh} Hr_{h} $ from Eqs. \eqref{ZEqnNum837340}, \eqref{ZEqnNum410484} and \eqref{ZEqnNum447542}, where $\alpha _{rh} \left(c\right)$ is a constant. For a limiting value of $c\approx 3.48$, $\alpha _{rh} {=1/\sqrt{2} } $ for an isothermal profile. After transforming the physical time \textit{t} to scale factor \textit{a}, Eq. \eqref{ZEqnNum979638} reads 
\begin{equation} 
\label{ZEqnNum251003} 
\frac{dr_{h} \left(t\right)}{dt} =\frac{3}{2} b_{rh} Hr_{h} \left(t\right)+Hr_{h} \left(t\right)\xi _{rh} \left(t\right),        
\end{equation} 
and
\begin{equation} 
\label{ZEqnNum400799} 
\frac{dr_{h} \left(t\right)}{d\ln a} =\frac{3}{2} b_{rh} r_{h} \left(t\right)+r_{h} \left(t\right)\xi _{rh} \left(t\right),        
\end{equation} 
where the multiplicative noise $r_{h} \left(t\right)\xi _{rh} \left(t\right)$ (proportional to $r_{h} $) describes the random evolution of halo size. The covariance of new noise $\xi _{rh} \left(t\right)$ satisfies 
\begin{equation} 
\label{eq:113} 
\left\langle \xi _{rh} \left(t\right)\xi _{rh} \left(t^{'} \right)\right\rangle =2D_{rh} {\delta \left(t-t^{'} \right)/H} , 
\end{equation} 
or equivalently the noise term $\xi _{gr} \left(t\right)$ in Eq. \eqref{ZEqnNum824858} satisfies
\begin{equation} 
\label{eq:114} 
\left\langle \xi _{gr} \left(t\right)\xi _{gr} \left(t^{'} \right)\right\rangle =4D_{rh} {t\delta \left(t-t^{'} \right)/\left(3b_{rh}^{2} \right)} ,       
\end{equation} 
where $D_{rh} ={\left(3\pi \alpha _{rh} \beta _{rh} \right)^{2}/2} $ is a dimensionless diffusion coefficient. The halo size $r_{h} \left(t\right)$ described by the geometric Brownian motion (Eq. \eqref{ZEqnNum400799}) has a lognormal probability distribution of ($r_{h0}$ is the initial halo size at starting time $t_i$)
\begin{equation} 
\label{ZEqnNum640133} 
\begin{split}
&P_{rh} \left(r_{h} ,t\right)=\frac{1}{r_{h} \sqrt{{8\pi D_{rh} \ln \left({t/t_{i} } \right)/3} } }\\
&\quad\quad \cdot \exp \left\{-\frac{\left(\ln \left({r_{h}/r_{h0} } \right)-\left(b_{rh} -{2D_{rh}/3} \right)\ln \left({t/t_{i} } \right)\right)^{2} }{{8D_{rh} \ln \left({t/t_{i} } \right)/3} } \right\},
\end{split}
\end{equation} 
with the \textit{m}th order moment of
\begin{equation}
\label{eq:116} 
\left\langle r_{h}^{m} \right\rangle =r_{h0}^{m} \left({t/t_{i} } \right)^{mb_{rh} +\frac{2}{3} m\left(m-1\right)D_{rh} } . 
\end{equation} 
The mean halo size grows linearly with time as $\left\langle r_{h} \left(t\right)\right\rangle =r_{h0} \left({t/t_{i} } \right)^{b_{rh} } \sim t^{b_{rh} } $, as expected.  The mode of the halo size grows as $r_{h0} \left({t/t_{i} } \right)^{b_{rh} -2D_{rh} } $ and the median halo size grows as $r_{h0} \left({t/t_{i} } \right)^{b_{rh} -{2D_{rh} /3} } $. Finally, the root mean square of halo size scales as $\left\langle r_{h}^{2} \right\rangle ^{{1/2} } =r_{h0} \left({t/t_{i} } \right)^{b_{rh} +{2D_{rh} /3} } $. Similarly, the halo mass ($m_{h} \propto r_{h} $) will also follow a lognormal distribution. 

\subsection{Stochastic model for particle distribution}
\label{sec:4.2}
To find the halo density profile, we need to derive the particle distribution function. The particle motion in halos is complicated as it is coupled to the varying halo size in previous section. This shares similarity with the derivation of diffusion equation for standard Brownian motion (see Appendix \ref{appendix:b} for a brief review). 

Let's consider the motion of a collisionless particle in a halo with varying size according to Eq. \eqref{ZEqnNum251003}. The goal is to derive the particle distribution function. The random position of that particle is 
\begin{equation} 
\label{ZEqnNum676188} 
r_{t} \left(t\right)={x_{t} \left(t\right)r_{s} \left(t\right)=x_{t} \left(t\right)r_{h} \left(t\right)/c},
\end{equation} 
where $r_{h} \left(t\right)$ is a stochastic time-varying halo size, $x_{t} \left(t\right)$ is the reduced position of that particle to the center of halo. For halos with a fixed size, $x_{t} \left(t\right)$ is expected to be a smooth function of time \textit{t} due to radial flow. The time variation of particle position $r_{t} \left(t\right)$ comes from both the time variation of $x_{t} \left(t\right)$ and the variation of halo size $r_{h} \left(t\right)$, i.e. $cdr_{t} =x_{t} \left(t\right)dr_{h} +r_{h} \left(t\right)dx_{t} $. The infinitesimal change $dr_{h} $ is presented in stochastic Eq. \eqref{ZEqnNum251003}. The infinitesimal change $dx_{t} $ for a fixed $r_{h} \left(t\right)$ is determined by the mean radial flow in Eq. \eqref{ZEqnNum278808},
\begin{equation}
u_{h} \left(x_{t} \right)=b_{rh} x_{t} \underbrace{-b_{rh} \frac{F\left(x_{t} \right)}{F^{'} \left(x_{t} \right)} }_{1}\quad \textrm{with} \quad b_{rh} =\left\langle \frac{\partial \ln r_{h} }{\partial \ln t} \right\rangle.   
\label{eq:118}
\end{equation}

\noindent Here, term 1 is the particle motion relative to the halo size change. Therefore, the infinitesimal change $dx_{t} $ for a fixed halo size $r_{h} \left(t\right)$ can be written as (see Appendix \ref{appendix:b} for stadnard Brownian motion)
\begin{equation} 
\label{ZEqnNum803708} 
r_{s} \left(t\right)\frac{dx_{tF} }{dt} =\frac{r_{s} \left(t\right)}{t} \left[u_{h} \left(x_{t} \right)-b_{rh} x_{t} +u_{h}^{*} \left(x_{t} \right)\right],       
\end{equation} 
and
\begin{equation} 
\label{ZEqnNum517814} 
r_{s} \left(t\right)\frac{dx_{tB} }{dt} =\frac{r_{s} \left(t\right)}{t} \left[u_{h} \left(x_{t} \right)-b_{rh} x_{t} -u_{h}^{*} \left(x_{t} \right)\right],       
\end{equation} 
for forward and backward change of $x_{t} \left(t\right)$ in time, respectively. Here $u_{h} \left(x_{t} \right)-b_{rh} x_{t} $ is the radial flow relative to the halo size change. Velocity $u_{h}^{*} \left(x_{t} \right)$ turns out to be the osmotic flow velocity acquired by particles in equilibrium to the external force. In Einstein's theory of Brownian motion, it has an origin from the osmotic pressure. Applying chain rule to Eq. \eqref{ZEqnNum676188} to obtain 
\[\frac{dr_{t} }{dt} =\frac{x_{t} \left(t\right)}{c} \frac{dr_{h} }{dt} +r_{s} \left(t\right)\frac{dx_{t} }{dt} \] 
and inserting Eq. \eqref{ZEqnNum251003} with $r_{s} \left(t\right)={r_{h} \left(t\right)/c} $ lead to the equation for particle position
\begin{equation} 
\label{ZEqnNum620785} 
\frac{dr_{t} }{dt} =\frac{r_{s} \left(t\right)}{t} \left[t\frac{dx_{t} }{dt} +b_{rh} x_{t} \right]+x_{t} \left(t\right)r_{s} \left(t\right)H\xi _{rh} \left(t\right).      
\end{equation} 
By inserting Eqs. \eqref{ZEqnNum803708} and \eqref{ZEqnNum517814} into Eq. \eqref{ZEqnNum620785}, the stochastic equations for $r_{t} \left(t\right)$ for forward and backward processes reads,
\begin{equation} 
\label{ZEqnNum544038} 
\frac{dr_{t} }{dt} =\underbrace{\frac{r_{s} \left(t\right)}{t} \left[u_{h} \left(x_{t} \right)+u_{h}^{*} \left(x_{t} \right)\right]}_{1}+\underbrace{\sigma \left(x_{t} \right)r_{s} \left(t\right)H\xi _{rh} \left(t\right)}_{2},      
\end{equation} 
\begin{equation} 
\label{ZEqnNum855118} 
\frac{dr_{t} }{dt} =\underbrace{\frac{r_{s} \left(t\right)}{t} \left[u_{h} \left(x_{t} \right)-u_{h}^{*} \left(x_{t} \right)\right]}_{3}+\sigma \left(x_{t} \right)r_{s} \left(t\right)H\xi _{rh}^{*} \left(t\right).      
\end{equation} 
The stochastic process $r_{t} \left(t\right)$ is not differentiable with respect to time \textit{t}, where the forward/backward velocities (the left/right side time derivatives of $r_{t} \left(t\right)$) can be different. The mean radial flow $u_{h} \left(x_{t} \right)$ (or the current velocity ) is the average of forward (term 1) and backward (term 3) velocities, while the osmotic flow $u_{h}^{*} \left(x_{t} \right)$ (or the fluctuation velocity) is the difference between forward and backward velocities that changes its sign in terms 1 and 3 (see Appendix \ref{appendix:b}). Both $u_{h} \left(x_{t} \right)$ and $u_{h}^{*} \left(x_{t} \right)$ contribute to the drift (terms 1 and 3) in Eqs. \eqref{ZEqnNum544038} and \eqref{ZEqnNum855118}, while the mean drift at given $x_{t} $ is the mean radial flow $u_{h} \left(x_{t} \right)$, as $u_{h}^{*} \left(x_{t} \right)$ is cancelled out. 

Noise terms $\xi _{rh} \left(t\right)$ and $\xi _{rh}^{*} \left(t\right)$ in term 2 represent the particle random motion due to a stochastic halo size $r_{h} \left(t\right)$, where $\xi _{rh} \left(t\right)$ is independent of $r_{t} \left(s\right)$ for $s\le t$ and $\xi _{rh}^{*} \left(t\right)$ is independent of $r_{t} \left(s\right)$ for $s\ge t$. The function $\sigma \left(x_{t} \right)$ indicates that the noise is of a multiplicative nature, i.e. the noise is dependent on the process $x_{t} $ itself. Here $\sigma \left(x_{t} \right)=x_{t} $ is expected (see Eq. \eqref{ZEqnNum620785}) because the halo size follows a geometric Brownian motion (Eq. \eqref{ZEqnNum251003}).

By comparing with Eqs. \eqref{eq:B2} and \eqref{eq:B3} for regular Brownian motion in Appendix \ref{appendix:b}, Eqs. \eqref{ZEqnNum544038} and \eqref{ZEqnNum855118} describe the random motion of collisionless particles with multiplicative noise due to the random halo size. This hints the halo internal structure (density profile) is highly correlated with inverse mass cascade. 

The corresponding Fokker-Planck equations (forward and backward in time) for probability $P_{r} \left(r,t\right)=P_{r} \left(x\left(t\right)\right)$ of particle position $r_{t}$ are used to describe the forward and backward processes,
\begin{equation} 
\label{ZEqnNum355357} 
\begin{split}
\frac{\partial P_{r} \left(r,t\right)}{\partial t} =&-\frac{r_{s} \left(t\right)}{t} \frac{\partial }{\partial r} \left[\left(u_{h} \left(x\right)+u_{h}^{*} \left(x\right)\right)P_{r} \right]\\
&+r_{s}^{2} \left(t\right)HD_{rh} \frac{\partial ^{2} }{\partial r^{2} } \left(\sigma ^{2} \left(x\right)P_{r} \right), 
\end{split}
\end{equation} 
\begin{equation} 
\label{ZEqnNum828202}
\begin{split}
\frac{\partial P_{r} \left(r,t\right)}{\partial t} =&-\frac{r_{s} \left(t\right)}{t} \frac{\partial }{\partial r} \left[\left(u_{h} \left(x\right)-u_{h}^{*} \left(x\right)\right)P_{r} \right]\\
&-r_{s}^{2} \left(t\right)HD_{rh} \frac{\partial ^{2} }{\partial r^{2} } \left(\sigma ^{2} \left(x\right)P_{r} \right).
\end{split}
\end{equation} 
Applying the chain rule from Eq. \eqref{ZEqnNum253473} and adding/subtracting Eqs. \eqref{ZEqnNum355357} to/from Eq. \eqref{ZEqnNum828202} lead to two independent equations for velocities $u_{h} \left(x\right)$ and $u_{h}^{*} \left(x\right)$, 
\begin{equation}
\label{ZEqnNum533047} 
b_{rh} x\frac{\partial P_{r} }{\partial x} =\frac{\partial }{\partial x} \left[u_{h} \left(x\right)P_{r} \right],         
\end{equation} 
\begin{equation} 
\label{ZEqnNum961031} 
u_{h}^{*} \left(x\right)=d_{r} \sigma ^{2} \left(x\right)\frac{\partial }{\partial x} \ln \left[\sigma ^{2} \left(x\right)P_{r} \left(x\right)\right],       
\end{equation} 
where $d_{r} =D_{rh} Ht={2D_{rh} /3} $. The integration of continuity Eq. \eqref{ZEqnNum533047} leads to the same expression as we have derived for $u_{h} \left(x\right)$ in Eq. \eqref{ZEqnNum278808}. For comparison, the osmotic velocity of standard Brownian motion has a dimensional form of $u_{h}^{*} =D{\nabla \rho /\rho } $ (i.e. related to the diffusion flux), where $D$ is the diffusivity and $\rho $ is the particle number density (See Appendix \ref{appendix:b} Eq. \eqref{ZEqnNum173816} for more details). 

As demonstrated in Section \ref{sec:3.3} (Eqs. \eqref{ZEqnNum621572} to \eqref{ZEqnNum676651}), the radial number density function $P_{r} \left(x\right)$ can be derived if the mean radial flow $u_{h} \left(x\right)$ is known. Similarly, number density $P_{r} \left(x\right)$ can be easily found for a given osmotic flow $u_{h}^{*} \left(x\right)$ and $\sigma ^{2} \left(x\right)$ from Eq. \eqref{ZEqnNum961031}, 
\begin{equation}
\label{ZEqnNum790424} 
P_{r} \left(x\right)=\frac{\alpha _{s} }{\sigma ^{2} \left(x\right)} \exp \left\{\frac{1}{d_{r} } \int \frac{u_{h}^{*} \left(x\right)}{\sigma ^{2} \left(x\right)} dx \right\},        
\end{equation} 
where $\alpha _{s} $ is a normalization constant for probability $P_{r} \left(x\right)$. For isothermal profile with constant radial number density $P_{r} \left(x\right)$, $u_{h} \left(x\right)=0$ and $u_{h}^{*} \left(x\right)=d_{r} {\partial \sigma ^{2} \left(x\right)/\partial x} $.

For a given function of $\sigma \left(x\right)$, a key relation between two velocities (current and osmotic flow) can be obtained from Eqs. \eqref{ZEqnNum533047} and \eqref{ZEqnNum961031},
\begin{equation} 
\label{ZEqnNum687681} 
u_{h}^{*} \left(x\right)=\frac{d_{r} \sigma ^{2} \left(x\right)}{x-u_{h} \left(x\right)} \frac{\partial u_{h} }{\partial x} +d_{r} \frac{\partial \sigma ^{2} \left(x\right)}{\partial x} .        
\end{equation} 
The closure problem of halo density profile is now equivalent to find an additional relation between two velocities $u_{h}^{*} \left(x\right)$ and $u_{h} \left(x\right)$. That relation combined with Eq. \eqref{ZEqnNum687681} will provide complete and consistent solutions of $u_{h} \left(x\right)$ and $u_{h}^{*} \left(x\right)$, and hence the halo density profile. Solutions for all other the relevant halo quantities can be obtained subsequently. 

However, unlike the simple closure $u_{h}^{} =-u_{h}^{*} $ for Brownian motion (see Appendix \ref{appendix:b}), it is much more complicated for dark matter flow due to the nature of long range gravitational interaction. More work is required along this line to better understand the fundamental mechanism behind a universal halo structure. A simple closure is proposed and discussed in Appendix \ref{appendix:b}.  

For a NFW profile, we have (from Eqs. \eqref{ZEqnNum533047} and \eqref{ZEqnNum961031})
\begin{equation}
\begin{split}
&P_{r} \left(x\right)=\frac{x}{\left(1+x\right)^{2} }, \quad u_{h} \left(x\right)=1+2x-\frac{\left(1+x\right)^{2} }{x} \ln \left(1+x\right)\\
&\textrm{and}\quad u_{h}^{*} \left(x\right)=d_{r} \frac{x\left(3+x\right)}{1+x}, 
\end{split}
\label{eq:130}
\end{equation}

\noindent where the integral of $P_{r} \left(x\right)$ diverges, a well-known difficulty of NFW profile. At this time, we will take a different route to model the halo internal structure by first identifying basic properties of the osmotic flow $u_{h}^{*} \left(x\right)$. A simple model of $u_{h}^{*} \left(x\right)$ is then proposed, followed by applying Eq. \eqref{ZEqnNum790424} for particle number density $P_{r} \left(x\right)$. 

Without loss of generality, let's assume a general power-law of $\sigma \left(x\right)=x^{\lambda _{r} } $ (we expect $\lambda _{r} =1$ though) and from Eq. \eqref{ZEqnNum961031}, 
\begin{equation} 
\label{eq:131} 
u_{h}^{*} \left(x\right)=d_{r} x^{2\lambda _{r} -1} \left(\frac{\partial \ln P_{r} }{\partial \ln x} +2\lambda _{r} \right).        
\end{equation} 
The derivative of $u_{h}^{*} \left(x\right)$ is
\begin{equation} 
\label{eq:132} 
\frac{\partial u_{h}^{*} \left(x\right)}{\partial x} =d_{r} x^{2\lambda _{r} -2} \frac{\partial ^{2} \ln P_{r} }{\partial \left(\ln x\right)^{2} } +\left(2\lambda _{r} -1\right)\frac{u_{h}^{*} \left(x\right)}{x} .       
\end{equation} 
The properties of $u_{h}^{*} \left(x\right)$ can be identified from above equations,
\begin{equation}
\begin{split}
&u_{h}^{*} \left(x=0\right)=0, \quad u_{h}^{*} \left(x=x_{0}^{*} \right)=0\\
&\textrm{when}\quad \left. \frac{\partial \ln P_{r} }{\partial \ln x} \right|_{x=x_{0}^{*} } =\left. \frac{\partial \ln \rho _{h} }{\partial \ln x} \right|_{x=x_{0}^{*} } +2=-2\lambda _{r}, 
\end{split}
\label{ZEqnNum477779}
\end{equation}
\noindent and
\begin{equation}
\begin{split}
&u_{h}^{*} \left(x=1\right)=2\lambda _{r} d_{r}\\
&\textrm{because} \quad \left. \frac{\partial \ln P_{r} }{\partial \ln x} \right|_{x=1} =\left. \frac{\partial \ln \rho _{h} }{\partial \ln x} \right|_{x=1} +2=0,
\end{split}
\label{ZEqnNum956240}
\end{equation}

\noindent where $x=1$ is the mode of probability function $P_{r} \left(x\right)$. Specifically, for $\lambda _{r} =1$ (From Eq. \eqref{ZEqnNum687681}),
\begin{equation} 
\label{ZEqnNum497738} 
\left. \frac{\partial u_{h}^{*} }{\partial x} \right|_{x=0} =\gamma _{r} =d_{r} \frac{2-\gamma _{h} }{1-\gamma _{h} } , \end{equation} 
where $\gamma _{h} =\left. \left({\partial u_{h} /\partial x} \right)\right|_{x=0} $ is the halo deformation parameter we defined before (Table \ref{tab:2}). 

Similar to the mean radial flow $u_{h}^{} \left(x\right)$, the osmotic flow $u_{h}^{*} \left(x\right)$ initially increases as $\gamma _{r} x$ and reaches a maximum, then decreases to zero at $x=x_{0}^{*} $ where the logarithmic slope of density $\rho _{h} $ is $-2-2\lambda _{r} $ (Eq. \eqref{ZEqnNum477779}). A simple but general model of $u_{h}^{*} \left(x\right)$ (expansion around $x=0$) with three free parameters satisfying all conditions in Eqs.  \eqref{ZEqnNum477779}, \eqref{ZEqnNum956240} and \eqref{ZEqnNum497738} can be written as,
\begin{equation}
\begin{split}
u_{h}^{*} \left(x\right)=\gamma _{r} x-\beta _{r} x^{1+\alpha _{r} }\quad \textrm{with} \quad \alpha _{r} >0.      
\label{ZEqnNum927116}
\end{split}
\end{equation}
\noindent Obviously the condition Eq. \eqref{ZEqnNum956240} requires,  
\begin{equation} 
\label{eq:137} 
\beta _{r} =d_{r} \left(\frac{2-\gamma _{h} }{1-\gamma _{h} } -2\lambda _{r} \right).         
\end{equation} 
The general solution of $P_{r}^{} \left(x\right)$ can be obtained from Eq. \eqref{ZEqnNum790424} with given $u_{h}^{*} \left(x\right)$ in Eq. \eqref{ZEqnNum927116},
\begin{equation}
P_{r} \left(x\right)=\alpha _{s} x^{-2\lambda _{r} } \exp \left\{\frac{x^{2-2\lambda _{r} } }{d_{r} } \left(\frac{\gamma _{r} }{2-2\lambda _{r} } -\frac{\beta _{r} x^{\alpha _{r} } }{2-2\lambda _{r} +\alpha _{r} } \right)\right\}    
\label{ZEqnNum714625}
\end{equation}
\noindent for $\lambda _{r} \ne 1$, where $\alpha _{s}$ is a normalization constant. While for $\lambda _{r} =1$,
\begin{equation}
P_{r} \left(x\right)=\frac{\alpha _{r} x^{{\gamma _{r} /d_{r} -2} } }{\Gamma \left[{\left(\gamma _{r} -d_{r} \right)/\left(\alpha _{r} d_{r} \right)} \right]} \left(\frac{\beta _{r} }{\alpha _{r} d_{r} } \right)^{\frac{\gamma _{r} -d_{r} }{\alpha _{r} d_{r} } } \exp \left\{-\frac{\beta _{r} x^{\alpha _{r} } }{\alpha _{r} d_{r} } \right\}.  
\label{ZEqnNum230459}
\end{equation}

The condition of maximum mean radial flow at $r=r_{s} $ ($\left. {\partial P_{r} /\partial x} \right|_{x=1} =0$ from Eq. \eqref{ZEqnNum273026}) requires $x=1$ is the mode of distributions $P_{r} \left(x\right)$, i.e. we will find maximum number of particles at $r=r_{s} $. For $\left. {\partial P_{r} /\partial x} \right|_{x=1} =0$ applied to Eq. \eqref{ZEqnNum714625}, we have the relation $2d_{r} \lambda _{r} =\gamma _{r} -\beta _{r} $. Specifically, for $\lambda _{r} =1$, relations between the drift and noise terms in Eqs. \eqref{ZEqnNum544038} and \eqref{ZEqnNum855118} are found as (analogy to the fluctuation-dissipation theorem)
\begin{equation}
2d_{r} =\gamma _{r} -\beta _{r} \quad \textrm{and} \quad \gamma _{r} =d_{r} \frac{2-\gamma _{h} }{1-\gamma _{h} },      
\label{eq:140}
\end{equation}

\noindent such that the particle distribution function (Eq. \eqref{ZEqnNum230459}) is reduced to a two-parameter distribution 
\begin{equation}
\label{ZEqnNum103079} 
P_{r} \left(x\right)=\frac{b_{r} {}^{a_{r} } }{\Gamma \left(a_{r} \right)\left(a_{r} -b_{r} \right)} \exp \left(-b_{r} x^{\frac{1}{a_{r} -b_{r} } } \right)x^{\frac{b_{r} }{a_{r} -b_{r} } }  
\end{equation} 
in terms of $a_{r} ={\left(\gamma _{r}-d_{r} \right)/\left(\alpha _{r} d_{r} \right)} $ and $b_{r} ={\beta _{r} /\left(\alpha _{r} d_{r} \right)} $, where $\alpha _{r} ={1/\left(a_{r} -b_{r} \right)}$. 

If we require $\gamma _{r} =\gamma _{h} $ for $x\to 0$ (i.e. the inward particle motion $u_{h} \left(x\right)-u_{h}^{*} \left(x\right)$ vanishes at the center of halo), we can determine that the constant $d_{r} ={1/6} $ (hence $D_{rh} ={1/4} $) with $\gamma _{h} ={2/3} $ from Eq. \eqref{ZEqnNum801048} that is required by Hubble flow at halo center. The other option is to require $\gamma _{r} =2\gamma _{h} $ for $x\to 0$ (the inward particle motion $u_{h} \left(x\right)-u_{h}^{*} \left(x\right)=-u_{h} \left(x\right)$, i.e. the inward mass flow to the halo center balances the outward mass flow). The constant $d_{r} ={1/3} $ (hence $D_{rh} ={1/2} $). The values of relevant parameters for different options are listed in Table \ref{tab:4}.   

\begin{table}
\caption{Values of relevant parameters for three possible options}
\begin{tabular}{p{0.1in}p{0.1in}p{0.15in}p{0.1in}p{0.05in}p{0.1in}p{0.15in}p{0.15in}p{0.4in}p{0.4in}} 
\hline 
$\gamma _{r} $ & $\gamma _{h} $  & $D_{rh} $ & $d_{r} $ & $\lambda _{r} $ & $\beta _{r} $ & $a_{r} $ & $b_{r} $ & \makecell{$u_{h}$\\$(x\to 0)$} & \makecell{$u_{h}^{*}$\\$(x\to 0)$} \\ \hline 
${1/2} $ & ${2/3} $ & ${3/16} $ & ${1/8} $ & $1$ & ${1/4} $ & ${3/\alpha _{r} } $ & ${2/\alpha _{r} } $ & ${2x/3} $ & ${x/2} $ \\ \hline 
${2/3} $ & ${2/3} $ & ${1/4} $ & ${1/6} $ & $1$ & ${1/3} $ & ${3/\alpha _{r} } $ & ${2/\alpha _{r} } $ & ${2x/3} $ & ${2x/3} $ \\ \hline 
${4/3} $ & ${2/3} $ & ${1/2} $ & ${1/3} $ & $1$ & ${2/3} $ & ${3/\alpha _{r} } $ & ${2/\alpha _{r} } $ & ${2x/3} $ & ${4x/3} $ \\ \hline 
\end{tabular}
\label{tab:4}
\end{table}

The function $F_{r} \left(x\right)$ is the fraction of particles with a distance smaller than a given \textit{x}, i.e. the cumulative distribution function of probability distribution $P_{r} \left(x\right)$, 
\begin{equation} 
\label{ZEqnNum768710} 
\begin{split}
&F_{r} \left(x=\frac{r}{r_{s} } \right)=\int _{0}^{x}P_{r} \left(y\right) dy=\frac{m_{r} }{m_{h} } \\
&=1-\frac{\Gamma \left(a_{r} ,b_{r} x^{{1/\left(a_{r} -b_{r} \right)} } \right)}{\Gamma \left(a_{r} \right)} =\frac{\gamma \left(a_{r} ,b_{r} x^{{1/\left(a_{r} -b_{r} \right)} } \right)}{\Gamma \left(a_{r} \right)},
\end{split}
\end{equation} 
where $\Gamma \left(x,y\right)$ and $\gamma \left(x,y\right)$ are the upper and lower incomplete Gamma functions, respectively. The corresponding halo density profile $\rho _{h} \left(x\right)$ is given by 
\begin{equation} 
\label{ZEqnNum495681} 
\begin{split}
\rho _{h} \left(x\right)&=\frac{m_{h} P_{r} \left(x\right)}{4\pi r_{s}^{3} x^{2} }\\
& =\frac{m_{h} }{4\pi r_{s}^{3} } \frac{b_{r} {}^{a_{r} } }{\Gamma \left(a_{r} \right)\left(a_{r} -b_{r} \right)} \exp \left(-b_{r} x^{\frac{1}{a_{r} -b_{r} } } \right)x^{\frac{3b_{r} -2a_{r} }{a_{r} -b_{r} } },
\end{split}
\end{equation} 
or equivalently with $\rho _{s} =\rho _{h} \left(x=1\right)$,
\begin{equation}
\rho _{h} \left(x\right)=\rho _{s} e^{b_{r} } \exp \left(-b_{r} x^{\frac{1}{a_{r} -b_{r} } } \right)x^{\frac{3b_{r} -2a_{r} }{a_{r} -b_{r} } }.    
\label{ZEqnNum239340}
\end{equation}

\noindent It can be verified that for inner profile $\rho _{h} \left(r<r_{s} \right)\propto r^{{\left(3\gamma _{h} -2\right)/\left(1-\gamma _{h} \right)} } $. The halo deformation rate parameter $\gamma _{h} $ is related to the new density profile as $\gamma _{h} ={b_{r} /a_{r} } $. The density profile in Eqs. \eqref{ZEqnNum495681} or \eqref{ZEqnNum239340} is general with four parameters ($\rho _{s}$, $r_{s}$, $a_{r}$ and $b_{r}$). For $\gamma _{h} ={b_{r} /a_{r} } ={2/3} $, Eq. \eqref{ZEqnNum239340} exactly reduces to the Einasto profile (Eq. \eqref{ZEqnNum305317}) with shape parameter $\alpha ={2/b_{r} } $. 

However, the density profile we proposed is a result of the random motion of collisionless particles in a halo with stochastically varying size according to Eq. \eqref{ZEqnNum251003}. The distribution $P_{r} \left(x\right)$ in Eq. \eqref{ZEqnNum103079} can be interpreted as the probability to find a particle at any position extending to infinity or the distribution of particles in an assembled halo (composite halo) incorporating all possible halos sizes. By contrast, the Einasto profile is for individual halos of finite size. Therefore, parameters of Eq. \eqref{ZEqnNum239340} can be different from the Einasto profile for individual halos with $\alpha\approx 0.2$ (as shown in Fig. \ref{fig:7} with $\alpha\approx 0.7$ for composite halos).

The number density $P_{r} \left(x\right)$ in Eq. \eqref{ZEqnNum103079} describes the radial density of collisionless particles from all halos with different sizes evolving according to Eq. \eqref{ZEqnNum251003}. The distribution $P_{r} \left(x={r/r_{s} } \right)$ is defined for all $x\ge 0$ extending to infinity. This density describes the probability distribution of all particles in all possible halos. It is different from the usual density profiles for individual halos with a finite size. 

\subsection{Density profiles from N-body simulation}
\label{sec:4.3}
In simulation, instead of working with the spherical averaged halo density profile for each individual halo with a finite size, we compute the density profile for a group of halos of same mass. The function $F_{r} \left(r\right)$ was computed for a halo group of size $n_{p} $ at a given scale factor \textit{a}, where $r$ is the distance of every particle in the halo group to the respective center of mass of the halo that it belongs to. In simulation, the cumulative function $F_{r} \left(r\right)$ is computed as the fraction of all particles in the same halo group with a distance smaller than $r$. The radial density profile $P_{r} \left(x={r/r_{s} } \right)$ (particle distribution probability) can be obtained by taking the derivative ${\partial F_{r} \left(x\right)/\partial x} $ (Eq. \eqref{ZEqnNum768710}). This procedure will significantly reduce the noise as the number of particles in the entire halo group is much greater than the number of particles in individual halos.

\begin{figure}
\includegraphics*[width=\columnwidth]{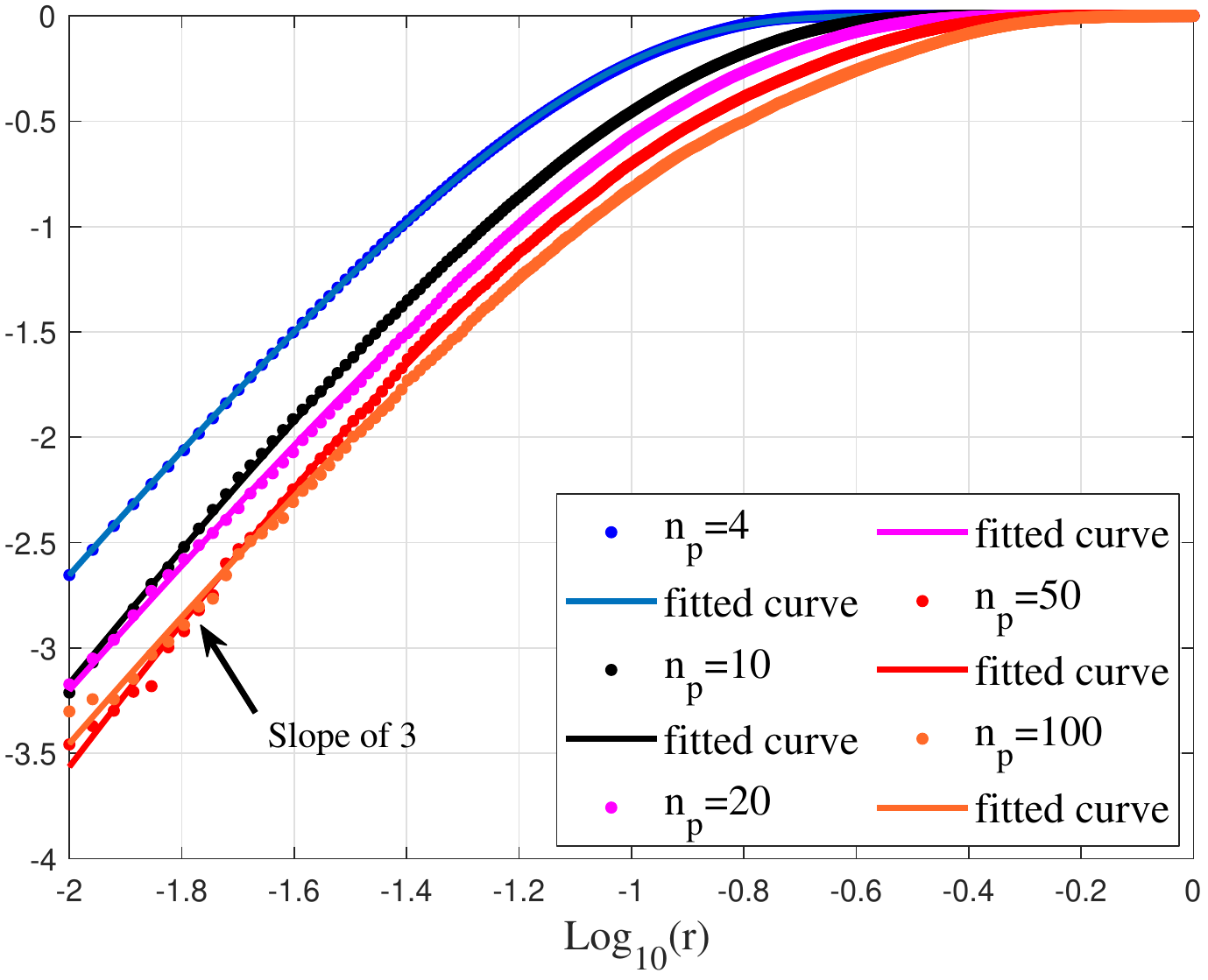}
\caption{The $log_{10}-log_{10}$ variation of the cumulative distribution function $F_{r} \left(r=xr_{s} \right)$ with the distance \textit{r} to the center of mass of the halo that particle belongs to. Function $F_{r} \left(r=xr_{s} \right)$ is computed based on all particles in a group of halos with same size $n_{p} $ at \textit{z}=0. Symbols of dot present the simulation data for five different sizes of halo groups. Solid lines plot the best fit of simulation data for each size of halo group using model (Eq. \eqref{ZEqnNum768710}) with three free parameters $r_{s} $, $a_{r} $, and $b_{r} $. The fitted values of these parameters are presented in Figs. \ref{fig:7} and \ref{fig:8}. The scaling $F_{r} \left(x={r/r_{s} } \right)\sim x^{3} $ for small \textit{x} clearly indicates the existence of a central core for composite halos.}
\label{fig:6}
\end{figure}

\begin{figure}
\includegraphics*[width=\columnwidth]{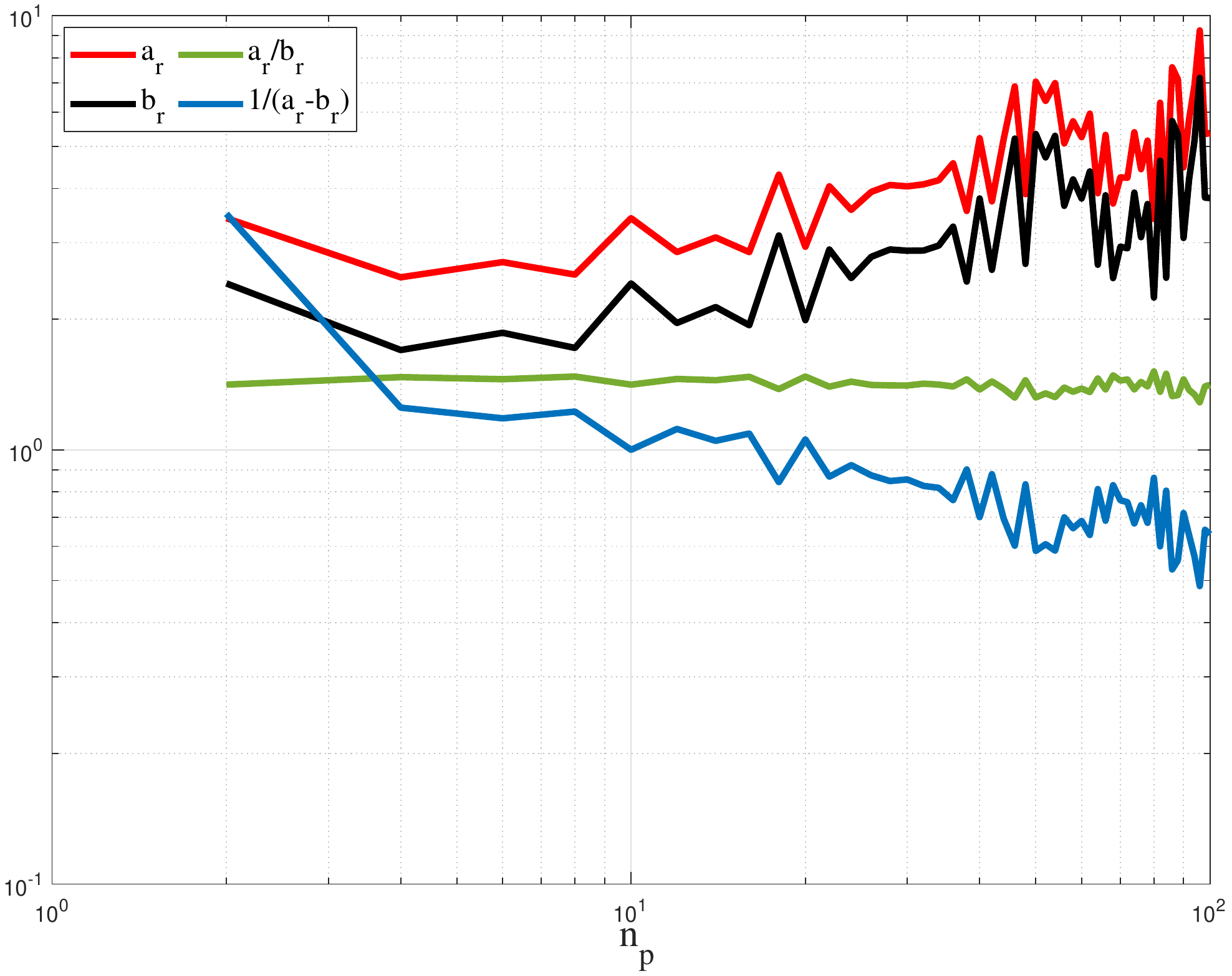}
\caption{The variation of fitted values of $a_{r} $ and $b_{r} $ for cumulative distribution function $F_{r} \left(r=xr_{s} \right)$ with halo group size $n_{p} $. Both values of $a_{r} $ and $b_{r} $ increase with the size $n_{p} $. However, the ratio of ${a_{r} /b_{r} } \approx {3/2} $ is found regardless of the halo size, as required by a finite density at halo center. The exponent parameter $\alpha _{r} ={1/\left(a_{r} -b_{r} \right)}=\alpha$ slowly decreases with the halo size from 1.2 to 0.7. This value is significantly larger than the shape parameter of $\alpha \approx 0.2$ for Einasto profile of individual halos with a finite size.}
\label{fig:7}
\end{figure}

\begin{figure}
\includegraphics*[width=\columnwidth]{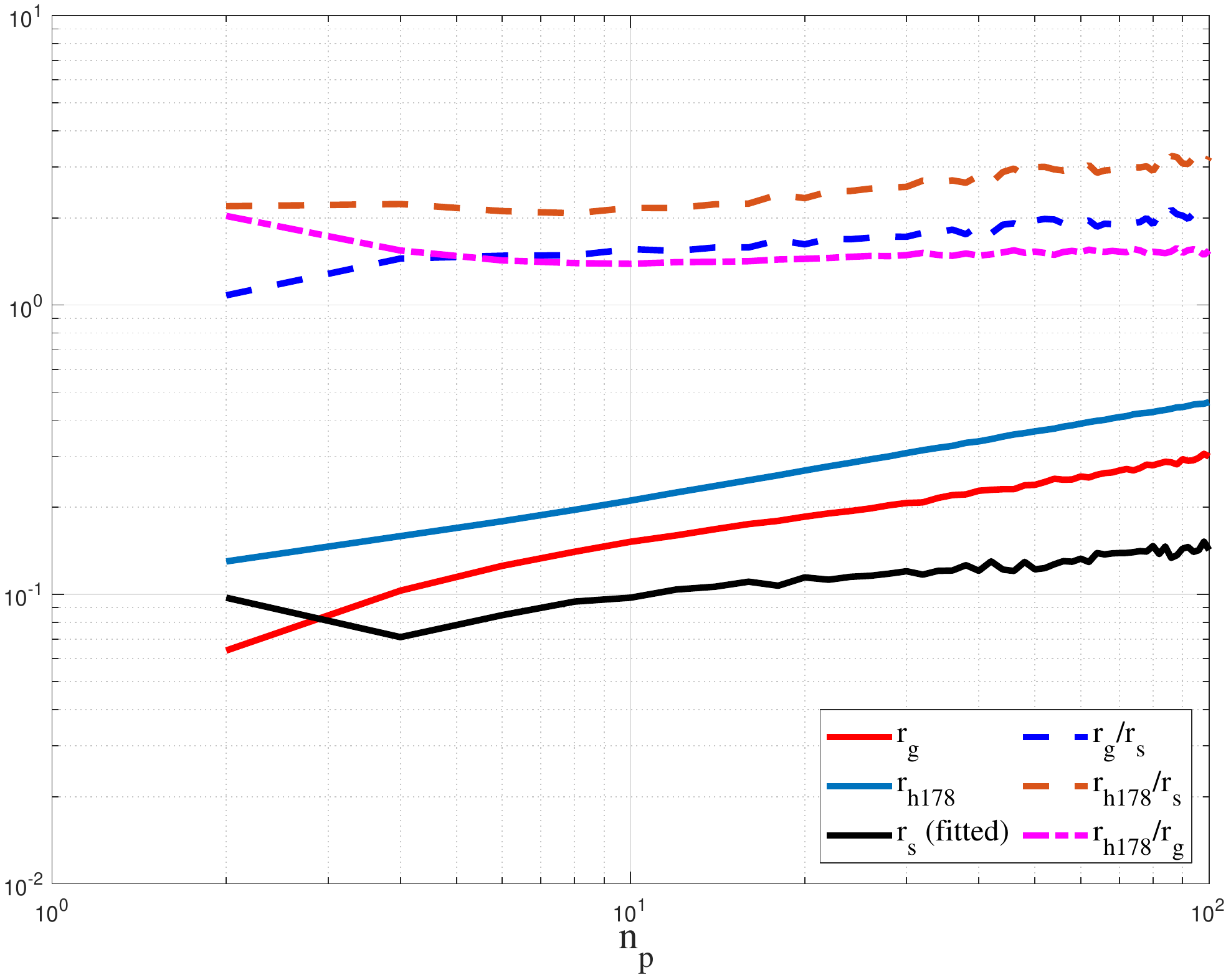}
\caption{The variation of three length scales (in the unit of Mpc/h) with halo group size $n_{p} $, i.e. the virial radius $r_{h\Delta } $ with $\Delta =178$ (Eq. \eqref{ZEqnNum125294}), the mean square radius $r_{g} $ (Eq. \eqref{ZEqnNum195815}, and the scale radius $r_{s} $ (fitted from the simulation data in Fig. \ref{fig:6}).}
\label{fig:8}
\end{figure}

Figure \ref{fig:6} plots the log-log variation of the radial cumulative distribution function $F_{r} \left(r=xr_{s} \right)$ for all particles in the same halo group of size $n_{p} $ at \textit{z}=0. Symbols plot the simulation data for five different sizes of halo groups, while the solid lines plot the best fit of the simulation data for each size of halo group using the proposed model Eq. \eqref{ZEqnNum768710} with three free parameters $r_{s} $, $a_{r} $, and $b_{r} $. The fitted values of three parameters varying with group size $n_{p} $ are presented in Figs. \ref{fig:7} and \ref{fig:8}. Clearly, the scaling $F_{r} \left(x={r/r_{s} } \right)\sim x^{3} $ for small \textit{x} indicates the existence of a central core at halo center (Table \ref{tab:2}). The proposed model Eq. \eqref{ZEqnNum768710} provides very good agreement with the simulation data for a wide range of halo group size.  

Figure \ref{fig:7} presents the variation of fitted values of $a_{r} $ and $b_{r} $ for $F_{r} \left(r=xr_{s} \right)$ with the halo group size $n_{p} $. Both values of $a_{r} $ and $b_{r} $ slowly increase with the halo size $n_{p} $. However, the ratio of ${a_{r} /b_{r} } \approx {3/2} $ is found for all halo sizes, regardless of the halo size, which is required by a finite density at halo center (Eq. \eqref{ZEqnNum239340}). Parameter $\alpha _{r} ={1/\left(a_{r} -b_{r} \right)}=\alpha$ slowly decreases with halo size from 1.2 to 0.7, which is significantly larger than shape parameter $\alpha \approx 0.2$ for Einasto profile of individual halos with finite size.

The \textit{k}th moment of distribution $P_{r} \left(x\right)$ can be easily found as
\begin{equation} 
\label{eq:145} 
M_{k} =\int _{0}^{\infty }P_{r} \left(x\right) x^{k} dx=\frac{\Gamma \left(a_{r} +\left(a_{r} -b_{r} \right)k\right)}{b_{r} {}^{\left(a_{r} -b_{r} \right)k} \Gamma \left(a_{r} \right)} .       
\end{equation} 
The momentum generating function of the distribution $P_{r} \left(x\right)$ reads
\begin{equation}
\label{eq:146} 
MGF\left(t\right)=\int _{0}^{\infty }P_{r} \left(x\right) e^{tx} dx=\sum _{k=0}^{\infty }\frac{t^{k} \Gamma \left(a_{r} +\left(a_{r} -b_{r} \right)k\right)}{k!b_{r} {}^{\left(a_{r} -b_{r} \right)k} \Gamma \left(a_{r} \right)}      
\end{equation} 
from which the halo mean square radius $r_{g} $ (the root mean square distance) reads
\begin{equation} 
\label{eq:147} 
r_{g} =r_{s} \sqrt{\frac{b_{r} {}^{-2\left(a_{r} -b_{r} \right)} \Gamma \left(3a_{r} -2b_{r} \right)}{\Gamma \left(a_{r} \right)} }  .        
\end{equation} 
Three characteristic length scales can be identified from the simulation data for group of halos of the same mass. The scale radius (halo core size) $r_{s} $ can be found by fitting Eq. \eqref{ZEqnNum768710} to the simulation data of $F_{r} \left(r=xr_{s} \right)$ in Fig. \ref{fig:6}. The mean square radius of halo group can be computed,
\begin{equation} 
\label{ZEqnNum195815} 
r_{g} =\sqrt{{\sum _{m=1}^{n_{h} }\sum _{k=1}^{n_{p} }\left(r_{km}^{2} \right)  /\left(n_{p} n_{h} \right)} } ,         
\end{equation} 
where $r_{km}$ is the distance of the \textit{k}th particle in \textit{m}th halo of the halo group to the center of that halo. Here $n_{h} $ is the number of halos in the group. The virial radius $r_{h\Delta } $ of a halo group can be found from the simulation data of $F_{r} \left(r=xr_{s} \right)$ as
\begin{equation}
\label{ZEqnNum125294} 
\frac{F_{r} \left(r_{h\Delta } \right)}{r_{h\Delta }^{3} } =\frac{4\pi \Delta _{c} \bar{\rho }\left(a\right)}{3m_{h} } ,
\end{equation} 
where $\bar{\rho }\left(a\right)$ is the mean background density at scale factor \textit{a}. 

Figure \ref{fig:8} plots the variation of three characteristic length scales (in the unit of Mpc/h) with the halo group size $n_{p} $, i.e. the virial radius $r_{h\Delta } $ with $\Delta _{c} =178$ (Eq. \eqref{ZEqnNum125294}), the mean square radius $r_{g} $ (Eq. \eqref{ZEqnNum195815}), and the scale radius $r_{s} $ (fitted from the simulation data in Fig. \ref{fig:6}). The ratios between three halo sizes are also presented. Note that all three length scales are defined based on the statistics of the entire halos group, instead of individual halos. 

Finally, all other relevant halo quantities can be obtained with number density $P_{r} \left(x\right)$ from Eq. \eqref{ZEqnNum103079}. Examples are the mean radial flow $u_{h} \left(x\right)$ from Eq. \eqref{ZEqnNum533047},
\begin{equation} 
\label{ZEqnNum980800} 
\begin{split}
u_{h} \left(x\right)&=x-\left(a_{r} -b_{r} \right)b_{r} {}^{-a_{r} }\\ 
&\cdot \exp \left(b_{r} x^{\frac{1}{a_{r} -b_{r} } } \right)x^{\frac{-b_{r} }{a_{r} -b_{r} } } \gamma \left(a_{r} ,b_{r} x^{\frac{1}{a_{r} -b_{r} } } \right).
\end{split}
\end{equation} 
The shifted potential $\phi _{h}^{*} \left(x,a\right)$ from Eq. \eqref{ZEqnNum269276},
\begin{equation} 
\label{ZEqnNum105452}
\begin{split}
&\phi _{h}^{*} \left(x,a\right)=\frac{Gm_{h} \left(a\right)}{\Gamma \left(a_{r} \right)r_{s} \left(a\right)}\\
&\quad\quad\quad \cdot \left[b_{r} {}^{a_{r} -b_{r} } \gamma \left(b_{r} ,b_{r} x^{\frac{1}{a_{r} -b_{r} } } \right)-\frac{1}{x} \gamma \left(a_{r} ,b_{r} x^{\frac{1}{a_{r} -b_{r} } } \right)\right].
\end{split}
\end{equation} 
The lower incomplete Gamma function has the properties that
\begin{equation}
\begin{split}
&\gamma \left(x,y\right)=\Gamma \left(x\right) \quad \textrm{for} \quad y\to \infty,\\ 
&\gamma \left(x,y\right)={y^{x}/x} \quad \textrm{for}\quad y\to 0, 
\end{split}
\label{eq:152}
\end{equation}

\noindent where the shifted potential $\phi _{h}^{*} \left(x,a\right)$ simplifies to 
\begin{equation} 
\label{eq:153} 
\phi _{h}^{*} \left(x\to \infty \right)=\frac{Gm_{h} \left(a\right)}{r_{s} \left(a\right)} \frac{b_{r}^{a_{r} -b_{r} } \Gamma \left(b_{r} \right)}{\Gamma \left(a_{r} \right)} ,        
\end{equation} 
\begin{equation} 
\label{eq:154} 
\phi _{h}^{*} \left(x\to 0\right)=\frac{Gm_{h} \left(a\right)}{r_{s} \left(a\right)} \frac{b_{r}^{a_{r} } \left(a_{r} -b_{r} \right)}{a_{r} b_{r} \Gamma \left(a_{r} \right)} x^{\frac{b_{r} }{a_{r} -b_{r} } } .       
\end{equation} 

\subsection{Equation of state for relative pressure and density}
\label{sec:4.4}
The velocity dispersion $\sigma _{nr}^{2} $ can be obtained with Eqs. \eqref{ZEqnNum807329} and $F_{r} \left(x\right)$ from \eqref{ZEqnNum768710}. It should be interesting to examine the equation of state (EOS) in the core region for small \textit{x}, where halo density can be well approximated (from Eq. \eqref{ZEqnNum239340} with $a_{r} ={3b_{r} /2}$) as
\begin{equation}
\begin{split}
&\rho_{h} \left(x\right)=\rho _{h} \left(0\right)\left(1-b_{r} x^{\frac{2}{b_{r} } } \right),\\
& \textrm{with} \\
&\rho _{h} \left(0\right)=\frac{m_{h} }{4\pi r_{s}^{3} } \frac{b_{r} {}^{a_{r} } }{\Gamma \left(a_{r} \right)\left(a_{r} -b_{r} \right)}.   
\end{split}
\label{ZEqnNum277257}
\end{equation}
\noindent From Eqs. \eqref{ZEqnNum253695} and \eqref{ZEqnNum594491}, pressure in core region is parabolic,
\begin{equation}
\label{ZEqnNum131959} 
p_{h} \left(x\right)=\rho _{h} \left(x\right)\sigma _{r}^{2} \left(x\right)=p_{h} \left(x=0\right)-\frac{1}{2} \frac{\rho _{h}^{2} \left(0\right)v_{cir}^{2} }{\bar{\rho }_{h} c^{2} } x^{2} .      
\end{equation} 
The equation of state in the core region for relative pressure and density can be finally written as (with Eq. \eqref{ZEqnNum277257}),
\begin{equation} 
\label{ZEqnNum609386} 
\left[p_{h} \left(0\right)-p_{h} \left(x\right)\right]=\frac{\left[\rho _{h}^{} \left(0\right)\right]^{2-b_{r} } v_{cir}^{2} }{2\left(b_{r} \right)^{b_{r} } \bar{\rho }_{h} c^{2} } \left[\rho _{h} \left(0\right)-\rho _{h} \left(x\right)\right]^{b_{r} }  
\end{equation} 
such that $\Delta p_{h} =K_{s} \left(\Delta \rho _{h} \right)^{b_{r} }$. Unlike the ideal gas with reference pressure and density being zero when molecules are infinitely far from each other, halos have their center pressure and density as a reference state where both gravitational and pressure forces vanish (gradient is zero). 

The relative pressure and density $\Delta p_{h} $ and $\Delta \rho _{h} $ to the center of halo satisfy the equation of state \eqref{ZEqnNum609386}. The parameter $b_{r} $ has the physical meaning as the exponent of equation of state. The regular polytropic equation of state $p_{h} \propto \rho _{h}^{1+{1/n} } $ for absolute pressure and density may not be applicable to dark matter halos. Figure \ref{fig:7} shows that the value of $b_{r} $ slightly increases from 1.5 to 3 with increasing halo group size. The NFW profile will not lead to such equation of state because of divergent pressure/density (Eqs. \eqref{ZEqnNum410484} and \eqref{ZEqnNum447542}).

Let's assume the equilibrium center pressure and density are $p_{h} \left(0\right)$ and $p_{h} \left(0\right)$, where both gravitational and pressure forces are not present. At any location in halo, the relative pressure $\Delta p_{hn} $

\noindent  and $\Delta \rho _{hn} $ (normalized and using Eqs. \eqref{ZEqnNum154562}, \eqref{ZEqnNum273026}, and \eqref{ZEqnNum482501}) can be written in terms of $F_{r} \left(x\right)$,
\begin{equation} 
\label{ZEqnNum973400} 
\begin{split}
&\Delta p_{hn} =\frac{p_{h} \left(0\right)-p_{h} \left(x\right)}{\left({r_{s}^{2} /t^{2} } \right)\left({m_{h} /4\pi r_{s}^{3} } \right)} \\
&=\frac{4\pi ^{2} }{F_{r} \left(1\right)} \int _{0}^{x}\frac{F_{r} \left(x\right)F_{r}^{'} \left(x\right)}{x^{4} } dx -\int _{0}^{x}\frac{F_{r}^{2} \left(x\right)F_{r}^{''} \left(x\right)}{x^{2} \left[F_{r}^{'} \left(x\right)\right]^{2} } dx  
\end{split}
\end{equation} 
and  
\begin{equation} 
\label{ZEqnNum131593} 
\Delta \rho _{hn} =\frac{\rho _{h} \left(0\right)-\rho _{h} \left(x\right)}{\left({m_{h} /4\pi r_{s}^{3} } \right)} =\left[\left. \frac{F_{r}^{'} \left(x\right)}{x^{2} } \right|_{x=0} -\frac{F_{r}^{'} \left(x\right)}{x^{2} } \right].   
\end{equation} 

\begin{figure}
\includegraphics*[width=\columnwidth]{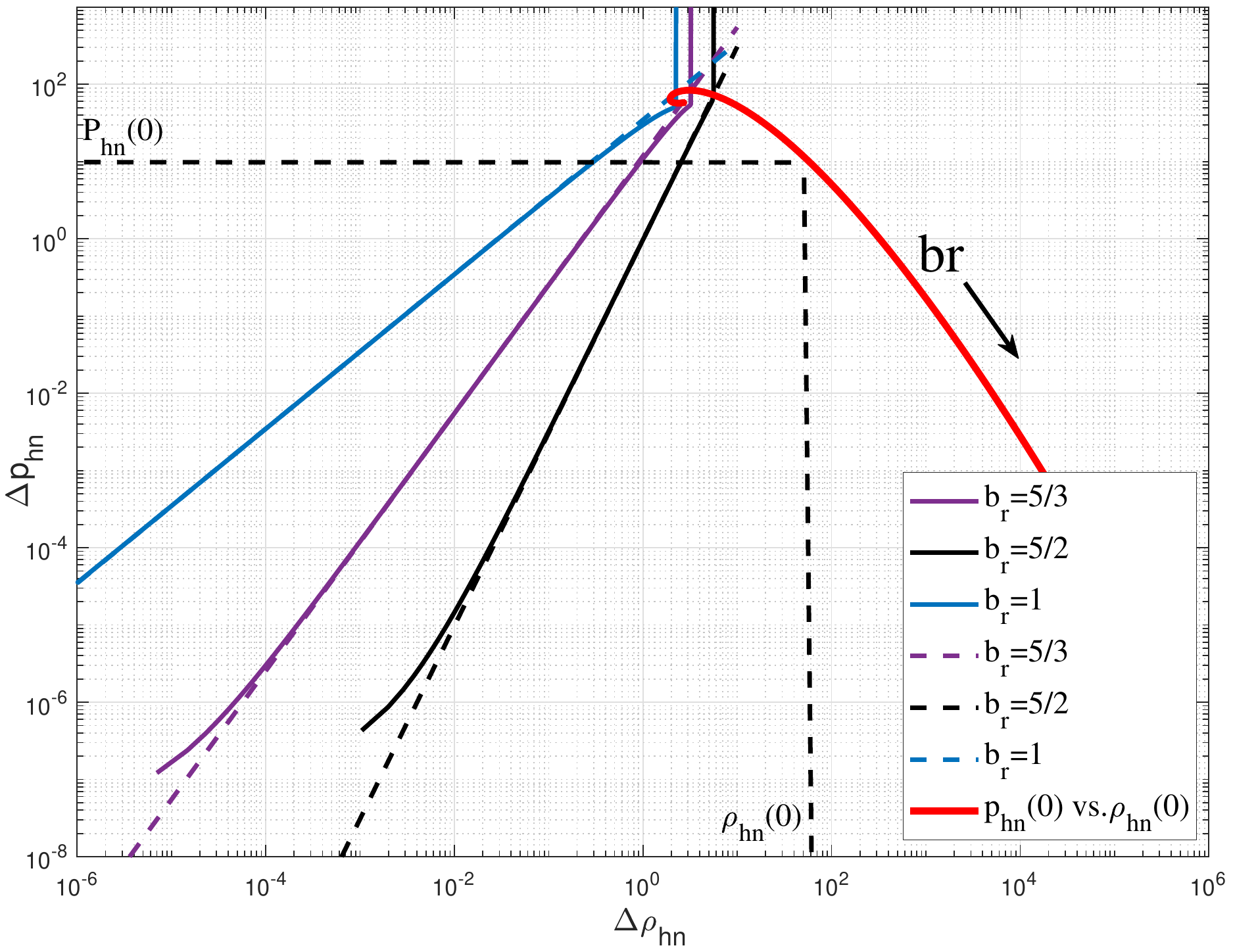}
\caption{The variation of relative pressure $\Delta p_{hn} $ with relative density $\Delta \rho _{hn} $ for different exponent $b_{r}$ using Eqs. \eqref{ZEqnNum973400} and \eqref{ZEqnNum131593}, where function $F_r(x)$ is from Eq. \eqref{ZEqnNum768710}. The proposed model yields an equation of state $\Delta p_{hn} \propto \left(\Delta \rho _{hn} \right)^{b_{r} } $ that can be clearly identified (solid lines). The equation of state \eqref{ZEqnNum805925} is plotted as dash lines for comparison. The relation (Eq. \eqref{ZEqnNum520886}) between center pressure $p_{hn} \left(0\right)$ and center density $\rho _{hn} \left(0\right)$ is presented as the red solid line for different $b_{r} $ (i.e. the maximum $\Delta p_{hn} $ vs. maximum $\Delta \rho _{hn} $). The center velocity dispersion $\sigma _{r0}^{2} $ can be identified accordingly.}
\label{fig:9}
\end{figure}

\noindent More specifically with $F_{r} \left(x\right)$ from Eq. \eqref{ZEqnNum768710} and $a_{r} ={3b_{r} /2} $, equation of state \eqref{ZEqnNum609386} reduces to
\begin{equation} 
\label{ZEqnNum805925} 
\Delta p_{hn} =\frac{2^{3-b_{r} } \pi ^{2} \left(b_{r} \right)^{3b_{r} -2-{3b_{r}^{2} /2} } }{3\gamma \left({3b_{r} /2} ,b_{r} \right)\Gamma \left({3b_{r} /2} \right)^{1-b_{r} } } \left(\Delta \rho _{hn} \right)^{b_{r} } ,       
\end{equation} 
where $\gamma \left(x,y\right)$ is the lower incomplete Gamma functions. 

Next we will derive the pressure, density, and velocity dispersion at halo center. Let's assume the equation of state \eqref{ZEqnNum609386} is valid for the entire range of \textit{x} extending to infinity. By setting $p_{h} \left(\infty \right)=0$ and $\rho _{h} \left(\infty \right)=0$, a simple relation (regardless of the value of $b_{r} $) between center pressure and density can be obtained from Eq. \eqref{ZEqnNum609386},
\begin{equation}
\label{ZEqnNum520886} 
p_{h} \left(0\right)=\frac{v_{cir}^{2}\rho _{h}^{2} \left(0\right)}{2\left(b_{r} \right)^{b_{r} } \bar{\rho }_{h} c^{2} } =\frac{2\pi Gr_{s}^{2} }{3\left(b_{r} \right)^{b_{r} } } \rho _{h}^{2} \left(0\right)=K_{r} \left(a\right)\rho _{h}^{2} \left(0\right),    
\end{equation} 
where the pre-factor $K_{r} \left(a\right)\approx 0.01aG\left(1Mpc/h\right)^{2}$ from simulation data should be independent of halo mass $m_{h} $. At the same redshift \textit{z} (or \textit{a}) and with $r_{s}^{} \propto m_{h}^{{1/3} } $ increasing with halo mass $m_{h} $, $b_{r} $ is also expected to be slightly increasing with halo mass $m_{h} $. On the other hand (from Eqs. \eqref{ZEqnNum520886} and \eqref{ZEqnNum277257}), 
\begin{equation} 
\label{ZEqnNum218090} 
\begin{split}
p_{h} \left(0\right)&=\frac{v_{cir}^{2} }{2\left(b_{r} \right)^{b_{r} } \bar{\rho }_{h} c^{2} } \rho _{h}^{2} \left(0\right)\\
&=\frac{Gm_{h} }{3r_{s} } \frac{\left(b_{r} \right)^{{b_{r} /2} -1} }{\Gamma \left({3b_{r} /2} \right)} \rho _{h} \left(0\right)=\sigma _{r0}^{2} \rho _{h} \left(0\right),
\end{split}
\end{equation} 
and
\begin{equation} 
\label{163} 
\sigma_{r0}^{2} \equiv \sigma _{r}^{2} \left(x=0\right)=\frac{Gm_{h} }{3r_{s} } \frac{\left(b_{r} \right)^{{b_{r} /2} -1} }{\Gamma \left({3b_{r} /2} \right)},
\end{equation} 
where $\sigma _{r0}^{2}$ is the center velocity dispersion and $\sigma _{r0}^{2} \propto m_{h}^{{2/3} } a^{-1} $. For $b_{r} ={5/3} $ (exponent for adiabatic process), $\sigma _{r0}^{2} \approx 26{r_{s}^{2} /t^{2} } $ compared to $\sigma _{r0}^{2} =2\pi ^{2} c^{2} {r_{s}^{2} /t^{2} } $ for isothermal profile (Fig. \ref{fig:4}). With $p_{h} \left(0\right)$ from Eq. \eqref{ZEqnNum520886}, the core size (where Hubble flow is dominant) $x_{c} =\left(b_{r} \right)^{{-b_{r} /2} } $ (in Eq. \eqref{ZEqnNum946518}) can be easily obtained by forcing $p_{h} \left(x\right)=0$ in Eq. \eqref{ZEqnNum131959}. The core size $x_{c} =1$ for $b_{r} =0$ (core size exactly equals scale radius) and decreases with $b_{r} $. For $b_{r} ={5/3} $, $x_{c} \approx 0.65$.  

Figure \ref{fig:9} summarizes the equation of state for relative pressure and density. With cumulative function $F_{r} \left(x\right)$ from Eq. \eqref{ZEqnNum768710}, the variation of relative pressure $\Delta p_{hn} $ with relative density $\Delta \rho _{hn} $ is plotted for different $b_{r} $ using Eqs. \eqref{ZEqnNum973400} and \eqref{ZEqnNum131593}. Clearly, equation of state follows a scaling law $\Delta p_{hn} \propto \left(\Delta \rho _{hn} \right)^{b_{r} } $ for most range of density, while deviation is only observed in the outer halo region with extremely low density. Analytical approximation (dash lines from Eq. \eqref{ZEqnNum805925}) is presented for comparison. The relation between center pressure and density is also plotted as red thick line. 

\begin{figure}
\includegraphics*[width=\columnwidth]{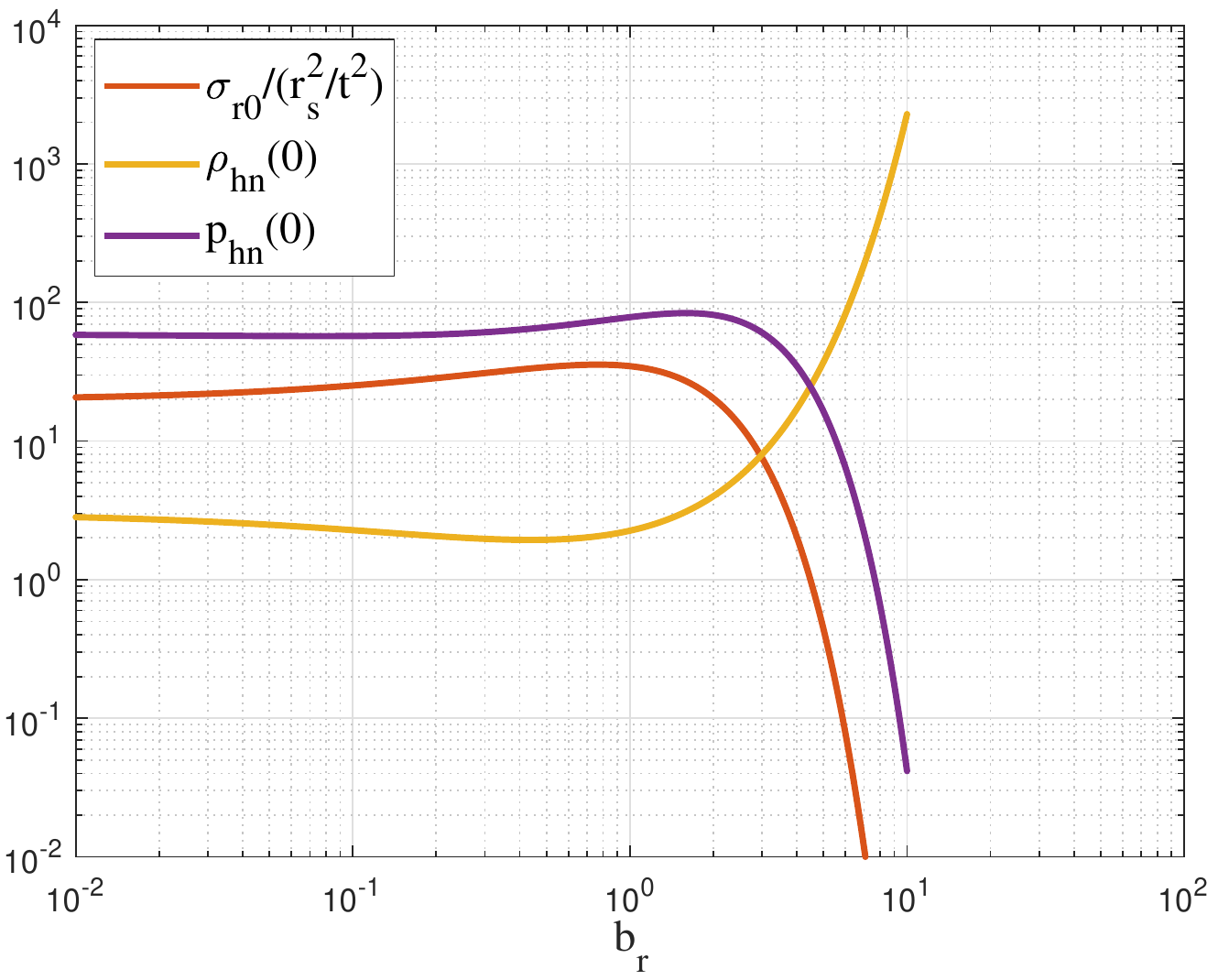}
\caption{The variation of normalized pressure $p_{h} \left(0\right)$ (Eq. \eqref{ZEqnNum218090}), density $\rho _{h} \left(0\right)$ (Eq. \eqref{ZEqnNum277257}), and dispersion ${\sigma_{r0}^{2}/(r_s^2/t^{2}})$ at halo center for different parameter $b_{r} ={2/\alpha } $, where $\alpha $ is the shape parameter in Einasto profile.}
\label{fig:10}
\end{figure}

Figure \ref{fig:10} plots the variation of center density $\rho _{h} \left(0\right)$, center pressure $p_{h} \left(0\right)$, and center dispersion $\sigma _{r0}^{2} $ with exponent $b_{r} $. For $b_{r} \to 0$, the velocity dispersion $\sigma _{r0}^{2} \approx 20{r_{s}^{2} /t^{2} } $. There exist minimum $\rho _{h} \left(0\right)$, maximum $p_{h} \left(0\right)$ and maximum $\sigma _{r0}^{2} $ at certain $b_{r}$.

\section{Conclusions}
\label{sec:5}
The gravitational collapse of dark matter is essentially a nonlinear self-gravitating collisionless fluid flow problem (SG-CFD). The inverse mass cascade is a unique feature of SG-CFD that shares many similarities with the  energy cascade in turbulence. Halos are intrinsically dynamical objects that mediate the mass cascade. This paper focus on the effect of inverse mass cascade on relevant halo properties and internal structures.

The halo internal structure is highly dependent on mass cascade. The continuous mass accretion creates a new layer of mass that deforms the original halo and creates a non-zero mean radial flow (Figs. \ref{fig:1} and \ref{fig:2}, and Eq. \eqref{ZEqnNum849591}, outflow for core region and inflow for outer region). The isothermal density profile is a natural result for halos with infinitely fast mass accretion and vanishing mean radial flow (Eq. \eqref{ZEqnNum278808}). The combined in- and out-flow lead to an extra length scale (the scale radius $r_{s}$) for density profile where the radial flow is at its maximum. A double-power-law density (Eqs. \eqref{ZEqnNum973033} and \eqref{ZEqnNum367971}) is proposed with inner density dominated by the halo deformation rate $\gamma _{h}$ and outer density controlled by a halo deformation parameter $\alpha _{h}$ that is dependent on halo concentration, and mass cascade parameters $\lambda$ and $\tau _{0}$ (Eq. \eqref{ZEqnNum412141}). The cusp-core controversy is related to the deformation rate (the gradient of mean radial flow) $\gamma _{h}$ at halo center. Slower deformation leads to a steeper core density profile (Eq. \eqref{ZEqnNum973033}). For large halos with extremely fast mass accretion and an expanding core, Hubble flow is expected at halo center that hints the existence of a central halo core with $\gamma _{h} ={2/3}$ (Eq. \eqref{ZEqnNum801048}).

The momentum and energy exchange between mean flow and velocity dispersion (random motion) is studied via the Jeans' equation (Eq. \eqref{ZEqnNum535400}) for spherical non-rotating isotropic halos. The radial flow is shown to enhance the radial dispersion in outer region (Fig. \ref{fig:4}). The closure problem for halo density profile can be reduced to the correct modeling of radial flow $u_{h} \left(x\right)$. The halo density profile can be derived for a given Taylor expansion of $u_{h} \left(x\right)$ around $x=0$. The critical halo concentration $c=3.48$ is obtained as a result of vanishing linear radial momentum for large halos (Fig. \ref{fig:3} and Eq. \eqref{ZEqnNum722864}). A complete analysis of the effects of radial flow on various halo energies is also presented (Eqs. \eqref{ZEqnNum811693}-\eqref{ZEqnNum765989}). Halo surface energy and surface tension are introduced for halos with finite size because extra energy is required to create the expanding surface (Eqs. \eqref{ZEqnNum267268} and \eqref{ZEqnNum212204}). An effective exponent $n_{e}$ of gravitational interaction is discovered with an estimate value of $n_{e} =-1.3$ (deviate from the usual exponent -1.0 for gravity) to reflect the effects of mass cascade and surface energy of halos (Eq. \eqref{ZEqnNum797157}). 

Halos are dynamically evolving due to inverse mass cascade. New stochastic models are formulated for the random evolution of halo size that follows a geometric Brownian motion (Eq. \eqref{ZEqnNum251003}). As a result, halo size follows a lognormal distribution (Eq. \eqref{ZEqnNum640133}). Stochastic models are also developed for the random motion of collisionless particles in halos with random size (Eqs. \eqref{ZEqnNum544038} and \eqref{ZEqnNum855118}). This model involves drift terms including both mean radial and osmotic flow ($u_{h} $ and $u_{h}^{*} $) and a multiplicative noise term due to the random halo size. The solution of that model leads to the relation between particle density distribution $P_{r} $ (the probability to find a particle at a given position) and the mean radial and osmotic flow (Eqs. \eqref{ZEqnNum533047} and \eqref{ZEqnNum961031}). It is demonstrated that the closure problem of halo density profile can be equivalently reduced to the correct modeling of either mean flow $u_{h} $, or osmotic flow $u_{h}^{*}$, or identifying an additional closure between $u_{h}$ and $u_{h}^{*}$ besides Eq. \eqref{ZEqnNum687681}. A simple closure between $u_{h} $ and $u_{h}^{*} $ is proposed for a self-consistent particle distribution function in Appendix \ref{appendix:b} (Eq. \eqref{eq:B7}).

In this work, a simple model of osmotic flow $u_{h}^{*} $ (Eq. \eqref{ZEqnNum927116}) is proposed such that the radial particle distribution function $P_{r} $ can be fully derived (Eq. \eqref{ZEqnNum103079}), as well as other relevant halo properties, including Eq. \eqref{ZEqnNum495681} for particle density distribution, Eq. \eqref{ZEqnNum980800} for radial flow, and Eq.\eqref{ZEqnNum105452} for shifted potential. The proposed model provides an excellent fit to the cumulative function of particle density $F_{r}$ that is computed for composite halos (group of halos of same sizes) from a N-body simulation. The model agrees with a very wide range of halo group sizes, where a central halo core exists with $F_{r} \left(x\right)\sim x^{3} $ due to the Hubble radial flow at halo center (Fig. \ref{fig:6}). With reference pressure and density defined at halo center where both gravitational and pressure forces are absent, equation of state for relative pressure and density is established based on this model (Fig. \ref{fig:9} and Eqs. \eqref{ZEqnNum609386} and \eqref{ZEqnNum805925}). The pressure, density, and velocity dispersion at halo center are also presented (Fig. \ref{fig:10} and Eqs. \eqref{ZEqnNum520886} and \eqref{ZEqnNum218090}).   

In short, inverse mass cascade is a fundamental feature of SG-CFD. Its effects on the structure formation and evolution remain an important topic. Some examples of future work are briefly discussed here. The concentration-mass relation (the mass dependence of \textit{c}) might be related to the mass dependence of $\alpha _{h}$ and/or $\lambda$, with Eq. \eqref{ZEqnNum412141} providing a relation between concentration $c$, deformation parameter $\alpha _{h} $ and mass cascade parameter $\lambda $ and $\tau _{0} $. Further study is also desired to identify a better closure for better understanding of the origin of universal halo structures.
 

\section*{Data Availability}
The data underlying this article are available on Zenodo \citep{Xu:2022-Dark_matter-flow-dataset-part1,Xu:2022-Dark_matter-flow-dataset-part2,Xu:2022-Dark_matter-flow-and-hydrodynamic-turbulence-presentation}. All data files are also available on GitHub \citep{Xu:Dark_matter_flow_dataset_2022_all_files}.

\bibliographystyle{mnras}
\bibliography{Papers}

\appendix
\section{Halo density profiles}
\label{appendix:a}
\subsection{The NFW density profile}
\label{sec:A1}
Although halos growing is a complex, hierarchical, and nonlinear process, the radial density profile $\rho _{h} \left(r\right)$ of halos can be robustly fitted by a simple double logarithmic function from cosmological \textit{N}-body simulations, e.g. NFW profile \citep{Navarro:1997-A-universal-density-profile-fr},
\begin{equation} 
\label{ZEqnNum832248} 
\rho _{h} \left(r\right)=\frac{m_{h} }{4\pi r^{3} } \frac{1}{F\left(c\right)\left(1+{r_{s} /r} \right)^{2} } =\frac{\bar{\rho }_{r} \left(r\right)}{3F\left({r_{s} /r} \right)\left(1+{r_{s} /r} \right)^{2} } . 
\end{equation} 
The logarithm slope of NFW density profile is
\begin{equation} 
\label{ZEqnNum658058} 
\frac{d\ln \rho _{h} }{d\ln r} =-\frac{1+3{r/r_{s} } }{1+{r/r_{s} } } ,          
\end{equation} 
where $r$ is the distance to center of halo, $m_{h} =m_{r} \left(r=r_{h} \right)$ is the total mass of a given halo. Here $m_{r} \left(r\right)$ is the total mass enclosed within radius \textit{r,} $\bar{\rho }_{r} \left(r\right)={m_{r} \left(r\right)/\left({4\pi r^{3} /3} \right)} $ is the mean density within radius \textit{r}, and $r_{h} $ is the virial radius in physical coordinates. The scale radius $r_{s} $ is defined as the radius where the density profile $\rho _{h} \left(r\right)$ changes its logarithmic slope from -1 for ${r/r} _{s} \ll 1$ to -3 for ${r/r} _{s} \gg 1$. The logarithmic slope is exactly -2 at the scale radius $r_{s} $. The halo concentration parameter $c={r_{h} /r_{s} } $ is a key ratio between the halo size and scale radius that reflects the halo structure. The function $F\left(x\right)$ can be found for NFW profile using Eq. \eqref{ZEqnNum632204},
\begin{equation} 
\label{ZEqnNum588557} 
F\left(x\right)=\ln \left(1+x\right)-\frac{x}{1+x} .         
\end{equation} 
The halo mass $m_{r} \left(r\right)$ within radius $r$ can be obtained by the integration of density,
\begin{equation} 
\label{eq:A4} 
m_{r} \left(r\right)=\int _{0}^{r}\rho _{h} \left(y\right)4\pi y^{2} dy= m_{h} \frac{F\left({r/r_{s} } \right)}{F\left({r_{h} /r_{s} } \right)} .        
\end{equation} 
The density at surface of halo $r=r_{h} $ can be obtained from Eq. \eqref{ZEqnNum832248},
\begin{equation} 
\label{ZEqnNum964954} 
\rho _{h} \left(r=r_{h} \right)=\frac{m_{h} }{4\pi r_{h}^{3} } \frac{1}{F\left(c\right)\left(1+{1/c} \right)^{2} } =\frac{\bar{\rho }_{h} }{3F\left(c\right)\left(1+{1/c} \right)^{2} } ,   
\end{equation} 
where $\bar{\rho }_{h} $ is the mean density of the entire halo. 

\subsection{The power-law density profile}
\label{sec:A2}
For comparison, another commonly used model is a power-law density $\rho _{h} \left(r\right)\sim r^{-m} $ with $m=2$ for isothermal density profile. We have
\begin{equation} 
\label{ZEqnNum859547} 
\rho _{h} \left(r\right)=\frac{\left(3-m\right)m_{h} }{4\pi r_{h}^{3} } \left(\frac{r}{r_{h} } \right)^{-m} =\frac{\left(3-m\right)}{3} \bar{\rho }_{r} \left(r\right), 
\end{equation} 
and the density at halo surface is
\begin{equation} 
\label{ZEqnNum789386} 
\rho _{h} \left(r=r_{h} \right)=\frac{\left(3-m\right)m_{h} }{4\pi r_{h}^{3} } ,         
\end{equation} 
where $m_{r} \left(r\right)=m_{h} \left({r/r_{h} } \right)^{3-m} $. For a power-law density, the halo density at radius $r$ is fully determined by the mean density $\bar{\rho }_{r} \left(r\right)$ of a sphere of radius $r$ ($\rho _{h} \left(r\right)\propto \bar{\rho }_{r} \left(r\right)$ from Eq. \eqref{ZEqnNum859547}), which is different from the NFW profile (See Eq. \eqref{ZEqnNum832248}). This difference reflects the effect of mass cascade on the density profiles. 

\subsection{The Einasto density profile}
\label{sec:A3}
The third popular density profile was first introduced by Einasto to describe the distribution of stars in Milky way \citep{Einasto:1984-Structure-of-Superclusters-and}, 
\begin{equation}
\rho _{h} \left(r\right)=\rho _{s} e^{{2/\alpha } } \exp \left[-\frac{2}{\alpha } \left(\frac{r}{r_{s} } \right)^{\alpha } \right], \quad \frac{d\ln \rho _{h} }{d\ln r} =-2\left(\frac{r}{r_{s} } \right)^{\alpha },
\label{ZEqnNum701176}
\end{equation}

\noindent where $r_{s} $ is the scale radius (same as NFW) that is defined as the location where the logarithmic slope is -2. The density $\rho _{s} $ is defined as the halo density at scale radius $r_{s} $. The shape parameter $\alpha $ is the exponent of the logarithmic slope in Eq. \eqref{ZEqnNum701176}. The mass $m_{r} \left(r\right)$ can be obtained by integration of density, 
\begin{equation} 
\label{eq:A9} 
m_{r} \left(r\right)=4\pi \rho _{s} r_{s}^{3} \frac{e^{{2/\alpha } } }{\alpha } \left(\frac{\alpha }{2} \right)^{{3/\alpha } } \left[\Gamma \left(\frac{3}{\alpha } \right)-\Gamma \left(\frac{3}{\alpha } ,\frac{2}{\alpha } \left(\frac{r}{r_{s} } \right)^{\alpha } \right)\right],     
\end{equation} 
or equivalently in terms of the total halo mass $m_{h} $, 
\begin{equation} 
\label{ZEqnNum807556} 
m_{r} \left(r\right)=m_{h} \frac{\Gamma \left({3/\alpha } \right)-\Gamma \left({3/\alpha } ,{2\left({r/r_{s} } \right)^{\alpha } /\alpha } \right)}{\Gamma \left({3/\alpha } \right)-\Gamma \left({3/\alpha } ,{2c^{\alpha } /\alpha } \right)} ,       
\end{equation} 
where $\Gamma \left(x,y\right)$ is a upper incomplete gamma function. The Einasto density profile can be equivalently expressed as,
\begin{equation} 
\label{ZEqnNum305317} 
\begin{split}
\rho _{h} \left(r\right)&=\frac{m_{h} }{4\pi r_{h}^{3} } \frac{c^{3} \exp \left[{-2\left({r/r_{s} } \right)^{\alpha } /\alpha } \right]\alpha \left({2/\alpha } \right)^{{3/\alpha } } }{\Gamma \left({3/\alpha } \right)-\Gamma \left({3/\alpha } ,{2c^{\alpha } /\alpha } \right)} \\
&=\frac{\bar{\rho }_{h} \left(r\right)r^{3} }{3r_{s}^{3} } \frac{\exp \left[{-2\left({r/r_{s} } \right)^{\alpha } /\alpha } \right]\alpha \left({2/\alpha } \right)^{{3/\alpha } } }{\Gamma \left({3/\alpha } \right)-\Gamma \left({3/\alpha } ,{2\left({r/r_{s} } \right)^{\alpha }/\alpha } \right)} .
\end{split}
\end{equation} 
The density at halo surface is 
\begin{equation} 
\label{ZEqnNum166295} 
\rho _{h} \left(r_{h} \right)=\frac{m_{h} }{4\pi r_{h}^{3} } \frac{c^{3} \exp \left({-2c^{\alpha } /\alpha } \right)\alpha \left({2/\alpha } \right)^{{3/\alpha } } }{\Gamma \left({3/\alpha } \right)-\Gamma \left({3/\alpha } ,{2c^{\alpha } /\alpha } \right)} .       
\end{equation} 

Same as NFW profile, the halo density at radius \textit{r} cannot be fully determined by the mean density $\bar{\rho }_{r} \left(r\right)$ in sphere of radius $r$ (Eq. \eqref{ZEqnNum305317}), which reflects the effect of mass cascade. The Einasto profile has three free parameters ($\rho _{s} $, $r_{s} $, and $\alpha $) compared to the NFW profile ($r_{s} $ and $c$) with two free parameters. However, if density at halo surface is required to be the same for two different profiles (Eqs. \eqref{ZEqnNum964954} and \eqref{ZEqnNum166295}), we will have an additional implicit relation between shape parameter $\alpha $ and halo concentration $c$ (Fig. \ref{fig:S1}),
\begin{equation} 
\label{ZEqnNum720481} 
c\left(\ln \left(1+c\right)-\frac{c}{1+c} \right)\left(1+c\right)^{2} =\frac{\Gamma \left({3/\alpha } \right)-\Gamma \left({3/\alpha } ,{2c^{\alpha } /\alpha } \right)}{\exp \left({-2c^{\alpha } /\alpha } \right)\alpha \left({2/\alpha } \right)^{{3/\alpha } } } .     
\end{equation} 

\section{A simple closure for self-consistent particle distribution}
\label{appendix:b}
The inverse mass cascade leads to the random variation of halo size that gives rise to the "diffusion" motion of collisionless particles (Eqs. \eqref{ZEqnNum544038}  and \eqref{ZEqnNum855118}). The osmotic velocity is the velocity acquired by particles to balance the external force and can be related to the flux of ``diffusion''. A simple model of the osmotic velocity $u_{h}^{*} \left(x\right)$ was proposed (Eq. \eqref{ZEqnNum927116}) to derive the particle probability function $P_{r}^{} \left(x\right)$ and halo density profile (Eqs. \eqref{ZEqnNum230459} and \eqref{ZEqnNum495681}). The relation between radial flow $u_{h}^{} \left(x\right)$ and osmotic velocity $u_{h}^{*} \left(x\right)$ is also presented in Eq. \eqref{ZEqnNum687681}. The distribution function $P_{r}^{} \left(x\right)$ is fully determined if an additional closure can be introduced between $u_{h}^{} \left(x\right)$ and $u_{h}^{*} \left(x\right)$, which is the focus of this section. 

In standard Brownian motion, a spherical particle of radius $a_{B} $ moving at a constant velocity $u_{h}^{} $ in a fluid of viscosity $\eta _{B} $ subject to a force $F_{B} $ can be described by the Stokes' law. Therefore, the local steady-state velocity $u_{h}^{} $ can be fully determined by the driving force $F_{B} $, i.e. the gradient of osmotic pressure $\Pi _{B} =\rho _{B} k_{B} T$ ($k_{B} $ is the Boltzmann constant and T is temperature), which is a localized short-range force. The current velocity $u_{h}^{}$ (from Stokes' law) and osmotic velocity $u_{h}^{*} $ can be written as,
\begin{equation} 
\label{ZEqnNum173816} 
\begin{split}
&u_{h} =\frac{F_{B} }{6\pi \eta _{B} a_{B} } =-\frac{1}{6\pi \eta _{B} a_{B} } \cdot \frac{1}{\rho _{B} } \frac{\partial \Pi _{B} }{\partial x} =-\frac{\mu _{B} }{\rho _{B} } \frac{\partial \Pi _{B} }{\partial x}\\ 
&\textrm{and}\\ 
&u_{h}^{*} =D_{B} \frac{\partial \ln \rho _{B} }{\partial x} ,    
\end{split}
\end{equation} 
where $\rho _{B} $ is the particle number density, $D_{B} $ is the particle diffusivity, and $\mu _{B} $ is the particle mobility. The stochastic equations for Brownian motion (forward and backward) read
\begin{equation} 
\label{eq:B2} 
\frac{dr_{t} }{dt} =\left[u_{h} \left(x_{t} \right)+u_{h}^{*} \left(x_{t} \right)\right]+\sqrt{2D_{B} } \xi \left(t\right),        
\end{equation} 
\begin{equation} 
\label{eq:B3} 
\frac{dr_{t} }{dt} =\left[u_{h} \left(x_{t} \right)-u_{h}^{*} \left(x_{t} \right)\right]+\sqrt{2D_{B} } \xi _{}^{*} \left(t\right).       
\end{equation} 
The corresponding Fokker-Planck equations read 
\begin{equation} 
\label{eq:B4} 
\frac{\partial P_{r} \left(x,t\right)}{\partial t} =-\frac{\partial }{\partial x} \left[\left(u_{h} \left(x\right)+u_{h}^{*} \left(x\right)\right)P_{r} \right]+D_{B} \frac{\partial ^{2} P_{r} }{\partial x^{2} } ,      
\end{equation} 
\begin{equation} 
\label{eq:B5} 
\frac{\partial P_{r} \left(x,t\right)}{\partial t} =-\frac{\partial }{\partial x} \left[\left(u_{h} \left(x\right)-u_{h}^{*} \left(x\right)\right)P_{r} \right]-D_{B} \frac{\partial ^{2} P_{r} }{\partial x^{2} } .      
\end{equation} 
The simple closure $u_{h}^{} =-u_{h}^{*} $ for Brownian motion is well known as the flux due to applied force $F_{B} $ must balance the diffusive flux. The Einstein relation $D_{B} =\mu _{B} k_{B} T$ is a direct result of this closure (Eq. \eqref{ZEqnNum173816}). For Brownian motion with closure $u_{h}^{} =-u_{h}^{*} $, the diffusion equation for particle probability $P_{r}^{} $ can be directly derived from Fokker-Planck equation (Eqs. \eqref{eq:B4} and \eqref{eq:B5}). 

However, it is different and much more complicated for halos formed in dark matter flow (SG-CFD). In contrast to Brownian motion, the osmotic velocity $u_{h}^{*} \left(r\right)$ may not be fully determined by the local current velocity $u_{h} \left(r\right)$ at location \textit{r} due to the long-range and non-local nature of the gravitational force. Let's first derive the radial linear momentum within an arbitrary radius \textit{r} for a given unknown function $F\left(x\right)$. With expressions of $u_{r} $ (from Eq. \eqref{ZEqnNum278808} and $\rho _{h} $ (from Eq. \eqref{ZEqnNum482501}), this can be obtained as,
\begin{equation} 
\label{ZEqnNum907656} 
\begin{split}
L_{hr} \left(r\right)&=\int _{0}^{r}u_{r} \left(r_{1} \right)4\pi r_{1}^{2} \rho _{h} \left(r_{1} ,a\right)dr_{1}  \\
&=\frac{m_{h} r_{s} }{tF\left(c\right)} \left(xF\left(x\right)-2\int _{0}^{x}F\left(x\right)dx \right).  
\end{split}
\end{equation} 
Due to the long-range interaction, the current velocity $u_{r} \left(r_{1} \right)$ of every single spherical shell with radius $r_{1} <r$ should contribute to the osmotic velocity $u_{h}^{*} \left(r\right)$ at radius \textit{r}. Therefore, $u_{h}^{*} \left(r\right)$ is proposed to be proportional to the mean radial velocity within a sphere of radius $r$,
\begin{equation} 
\label{eq:B7} 
u_{h}^{*} \frac{r_{s} }{t} =\lambda _{h} \frac{L_{hr} \left(r\right)}{m_{r} \left(r\right)} =\frac{\lambda _{h} }{m_{r} \left(r\right)} \int _{0}^{r}u_{r} \left(r_{1} \right)4\pi r_{1}^{2} \rho _{h} \left(r_{1} ,a\right)dr_{1}  ,      
\end{equation} 
where $\lambda _{h} $ is a proportional constant and $m_{r} \left(r\right)$ is the halo mass within radius \textit{r} (Eq. \eqref{ZEqnNum632204}). The final expression for osmotic velocity reads (using Eq. \eqref{ZEqnNum907656}), 
\begin{equation} 
\label{ZEqnNum947079} 
u_{h}^{*} \left(x\right)=\lambda _{h} \left[x-\frac{2}{F\left(x\right)} \int _{0}^{x}F\left(y\right)dy \right].       \end{equation} 
A third order differential equation for $F\left(x\right)$ can be obtained by combining two closures Eqs. \eqref{ZEqnNum947079} and \eqref{ZEqnNum687681},
\begin{equation} 
\label{ZEqnNum174945} 
\begin{split}
F^{'''} \left(x\right)&+\frac{F^{''} \left(x\right)F^{'} \left(x\right)}{F\left(x\right)} +\frac{2}{x} F^{''} \left(x\right)-\frac{\left[F^{''} \left(x\right)\right]^{2} }{F^{'} \left(x\right)}\\
& +\left(2+\frac{1}{d_{h} } \right)\frac{F^{'} \left(x\right)}{x^{2} } +\left(2-\frac{1}{d_{h} } \right)\frac{\left[F^{'} \left(x\right)\right]^{2} }{xF\left(x\right)} =0, 
\end{split}
\end{equation} 
where $\sigma \left(x\right)=x^{\lambda _{r} } =x$ is used with $\lambda _{r} =1$. Parameter $d_{h} ={d_{r} /\lambda _{h} } $ lumps $d_{r} $ and $\lambda _{h} $ together. The associated boundary conditions are
\begin{equation} 
\label{eq:B10} 
F\left(0\right)=0, \quad \textrm{and}\quad F\left(\infty \right)=1,         
\end{equation} 
\begin{equation}
F^{''} \left(1\right)=0,\quad \textrm{and} \quad {\mathop{\lim }\limits_{x\to 0}} \frac{\partial u_{h} }{\partial x} ={\mathop{\lim }\limits_{x\to 0}} \frac{F\left(x\right)F^{''} \left(x\right)}{\left[F^{'} \left(x\right)\right]^{2} } =\gamma _{h}.
\label{eq:B11}
\end{equation}
\noindent Complete solution of $F\left(x\right)$ from Eq. \eqref{ZEqnNum174945} gives rise to the halo density profile and all other relevant quantities. By introducing a set of new variables $y_{1} $, $y_{2} $, and $x_{1} $, where
\begin{equation}
y_{1} =\ln F\left(x\right), \quad y_{2} =\ln F^{'} \left(x\right),\quad \textrm{and} \quad x_{1} =\ln x,      
\label{eq:B12}
\end{equation}

\noindent the original Eq. \eqref{ZEqnNum174945} can be equivalently reduced to two coupled equations
\begin{equation} 
\label{ZEqnNum715788} 
\frac{\partial ^{2} y_{2} }{\partial x_{1}^{2} } +\frac{\partial y_{2} }{\partial x_{1}^{} } \left(\frac{\partial y_{1} }{\partial x_{1}^{} } +1\right)+\left(2-\frac{1}{d_{h} } \right)\frac{\partial y_{1} }{\partial x_{1}^{} } +\left(2+\frac{1}{d_{h} } \right)=0,      
\end{equation} 
\begin{equation} 
\label{eq:B14} 
\frac{\partial y_{1} }{\partial x_{1}^{} } =\exp \left(y_{2} -y_{1} +x_{1} \right),         
\end{equation} 
with corresponding boundary conditions,
\begin{equation}
\left. \frac{\partial y_{1} }{\partial x_{1}^{} } \right|_{-\infty } =\frac{1}{1-\gamma _{h} } \quad \textrm{and} \quad \left. y_{1} \right|_{\infty } =0,      
\label{eq:B15}
\end{equation}
\begin{equation}
\left. \frac{\partial y_{2} }{\partial x_{1}^{} } \right|_{-\infty } =\frac{\gamma _{h} }{1-\gamma _{h} } \quad \textrm{and} \quad \left. \frac{\partial y_{2} }{\partial x_{1}^{} } \right|_{0} =0.     
\label{eq:B16}
\end{equation}
\noindent For small $x_{1} \to 0$ with power-law solution $F\left(x\right)\propto x^{n_{1} } $,  
\begin{equation}
\begin{split}
&\frac{\partial y_{2} }{\partial x_{1}^{} } \equiv \frac{\partial y_{1} }{\partial x_{1}^{} } -1, \quad \frac{\partial ^{2} y_{2} }{\partial x_{1}^{2} } =0,\\
&\text{and}\quad n_{1} =\frac{1}{2} \left[\left(\frac{1}{d_{h} } -2\right)\pm \sqrt{\frac{1}{d_{h} } \left(\frac{1}{d_{h} } -8\right)} \right],   
\end{split}
\label{eq:B17}
\end{equation}

\noindent where we have $d_{h} ={1/8} $ for $n_{1} =3$ (Table \ref{tab:4}), which corresponds to density profile with a central core. For large $x_{1} \to \infty $ with $F\left(x\right)\to 1$ or $y_{1} \to 0$ and ${\partial y_{1} /\partial x_{1} } \ll 1$, the simplified equation and a power-law solution of $F^{'} \left(x\right)\propto x^{n_{2} } $ can be obtained (from Eq. \eqref{ZEqnNum715788}),
\begin{equation}
\begin{split}
\frac{\partial ^{2} y_{2} }{\partial x_{1}^{2} } +\frac{\partial y_{2} }{\partial x_{1}^{} } +\left(2+\frac{1}{d_{h} } \right)=0 \quad \textrm{and} \quad n_{2} =-\left(2+\frac{1}{d_{h} } \right).     
\end{split}
\label{eq:B18}
\end{equation}
\noindent With $d_{h} ={1/8} $,  $F^{'} \left(x\right)\propto x^{-10} $ for $x_{1} \to \infty $. 

More study is required to identify other possible non-local closures between $u_{h}^{} \left(x\right)$ and $u_{h}^{*} \left(x\right)$. The other option is to use the equation of state for relative pressure and density as a simple closure. For virialized halos with vanishing radial flow (no term 2 in Eq. (\ref{eq:61})), the hydrostatic equilibrium equation 
\begin{equation} 
\label{eq:B19} 
\frac{d}{dr} \left(\frac{r^{2} }{\rho _{h} } \frac{dp_{h} }{dr} \right)=-4\pi Gr^{2} \rho _{h} \left(r\right) 
\end{equation} 
can be used to relate the pressure to density. Assuming equation of state \eqref{ZEqnNum609386} is valid for entire virialized halo and inserting it into the hydrostatic equilibrium equation,
\begin{equation} 
\label{eq:B20} 
K_{s} b_{r} \frac{d}{dr} \left(r^{2} \frac{\left(\rho _{h} \left(0\right)-\rho _{h} \right)^{b_{r} -1} }{\rho _{h} } \frac{d\rho _{h} }{dr} \right)=-4\pi Gr^{2} \rho _{h} .      
\end{equation} 
This model leads to an isothermal density profile $\rho _{h} \propto r^{-2} $ for large \textit{r} with $\rho _{h} \left(r\right)\to 0$ and an Einasto profile for small \textit{r} with a central core. Equation \eqref{ZEqnNum805925} may also be modified with a density dependent exponent of equation of state,
\begin{equation} 
\label{eq:B21} 
\Delta p_{hn} =K_{1} \left(\Delta \rho _{hn} \right)^{b_{r} \left(\frac{\rho _{h} \left(x\right)}{\rho _{h} \left(0\right)} \right)^{n} } .         
\end{equation} 
With this closure and Eqs. \eqref{ZEqnNum973400} and \eqref{ZEqnNum131593} for $\Delta p_{hn} $ and $\Delta \rho _{hn} $, the unknow function $F\left(x\right)$ can be fully determined. Further study is needed for a self-consistent particle distribution function that will provide fundamental understanding of halo internal structures. 

\label{lastpage}
\end{document}